\documentclass[12pt]{iopart}
\usepackage{epsfig,rotate}
\usepackage{graphicx}
\newcommand{\nab}{\mbox{\boldmath $\nabla$}}
\newcommand{\tim}{\mbox{\boldmath $\times$}}

\newcommand{\ls}{\raisebox{-.8ex}{$\buildrel{\textstyle<}\over\sim$}}

\usepackage{epsfig}

\begin{document}

\title[Planet formation and migration] {Planet formation and migration}

\author{John C B Papaloizou \dag \ddag   
\; and Caroline Terquem \S $\parallel$ \P
\footnote[6]{To
whom correspondence should be addressed (terquem@iap.fr)}
}

\address{\dag\ Astronomy Unit, School of Mathematical Sciences, Queen
Mary, University of London, Mile End Road, London E1 4NS, UK}

\address{\ddag  Department of applied Mathematics and Theoretical Physics,
Centre for Mathematical Sciences, Wilberforce Road, Cambridge CB3 0WA, UK}

\address{\S Institut d'Astrophysique de Paris, UMR7095 CNRS,
Universit\'e Pierre \& Marie Curie--Paris~6, 98bis boulevard Arago,
75014 Paris, France}

\address{$\parallel$ Universit\'e Denis Diderot--Paris~7, 2 Place
Jussieu, 75251 Paris Cedex 5, France}

\address{\P Institut Universitaire de France}

\begin{abstract}
We review the observations of extrasolar planets, ongoing developments
in theories of planet formation, orbital migration, and the evolution
of multiplanet systems.

\end{abstract}



\maketitle

\section{Introduction}

At the time of the writing of this review, the tenth anniversary of
the discovery of the first extrasolar planet around a solar--type star
is being celebrated.  When this detection was announced, it came as a
shock, not because an extrasolar planet was at last found around a
star like our Sun --- nobody really had doubts that such a detection
would occur one day or another --- but because of the very short
period of this planet, 51~Pegasi.  It orbits its parent star in only
4~days!  It is 100 times closer to the star than Jupiter is from the
Sun.  The temperatures at these distances from the star are so high
that the material needed to assemble the core of such planets cannot
be found in a solid phase there.  The implication is that the planet
formed further out and moved in.  It did not take long for the
theorists to propose an explanation: tidal interactions between the
planet and the disk in which it formed led to orbital migration.  The
theory of such processes had already been developed, about 15~years
before the discovery of 51~Pegasi, and orbital migration of planets
had been predicted.  The real surprise, then, was that nobody ever
predicted that planets might be found on very short period orbits.

As of today, about 150 planets have been detected.  Their orbital
chatacteristics suggest that planet/disk tidal interactions is very
important in determining the characteristics of planetary systems.
This review is divided in four sections.  The first section is devoted
to the observations of extrasolar planets.  We first present the
different methods used to detect planets, and review the statistical
properties of the systems observed so far.  In the second section, we
review the theories of planet formation.  We describe the different
stages that lead to the build--up of a terrestrial planet or a
planetary core (grain sedimentation, obtention of planetesimals,
runaway accumulation), and the process by which an enveloppe of gas
may be captured to produce a giant planet.  We also discuss giant
planet formation by gravitational instabilities.  The third section is
devoted to disk/planet interactions and the different types of orbital
migration.  Finally, in the fourth section, we review some results
about multiplanet systems and their interactions.

\section{Properties of exoplanet systems}

The first object outside our solar system with a mass in the range of
planetary masses was detected in 1992 around the millisecond pulsar
PSR1257+12a (Wolszczan \& Frail 1992).  Three planets of 0.02, 4.3
and 3.9 Earth masses orbiting at 0.19, 0.36 and 0.46 astronomical units
(AU) from the pulsar, respectively, were actually found.  Another
object of about 100 Earth masses with a separation of 40~AU was
reported more recently (Joshi \& Rasio 1997, Wolszczan 1996).
In 1993, Backer \etal announced the detection of a 2.5 Jupiter mass
(M$_{\rm J}$) planet orbiting the millisecond pulsar 1620--26 at a
distance of 23~AU.  This pulsar is in a binary system with a white
dwarf companion (Thorsett \etal 1999) and a separation of about 40~AU.
It is not clear whether these planets formed before the explosion of
the parent supernova or after, either from the material which was
ejected during the explosion or, in the case of the pulsar 1620--26,
in the disk of material transferred from the white dwarf.

The first extrasolar planet around a sun--like star (51~Pegasi~b) was
reported in 1995 by Mayor \& Queloz.  It was detected from
Haute--Provence Observatory, and immediately confirmed by Marcy \&
Butler (1995), who observed it from Lick Observatory (see the review
by Marcy~\& Butler~1998 for more details about the detection of the
first eight extrasolar planets) .  Since then, there has been an
uninterrupted flow of detections.  As of today (16 March 2005), 152
planets in 134 planetary systems (including 14 multiple planet
systems) have been reported around solar type stars.  The lightest
object has a mass of 0.042~M$_{\rm J}$.  We will focus here on the
objects with masses smaller than 13~M$_{\rm J}$, as heavier objects
are usually considered to be brown dwarfs.  This review is concerned
with extrasolar planets around solar type stars, so we will not
discuss further planets around pulsars.

\subsection{Detection methods}

\subsubsection{Radial velocity: \/ \\ \\}

So far, almost all the planets have been detected by this technique,
which consists in measuring the Doppler shift due to the motion of the
star around the center of mass of the system.  The velocity of the
star projected along the line of sight (or {\em radial velocity}) is,
to within a constant, $v=K  f(t)$, where  the function 
$f(t)$  varies periodically
with time $t$ around the orbit  and has zero mean  and

\begin{equation}
K= \left(\frac{2 \pi G }{T} \right)^{1/3} \frac{M_p \sin i}
{\left( M_p + M_\star \right)^{2/3}} \frac{1}{\sqrt{1-e^2}}.
\label{eqK}
\end{equation}

\noindent Here $G$ is the constant of gravitation, $T$ is the orbital
period, $M_\star$ is the mass of the star, $M_p$ is the planet mass,
$i$ is the angle of the line of sight with respect to the
perpendicular to the orbital plane and $e$ is the eccentricity of the
orbit.  The observation of $v$ as a function of time gives $T$
directly.  If the orbit is circular, $f(t)$ is a sinusoid with unit amplitude.
For non circular orbits the amplitude can reach $1+e$
 and the departure from a sinusoid 
allows determination of the eccentricity
$e$.  Given $T$ and $e$, one can deduce $M_p \sin i$ from the
measurement of $K$ ($M_\star$ being known).  Note that since $M_p \ll
M_\star$, the semi--major axis $a$ can be calculated from $T$.

Since $i$ is unknown, only the lower limit $M_p \sin i$ for the planet
mass, or {\em projected mass}, can be obtained.  This was of course an
important matter of debate when the very first planet was detected.
However, for a population of about 150 objects, the ratio of the real
mass to the projected mass is on average on the order of unity. The
expression for $K$ shows that only relatively massive planets with
short periods can be detected with the radial velocity method.  Using
$M_p \ll M_\star$, we can rewrite $K$ and $T$ in the form:

\begin{equation}
K \; ({\rm m \; s}^{-1}) = 28.4 \left( \frac{T}{1 {\rm \; yr}}\right)^{-1/3}
\left( \frac{M_p \sin i}{{\rm M}_{\rm J}} \right) \left(
\frac{M_\star}{{\rm M}_\odot} \right)^{-2/3} \frac{1}{\sqrt{1-e^2}} ,
\end{equation}

\begin{equation}
T \; ({\rm yr}) = \left( \frac{a}{1 {\rm \; AU}} \right)^{3/2}
\left( \frac{M_\star}{{\rm M}_\odot} \right)^{-1/2}.
\end{equation}

The detection limit of the instruments that are currently being used
is about 1~m~s$^{-1}$.  In principle, Jupiter mass planets could be
detected rather far from their parent star.  However, the detection is
confirmed only if the planet if monitored for at least one orbital
period.  Given that surveys began about 10~years ago, only planets
with semi--major axes less than about 4--5~AU can be observed.  Note
that a Jupiter mass planet with $a=5$~AU around a solar mass star
induces a velocity $v \simeq 11$~m~s$^{-1}$.

For an Earth mass planet with $a=1$~AU around a solar mass star, $v
\simeq 0.1$~m~s$^{-1}$.  It is not clear whether such a precision
could be reached.  Such a signal would have to be extracted from the
``noise'' at the stellar surface, i.e. random Doppler shifts due to
atmospheric oscillations, stellar spots, or long period effects caused
by magnetic cycles.  Note that the distance to the star is not a
limiting factor as long as enough photons can be collected.

Several long--term radial velocity surveys for planets are ongoing in
different countries  (for a list see Perryman et al. 2005).  They
all use ground--based telescopes. New detections are regularly
announced.  So far all the planets found with this technique are
within about 100~parsecs (pc) from the Earth.

\subsubsection{Transits: \/ \\ \\}

The first transit of an extrasolar planet (HD~209458~b) was detected
in September 1999 (Charbonneau \etal 2000).  The planet, which is on
an orbit with $a= 0.045$~AU, had already been observed by the radial
velocity method.  The detection of the transit, in addition to proving
beyond doubt that the Doppler shift was indeed caused by a planet,
enabled the planet mass (0.63 M$_{\rm J}$) to be determined, as $\sin
i$ has to be close to~1 for a transit to be seen.  It also enabled the
planet radius (1.3  Jupiter radius) to be measured, as the
relative variation of the stellar flux during the transit is:

\begin{equation}
\frac{\Delta F_\star}{F_\star} = \left( \frac{R_p}{R_\star} \right)^2 ,
\end{equation}

\noindent where $R_p$ is the planet radius and $R_\star$ is the
stellar radius.  For Jupiter orbiting the Sun, $\Delta F_\star /
F_\star = 1$\%, whereas for the Earth it is 0.01\%.  From the ground,
the accuracy is about 0.1\% and is limited by the variable extinction
of the atmosphere.  From space, the accuracy reaches $10^{-4}$~\% so
that terrestrial planets could be detected without any problem.
However, the probability of detecting a transit, $R_\star/a$, is
rather small.  For $R_\star=1$~R$_\odot$, this probability is about
10\% for $a=0.05$~AU, i.e. for the planets on the tightest orbits
known to date.  Therefore, a large number of stars have to be observed
for a significant number of transits to be detected.

Note that the mean mass density of HD~209458~b can be calculated from
its mass and radius.  It is found to be about 400~kg~m$^{-3}$,
significantly smaller than the mean density of Saturn
(700~kg~m$^{-3}$), the least dense of the planets of our solar system.
This confirms that this planet is made mainly of gas.  The atmosphere
of the planet was further detected by HST, which observed additional
sodium absorption due to the light passing through the planet
atmosphere as it transited across the star (Charbonneau \etal 2002).
More recently, atomic hydrogen, oxygen and carbon have also been
detected in the extended envelope around the planet (Vidal-Madjar
\etal 2003, 2004), suggesting the atmosphere is evaporating.

The first planet found by the transit method, TrES--1, was detected by
the STARE (STellar Astrophysics \& Research on Exoplanets) program
(Alonso \etal 2004), and confirmed 8~days later by an amateur
astronomer in Belgium.  Since then, 5 more planets have been found by
this method by OGLE (Optical Gravitational Lensing Experiment).  All
these planets have lately been confirmed by radial velocity
measurements.  Their separation from the central star ranges from
0.0225 to 0.047~AU.  The planets found by OGLE are about 1,500~pc
away, much further away than the planets found by the radial velocity
method.

Several ground--based surveys for planet transits are ongoing.  A few
space missions dedicated at least in part to the planet search are
also being developed.  The first experiment, which will be launched in
2006, is COROT (for COnvection ROtation and planetary Transits), a
mini-satellite developed by the French National Space Agency (CNES).
Its precision is 0.03\%, so that it will detect only planets with a
radius larger than twice the Earth radius.  It will monitor about
60,000 stars during 2.5 years.  The next experiment is KEPLER, which
will be launched in 2007 by NASA.  Its precision is 10$^{-3}$~\% and
it will monitor 10$^5$ stars during 4~years.  It is expected to find
at least 50~terrestrial planets.

\subsubsection{Astrometry: \/ \\ \\}

The radial velocity method described above uses the motion of the star
{\em along} the line of sight.  It is also possible to measure the
motion of the star {\em perpendicular} to the line of sight and to
derive the characteristics of the planet and its orbit from it.  If
the planet is on a circular orbit, the angular motion of the star is:

\begin{equation}
\alpha \; ({\rm arcsec}) = \frac{M_p}{M_\star} \frac{a/1 \; {\rm AU}}
{d/1 \; {\rm pc}} \; ,
\end{equation}

\noindent where $d$ is the distance to the star, measured in parsecs.
Since $d$ and $M_\star$ are known and $a$ is given by the periodicity
of the star motion, $M_p$ can be derived from the measure of $\alpha$.
The motion of the Sun caused by Jupiter and seen from a distance of
10~pc is 500~$\mu$as, whereas that due to the Earth, and seen from the
same distance, is 0.3~$\mu$as.

>From the ground, the interferometers provide the best astrometric
accuracy.  It is about 1~mas for the VLTI and KECK in the near~IR, and
should reach 10~$\mu$as in the near future over narrow fields of view.
Surveys for planets with these interferometers will begin in the very
near future.

For space observations, the accuracy will be on the order of
1~$\mu$as.  Therefore, since the stars that are being observed are at
least at a distance of a few parsecs, it will not be possible to
detect terrestrial planets in habitable zones with this method, which
is better suited for the observation of giant planets. Note that
astrometric measurements are easier for longer period systems.  This
method and the radial velocity method are thus complementary.

To date, only one space mission, Hipparcos (which was launched by ESA
in 1989 and operated until 1993), was dedicated to astrometric
measurements.  Although it was not meant to detect planets, it did
observe stars around which planets were later observed.  Its
measurements were therefore used to constrain the mass of some of the
planets discovered by the radial velocity method.

Two space experiments are currently being developed.  SIM (Space
Interferometry Mission) is scheduled for launch by NASA in 2010.  Its
accuracy will be 1~$\mu$as.  It will not observe a very large number of
stars but will focus in particular on  the multiple
systems already detected from the ground to get precise measurement of
the orbital characteristics.  GAIA, which is scheduled for launch by
ESA in 2011, will have an accuracy of 2---10~$\mu$as and will observe
$\sim 10^5$ stars within a distance of about 200~pc from the Sun
during 5~years.  Simulations indicate that GAIA could detect between
10,000 and 50,000 giant planets.

\subsubsection{Microlensing: \/ \\ \\}

This technique relies on a background star to provide a source of
light. Then, foreground stars act as gravitational lenses when they
pass in front of the background star and cause the source star to
suddenly brighten by as much as 1000 times.  This typically lasts for
a few weeks or months.  While this is the normal pattern of a
microlensing event, things are substantially different when the
lensing (foreground) star has a smaller companion.  If a planet is
positioned close enough to the lensing star, it will indeed perturb
the light curve by adding two successive spikes of brightness,
superimposed upon the regular pattern of the microlensing event (Mao
\& Paczynski 1991, Gould \& Loeb 1992, Bennett \& Rhie 1996).
Microlensing is most sensitive to planets at a separation $\sim
1$--5~AU from the lens star.  In addition, effects on light curves are
large even for planets of less than an Earth mass.  This technique is
therefore well suited for finding terrestrial planets in habitable
zones.  An inconvenient aspect of the microlensing technique is that
the events do not repeat.

Several collaborations, from around the world, are presently searching
for planets with the microlensing technique.  Two of them (OGLE and
MOA) are survey programs which monitor large number of stars.  The two
other projects (PLANET and MicroFUN) pick up on the alerts that are
issued by the survey collaborations when a microlensing event is
detected and do a follow--up monitoring of the event.

One planet (OGLE235--MOA53) has been found in 2004 simultaneously by
the MOA and OGLE collaborations (Bond \etal 2004).  Its mass and
semi--major axis are about 2~M$_{\rm J}$ and 3~AU, respectively.  It
is located at a distance of 5.2~kpc from Earth, which is much too far
away for the radial velocity method to be used.

So far, the microlensing programs have mainly put limits on the
percentage of stars harboring planets within a given range of
separations.  Analysis of 5~years of PLANET photometry indicates that
less than 33\% of the M~dwarfs in the Galactic bulge can have
Jupiter--mass companions within a distance between 1.5 and 4~AU.  The
upper limit is raised to 45\% if the range of separations considered
is 1--7~AU (Gaudi \etal 2002).

A space--based gravitational microlensing program (GEST, for Galactic
Exoplanet Survey Telescope) is currently being studied by NASA. GEST,
which would be able to detect planets with masses as low as that of
Mars, would survey more than 100 million stars.  It would be expected
to find about 100 Earth mass planets.

\subsubsection{Direct imaging: \/ \\ \\}

The direct detections of the transiting planets HD~209458~b (Deming et
al. 2005) and TrES--1 (Charbonneau \etal 2005) were reported
simultaneously in March 2005.  These were the very first observations
of photons emitted by extrasolar planets.  The detections were made by
the Spitzer Space Telescope by observing the decrement in infrared
flux during the secondary eclipses, when the planets passed behind
their parent star.  These observations enable the brightness
temperature of the planets at a given wavelength to be calculated and,
assuming some model for the atmosphere, a blackbody temperature to be
estimated.

In general, direct imaging of extrasolar planets will enable long
period objects to be detected, and therefore will complement radial
velocity surveys.  Very low mass non transiting objects can currently
be detected only if they are isolated (e.g., free floating planets).
Bound planets are much more difficult to observe because of the very
high star to planet flux ratio.  The contrast is smaller in the IR,
which is therefore the most favourable range of wavelengths for planet
imaging.  The light coming from the star can in addition be blocked by
using a coronograph or nulling techniques on interferometers.  To
limit the halo of the star so that it is not brighter than the planet,
both turbulence and defects of the instrument have to be corrected
with unprecedented precision, so a high--performance adaptive optics
system is needed.  The latest developments in adaptive optics (the
first light from NAOS, a system installed by ESO on the VLT, was
received in 2001) suggest that a system for direct detection of
extrasolar planets may be available on 8--10 meter ground--based
telescopes within 5 years.

Several ground--based projects are currently being developed or under
study.  Most of them rely on interferometers to image planets.  There
are also several space based projects, among which are Darwin (studied
by ESA) and Terrestrial Planet Finder (studied by NASA).  These space
interferometers would be able to detect Earth--like planets in
habitable zones and to perform spectroscopic observations of their
atmosphere.  Given the ambitious nature of both projects, NASA and ESA
may collaborate on the final mission.

\subsection{Statistical properties of observed extrasolar planets}

So far, about 5\% of the stars monitored in radial velocity surveys
have been found to harbor at least one planet.  This is a lower limit
of course, as low--mass and/or long--period objects cannot be detected
with the current techniques.

\subsubsection{Mass and separation distributions: \/ \\ \\}  
\label{mass_sep}

\begin{figure}
\begin{minipage}[b]{\textwidth}
\parbox[t]{8cm}{
\includegraphics[width=8cm]{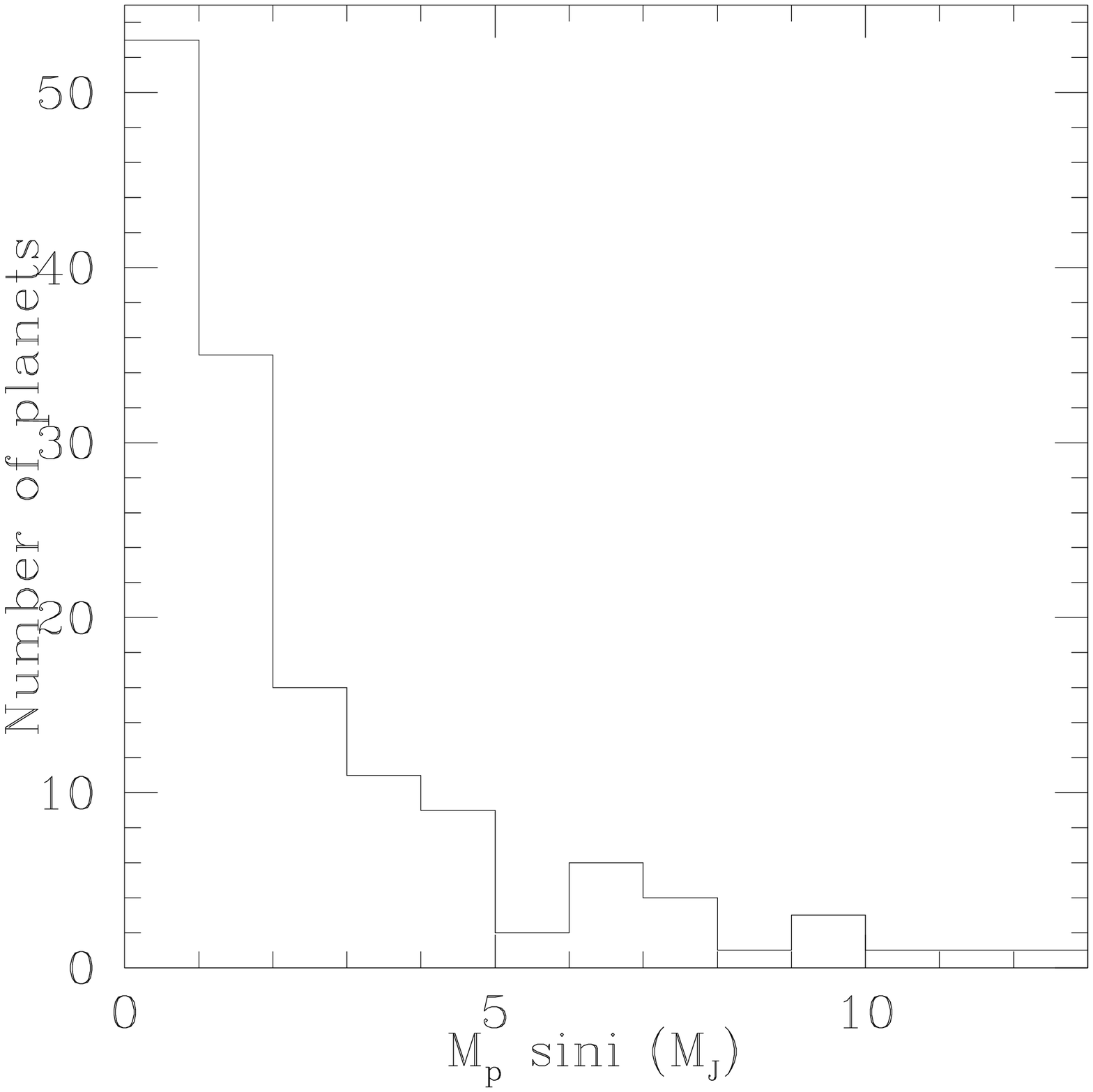}  
}
\hfill
\parbox[t]{8cm}{
\includegraphics[width=8cm]{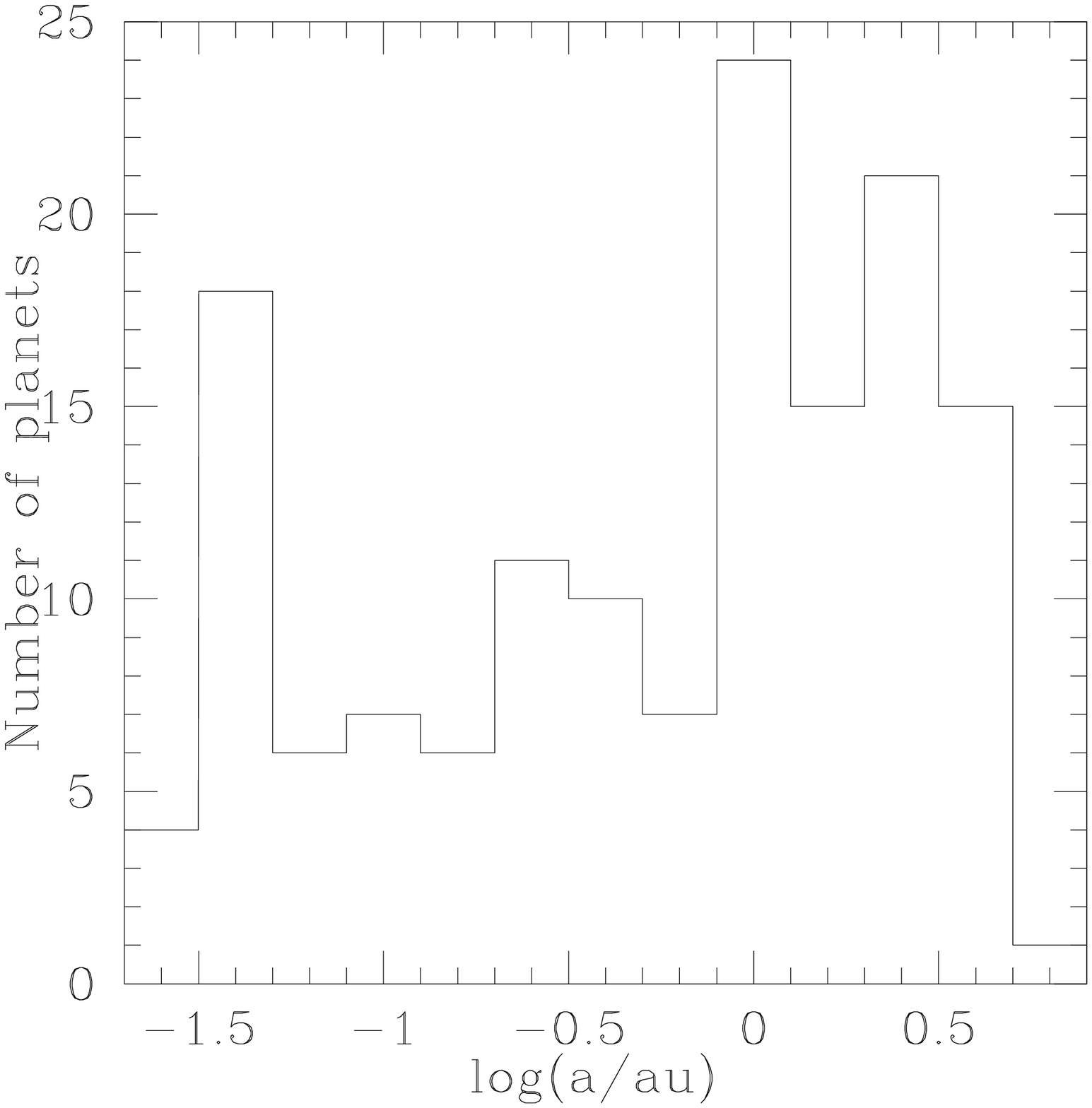}
}
\\ \mbox{}
\caption{\label{fig_stath} Distribution of projected mass $M_p \sin i$
(in M$_{\rm J}$, {\em left panel}) and semi--major axis $a$ (in AU,
logarithmic scale, {\em right panel}) for the known extrasolar
planets.  }
\end{minipage}
\end{figure}

The histograms of the projected mass $M_p \sin i$ and semi--major axis
$a$ of the known extrasolar planets are shown in
Figure~\ref{fig_stath}.

The lightest object has a mass of 0.042~M$_{\rm J}$ or 13.3~Earth
masses, which is very close to the mass of Uranus.  Although the
radial velocity surveys are biased towards massive planets, 37\% of
the planets detected so far have a mass smaller than 1~M$_{\rm J}$ and
only 13\% have a mass larger than 5~M$_{\rm J}$.  {\em Within a few~AU
from the central star, there is a deficit of massive planets.}

The distribution of $a$ shows that there is a significant number of
planets (12\%) with $a$ between 0.035 and 0.05~AU, i.e. an orbital
period between 2.5 and 4~days (the so--called ``hot Jupiters'').  Only
4~objects have a smaller period, which suggests that {\em there is a
pile--up at a rather well--defined separation ($\sim 0.4$~AU) or
period ($\sim 3$~days).} Only very few objects with a mass larger than
that of Uranus seem to be able to ``leak'' through what looks like a
barrier.  Since observations are biased towards short--period planets,
as noted by, e.g., Marcy et al. (2005), {\em there is clear rise of
the distribution with separation.}  Beyond $\sim 0.4~AU,$ the data
appears to be subject to $\sqrt{N}$ noise with dependence on binning,
an approximate smooth representation for the number of planets $N(a)$
with semi--major axis less than $a$ in $AU$ is $d N/d(\log a) =50\log
a +95.$ According to this, the number expected within an interval of
$0.2$ in $\log a$ is $\sim 9$ or 19 at $\log a =-1$ or 0,
respectively.

\begin{figure}
\begin{center}
\includegraphics[width=8.cm]{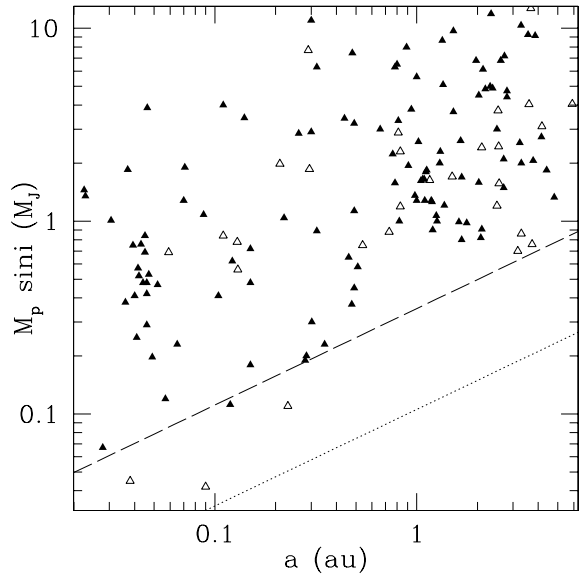}
\caption{\label{fig_stat1} Projected mass $M_p \sin i$ (in M$_{\rm
J}$) {\em vs.} semi--major axis $a$ (in AU) on logarithmic scales for
the known extrasolar planets. Planets in multiple systems are
indicated by open symbols. The dotted and dashed lines represent the
detection limit for planets on circular orbits and an accuracy of
3~and 10~m~s$^{-1}$, respectively (eq.~[\ref{eqK}] with $e=0$).}
\end{center}
\end{figure}

Figure~\ref{fig_stat1} shows the projected mass $M_p \sin i$ {\em vs.}
semi--major axis $a$. Planets in multiple systems are indicated by
open symbols.  {\em There is a deficit of sub--Jupiter mass planets at
separations larger than $\sim 0.5$~AU.}  More precisely, there is no
planet with $M_p \sin i < 0.7$~M$_{\rm J}$ at $a>0.5$~AU.  Since about
10\% of the planets have a mass lower than 0.7~M$_{\rm J}$, the
absence of such planets at separations larger than 0.5~AU, where 57\%
of the planets are, is statistically significant.  The dotted and
dashed lines in figure~\ref{fig_stat1} indicate the detection limit
for planets on circular orbits and an accuracy of 3~and 10~m~s$^{-1}$,
respectively.  It is not clear that the deficit of low--mass planets
at larger $a$ is the result of observational biases (see also Udry et
al. 2003).  {\em There is also a deficit of high mass planets at
separations smaller than $\sim 0.3$~AU.}  More precisely, there is no
planet with $M_p \sin i > 4$~M$_{\rm J}$ at $a < 0.29$~AU.  As
observations favor massive planets with short--periods, this deficit
is not due to an observational bias. It has further been pointed out
that planets more massive than 2~M$_{\rm J}$ at $a < 4$~AU are
actually members of binary star systems (Zucker \& Mazeh~2002).

\subsubsection{Eccentricity distribution: \/ \\ \\} 

\begin{figure}
\begin{minipage}[b]{\textwidth}
\parbox[t]{8cm}{
\includegraphics[width=8cm]{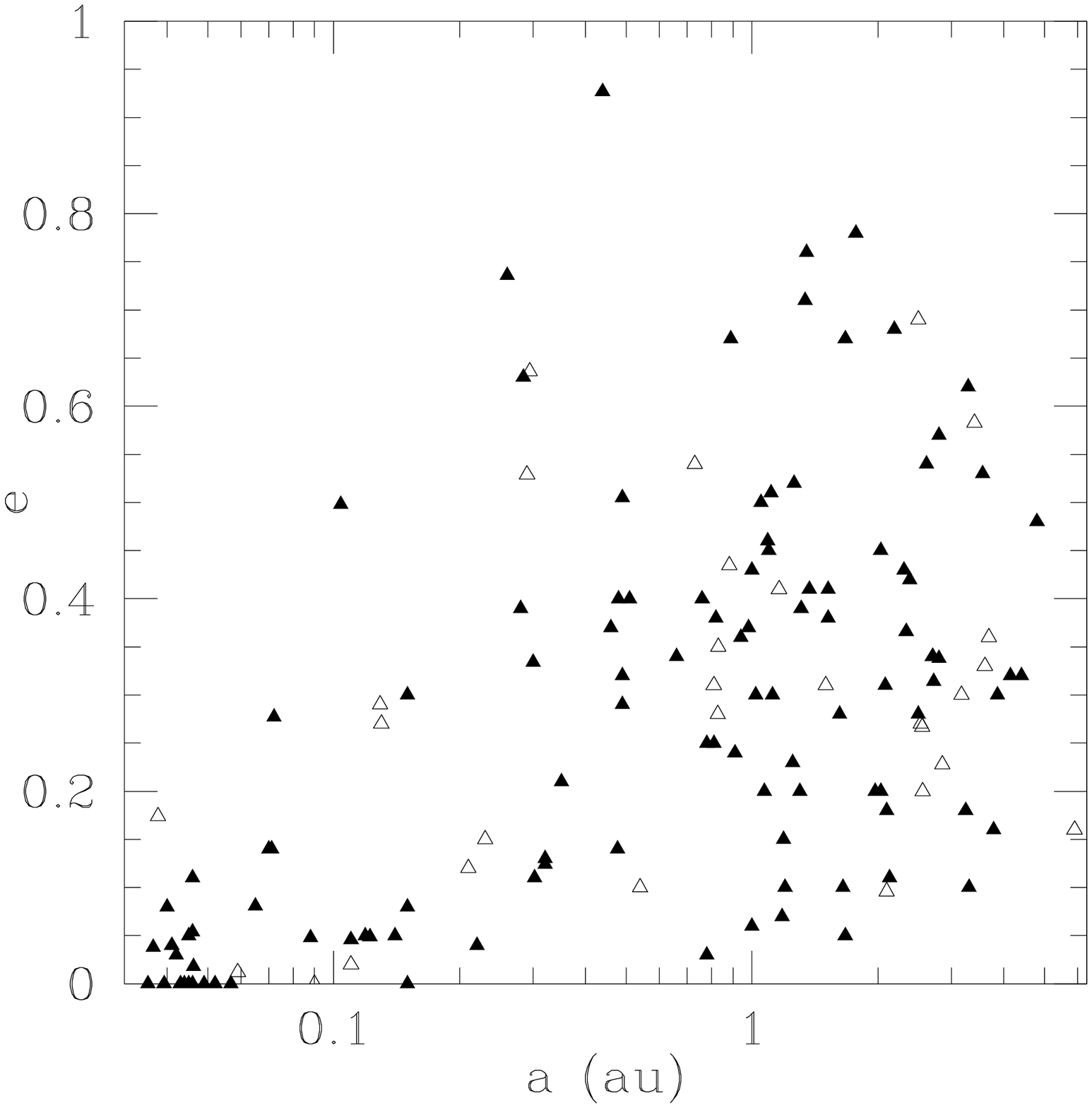}
}
\hfill
\parbox[t]{8cm}{
\includegraphics[width=8cm]{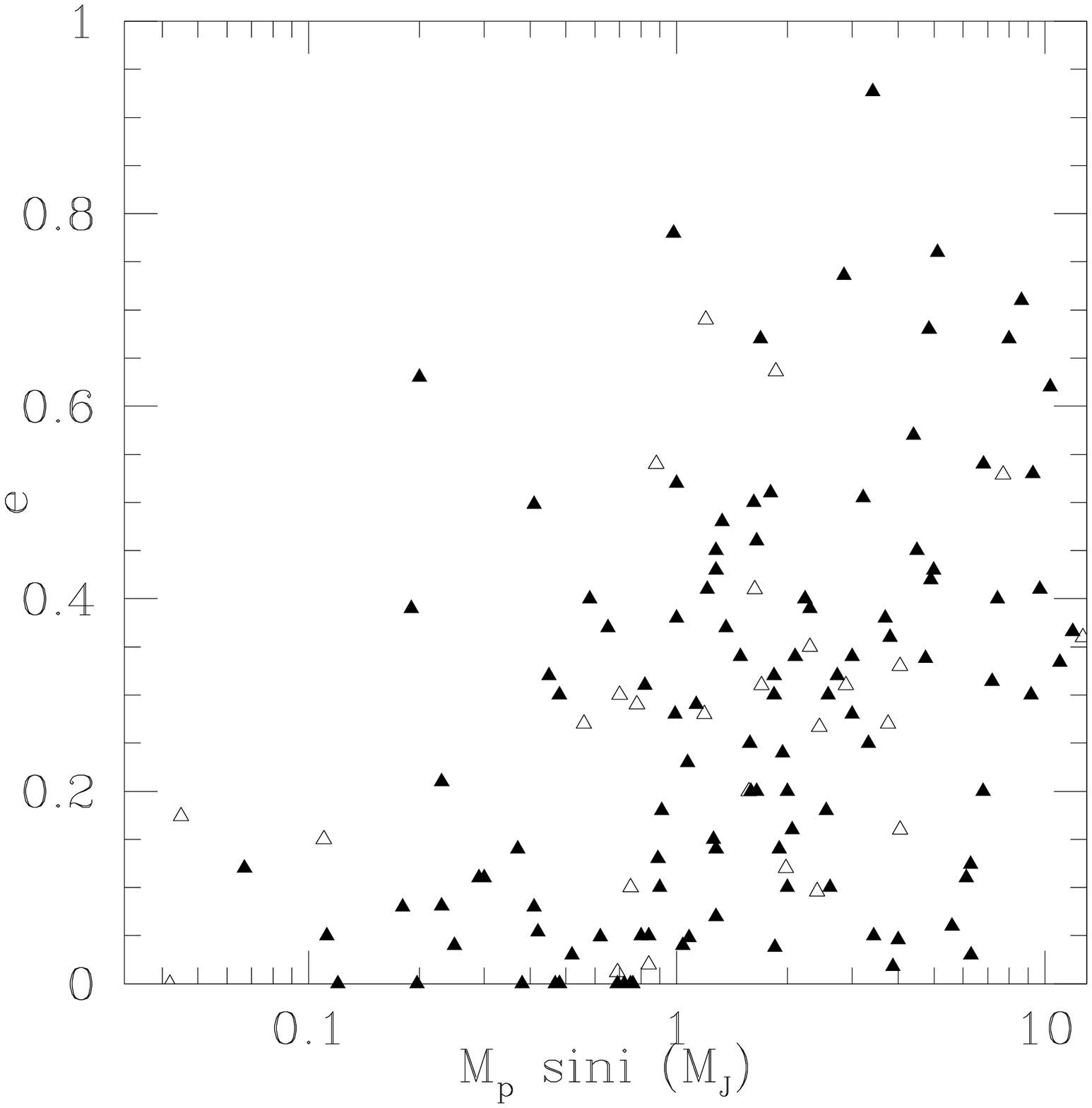}
}
\\ \mbox{}
\caption{\label{fig_state} Eccentricity $e$ {\em vs.} semi--major axis
$a$ (in AU) on  a logarithmic scale ({\em left panel}) and projected mass
$M_p \sin i$ (in M$_{\rm J}$) on a  logarithmic scale ({\em right panel})
for the known extrasolar planets. Planets in multiple systems are
indicated by open symbols.  }
\end{minipage}
\end{figure}

Figure~\ref{fig_state} shows the eccentricity $e$ {\em vs.}
semi--major axis $a$ and projected mass $M_p \sin i$.
  
\noindent Planets with $a< 0.05$~AU have almost circular orbits, which
is consistent with tidal circularization. At larger separations, {\em
the eccentricity varies from 0 up to 0.93.}  This is in sharp contrast
with the planets of our solar system, which are on nearly circular
orbits.  The upper limit of 0.93 may be special, as it may be due to
the secular perturbation of the planet HD~80606~b by a stellar
companion on an inclined orbit (Wu \& Murray 2003). Nonetheless,
eccentricities in the range 0--0.8 are common.  Note that higher
eccentricities seem to be obtained for more distant planets. The
eccentricity distribution of planets  in systems of multiple
planets is similar to that of planets in single systems.

Figure~\ref{fig_state} shows that higher eccentricities are obtained
for more massive planets.  This is consistent with the trend
noted above of higher eccentricities at larger separations together
with the fact that more massive planets are at larger separations.
 But note that the observed distributions could be produced either
by physical processes tending to produce higher eccentricities for
higher masses or higher eccentricities at larger orbital separations
or a combination of the two.  A suggested process of the former type
is eccentricity generation through disk planet interaction (
Artymowicz 1992) and one of the latter type is orbital circularization
close to the central star. Here again, the eccentricity distribution
of planets in  systems of multiple planets is similar to that
of planets in single systems.

Halbwachs \etal (2005) suggest that the differences between the
eccentricity distributions of planets and stellar binaries is an
indication that these two classes of objects have different formation
mechanisms.

\subsubsection{Metallicity of stars harboring planets: \/ \\ \\}

\begin{figure}
\begin{center}
\includegraphics[width=8.cm]{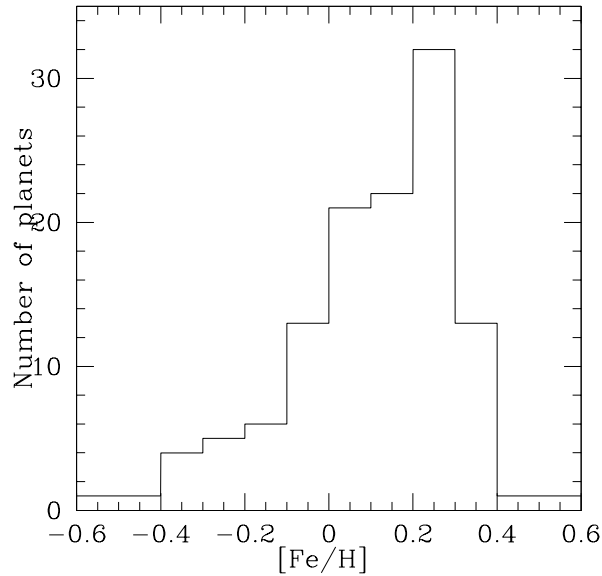}
\caption{\label{fig_hist3} Distribution of metallicity [Fe/H] for the
stars harboring planets.  }
\end{center}
\end{figure}

Figure~\ref{fig_hist3} shows the metallicity distribution of the stars
around which planets have been found.  The metallicity of most of the
field dwarfs with no known companion lies in the range -0.3--0.2
(Santos \etal 2001).  The planet host stars are therefore metal rich,
and the planetary frequency is clearly increasing with metallicity.
This metallicity excess appears to have a ``primordial'' origin
(Pinsonneault \etal 2001; Santos \etal 2001; Santos \etal 2002;
Sadakane \etal 2002),  rather than being due to the infall of
planetary material onto the star which might result in observed heavy
element polution for stars in the mass range 1--1.3~M$_{\odot}$
(e.g. Sandquist et al. 1998, 2002).  The favored explanation for the
high metallicity of planet host stars is therefore that planets form
more easily in a metal rich environment.

Note that no short--period planets were found in the metal poor
globular cluster 47~Tucanae in which about 34,000 stars were monitored
by HST.  Simulations had shown that about 17 hot Jupiters should have
been detected by photometric transit signals if the frequency of these
objects were the same in the solar neighborhood and in 47~Tucanae
(Gilliland \etal 2000).

\begin{figure}
\begin{minipage}[b]{\textwidth}
\parbox[t]{5.cm}{
\includegraphics[width=5.cm]{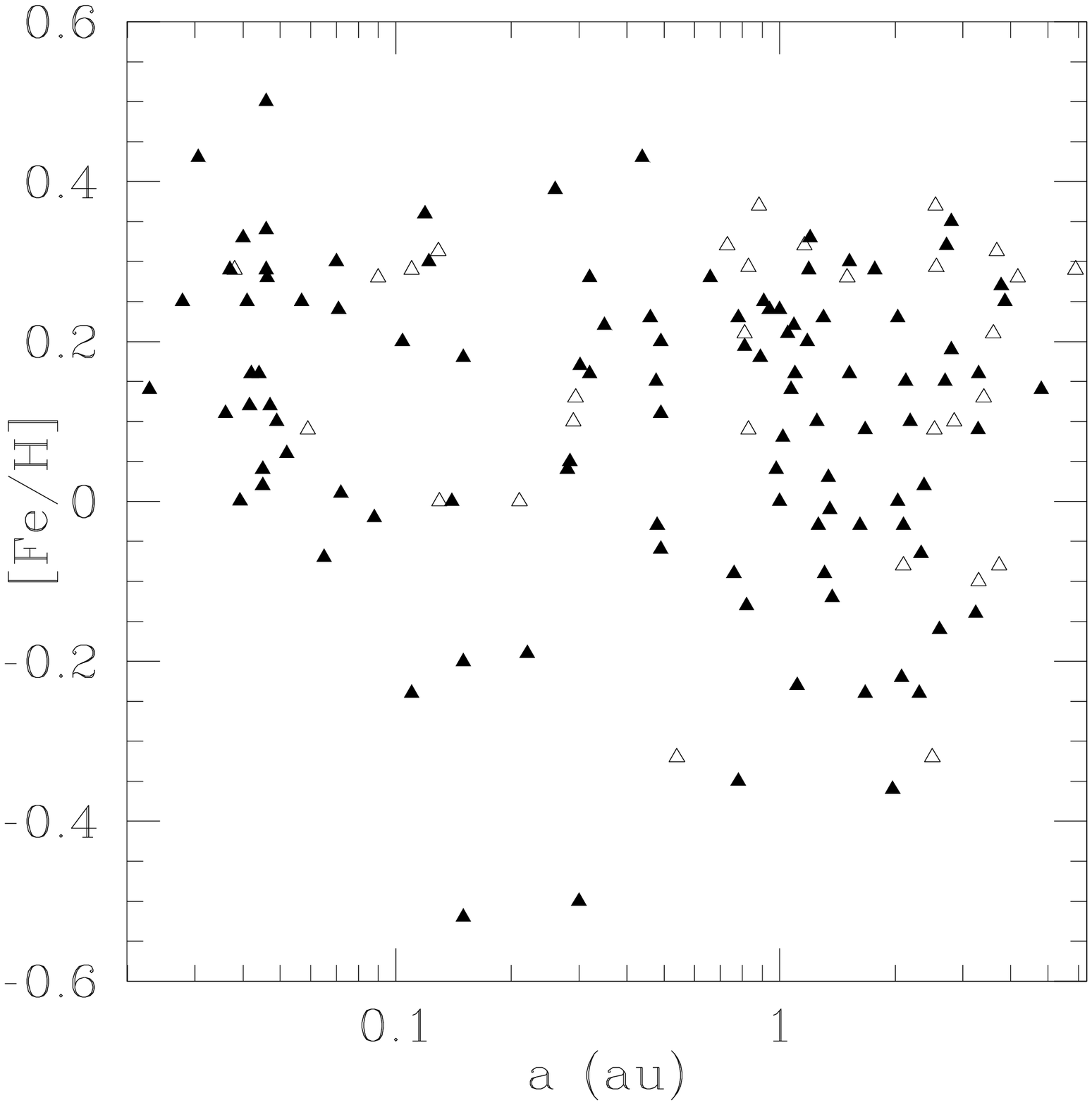}
}
\hfill
\parbox[t]{5.cm}{
\includegraphics[width=5.cm]{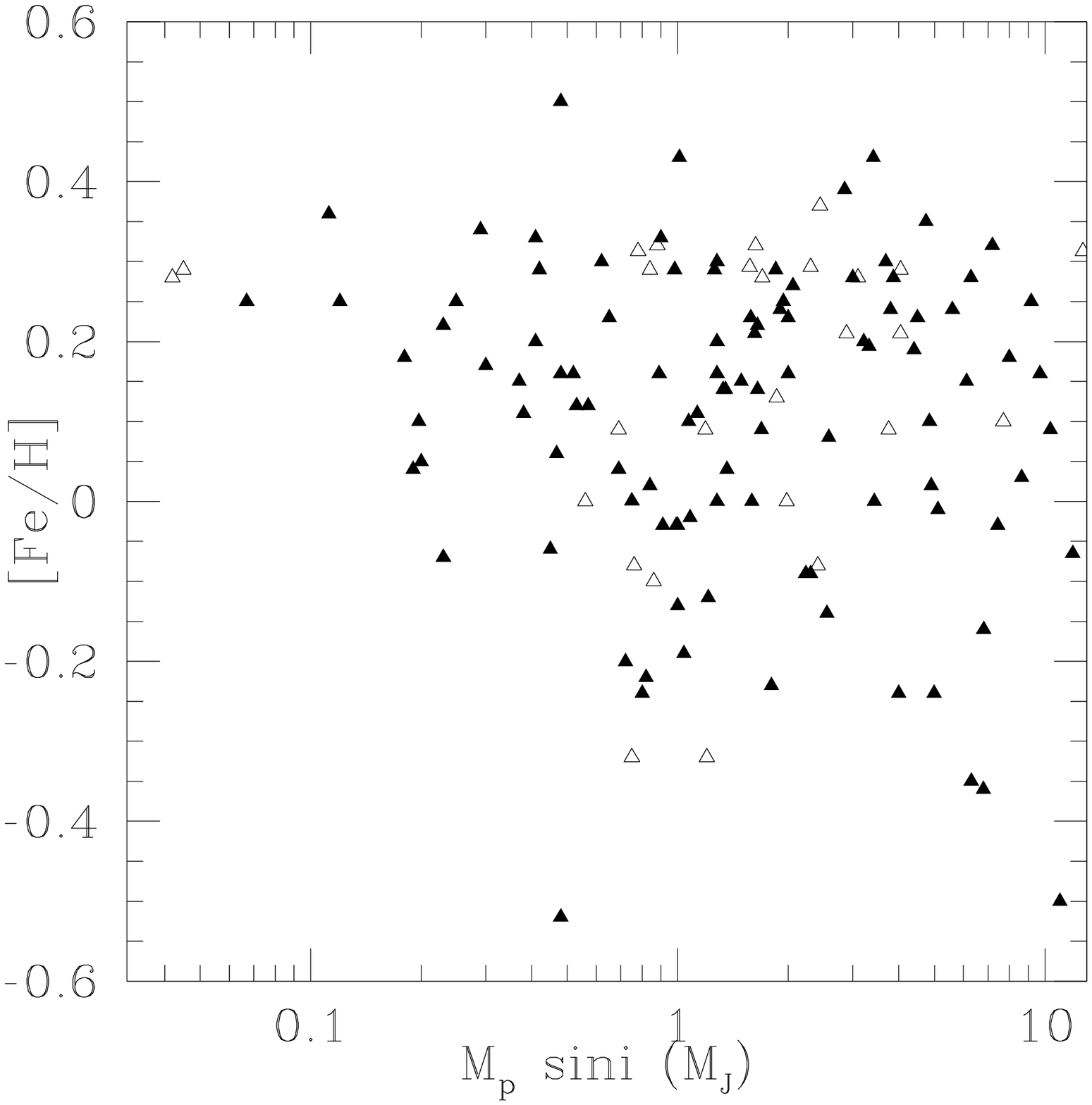}
}
\hfill
\parbox[t]{5.cm}{
\includegraphics[width=5.cm]{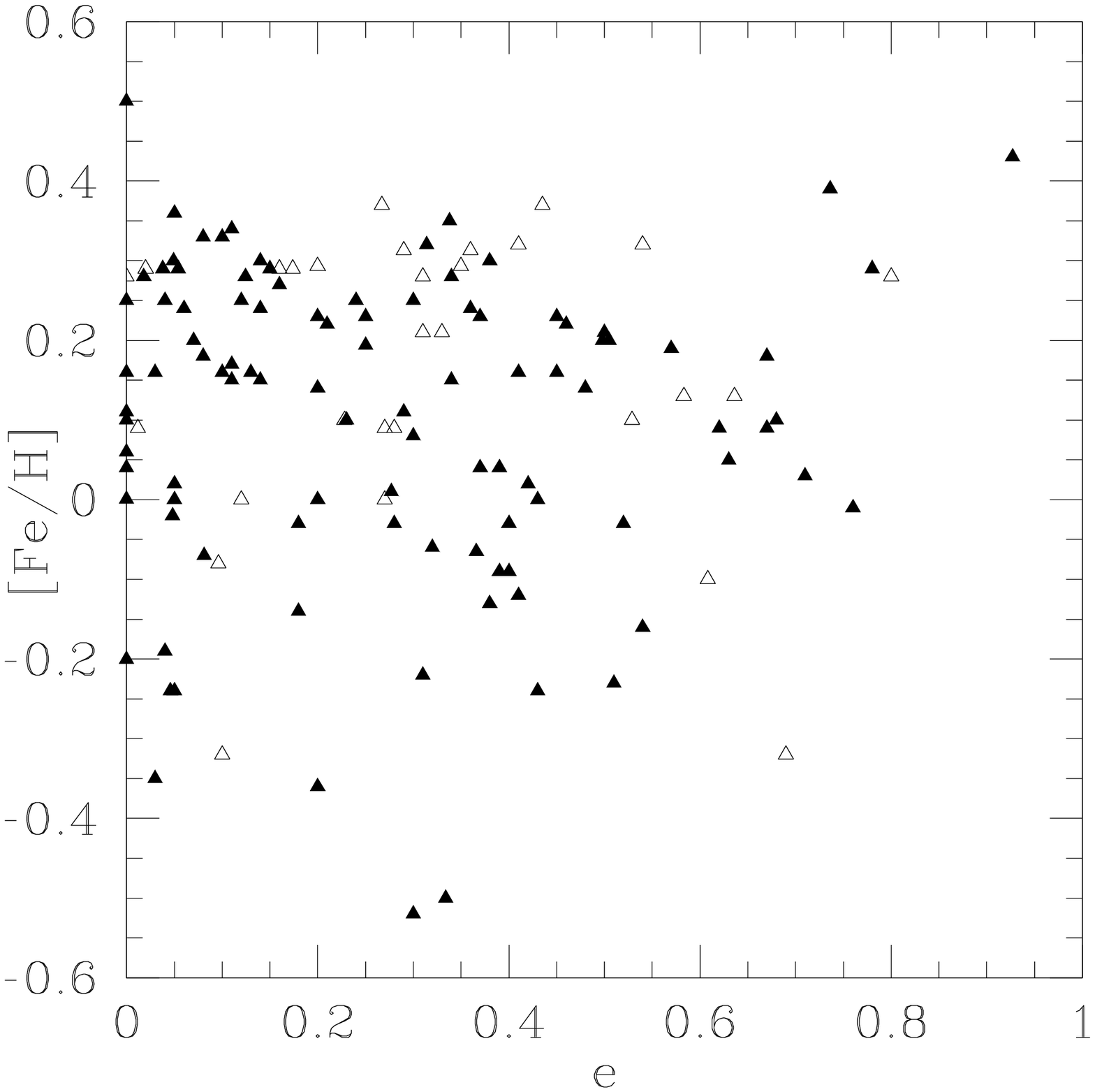}
}
\\ \mbox{}
\caption{\label{fig_statm} Stellar metallicity [Fe/H] {\em vs.}
semi--major axis $a$ (in AU) in logarithmic scale ({\em left panel}),
projected mass $M_p \sin i$ (in M$_{\rm J}$) in logarithmic scale
({\em middle panel}) and eccentricity $e$ ({\em right panel}) for the
known extrasolar planets. Planets in multiple systems are indicated by
open symbols. }
\end{minipage}
\end{figure}

Figure~\ref{fig_statm} shows the stellar metallicity {\em vs.}  $a$,
$M_p \sin i$ and $e$.  Sozzetti (2004) has found that there is a weak
correlation (at the 2 to 3$\sigma$ level) between the stellar
metallicity and the orbital period of the system.  This correlation
becomes stronger when only single stars with one detected planet are
considered.

\subsubsection{Planets in multiple stellar systems: \/ \\ \\}

As noted above, the most massive short--period planets are all members
of stellar binaries (Zucker and Mazeh 2002).  It also appears that
planets orbiting a component of a multiple star system tend to have a
very low eccentricity when their period is shorter than about 40~days,
i.e. when $a$ is smaller than about 0.2~AU (Eggenberger \etal 2004).
 
\subsubsection{Free floating planets: \/ \\ \\}

Recently, a population of about 30 very young (a few Myr)
free--floating objects with planetary masses (from a few to 13~M$_{\rm
J}$) has been found in the Sigma Orionis open cluster (Zapatero Osorio
\etal 2000) and in the Trapezium (Lucas and Roche 2000).  The lightest
of these objects is S~Ori~70 (Zapatero Osorio \etal 2002).  Its mass
is derived from evolutionary models and is estimated to be 3~M$_{\rm
J}$.  However, Burgasser \etal (2004) claim that this object may
actually not be a member of the $\sigma$ Orionis cluster, in which
case its mass may be much higher.

B\'ejar \etal (2001) have pointed out that if the distribution of
stellar and substellar masses in $\sigma$ Orionis is representative of
the Galactic disk, older and much lower luminosity free--floating
objects with masses down to a few M$_{\rm J}$ should be abundant in
the solar vicinity, with a density similar to M--type stars.

The discovery of these free--floating objects has brought some
confusion in the definition of planets, as they are not orbiting a
star and they probably have not formed through the standard
core--accretion scenario (see below).

\section{Theories of planet formation}

Terrestrial planets are believed to be formed via solid body accretion
of km--sized objects, which themselves are produced as a result of the
sedimentation and collisional growth of dust grains in the
protoplanetary disk (see Lissauer 1993 and references therein).  This
theory comes from the idea already proposed in the 19th century that
the Earth and the other terrestrial planets were formed from
meteoritic material.  At the beginning of the 20th century, this idea
was supported by the geologist T.~C.~Chamberlin who called
``planetesimals'' the solid objects that had accumulated to form the
planets.  But it is only in the 1960s that the theory was developed in
a quantitative way by V.~Safronov (1969), who calculated in details
the different stages of terrestrial planet formation. This pioneering
work, originally written in Russian, was published in English in 1972
under the title ``Evolution of the Protoplanetary Cloud''.  Since
then, the theory has been further developed, in particular by
G.~W.~Wetherill and collaborators who have used intensive numerical
simulations to study planetesimal accumulation.

As far as giant planets are concerned, two theories have been
proposed.  According to the first theory, giant planets are formed
through collapse and fragmentation of protostellar disks.  This
scenario is related to that proposed by Laplace in 1796.  In its
modern version, it has been developed by G.~Kuiper in the 1950s and by
A.~G.~W.~Cameron in the 1960s and 1970s.  It is still being studied
today.  The second theory, proposed by Cameron in 1973, is called the
``core accretion model''.  A solid core is first assembled in the same
way as terrestrial planets.  Once the core becomes massive enough to
gravitationally bind the gas in which it is embedded (typically at
around a tenth of an Earth mass  (see section~\ref{sec:formgiant}
below), a gaseous envelope begins to form around the core.

\subsection{Terrestrial planet and planetary core formation}

In the formation of terrestrial planets or planetary cores, several
stages can be considered.  Firstable, micron sized dust grains that
are present when the disk forms sediment towards the disk midplane.
These grains are subject to both the drag force exerted by the gas and
the gravitational force from the central star.  Accumulation of grains
during this first stage leads to cm/m sized particles.  It is still
not clear how these objects further grow to become 0.1--1~km sized
planetesimals.  It is likely that growth occurs through collisions and
accumulation.  The planetesimals are massive enough that mutual
gravitational interactions have to be taken into account in computing
their evolution.  Further collisions and accumulation lead to
terrestrial planets or the core of giant planets.  These different
processes are described in details below.

\subsubsection{Growth from dust to cm/m sized particles: \\ \\}

When the disk forms out of the collapse of a molecular cloud, it
contains dust grains which size is that of the interstellar grains,
i.e. 0.1--1~$\mu$m.  Under the action of the gravita\-tional force
exerted by the central star and the drag force exerted by the gas,
when there is no turbulence present, these grains tend to undergo
damped oscillations around the disk midplane, toward which they
eventually sediment.  Simultaneously, they drift toward the disc
center.  This is because the coupling with the gas forces them to
rotate at a slightly sub--Keplerian velocity.  Since they are not
supported by pressure forces, they consequently cannot stay on a fixed
orbit.

For typical disk parameters, the gas mean free path around the disc
midplane is $\sim$~1~m.  This is very large compared to the size of
the grains we are interested in here.  In this regime, the drag force
exerted by the gas on the grains is $\sim - \pi r_d^2 \rho c_s \left(
{\bf v} - {\bf v}_d \right)$ (Epstein law; e.g., Weidenschilling
1977), where $r_d$ is the radius of the grains, which are assumed to
be spherical, $c_s$ is the thermal velocity of the gas (i.e. the sound
speed), $\rho$ is the mass density of gas in the disk, ${\bf v}_d$ is
the velocity of the grains and ${\bf v}$ that of the gas.  This
formula is valid provided $\left| {\bf v} - {\bf v}_d \right| \ll
c_s$, which is the case here as can be checked {\em a posteriori}.  We
define the time: $$\tau_e \equiv \frac{m_d}{\pi r_d^2 \rho c_s},$$
with $m_d$ being the mass of a dust grain.  This is the timescale it
takes for the velocity of a grain subject only to the drag force to
decrease by a factor $e$.  For $r_d=1$~$\mu$m, $\rho \sim
10^{-10}$~g~cm$^{-3}$, $c_s \sim 10^5$~cm~s$^{-1}$ (which corresponds
to a temperature of about 700~K), which are typical conditions at
about 1~AU from the central star, and assuming the mass density of the
grains to be 3~g~cm$^{-3}$ (which is comparable to the mass density of
the Earth's crust and that of the asteroids), we get $\tau_e \sim
10$~s.

For the gas, we write ${\bf v}=(v_r,v_{\varphi},v_z)$ in cylindrical
coordinates and, for the dust, ${\bf v}_d=(v_{r,d},v_{\varphi,
d},v_{z,d})$ .  We assume the motion of the grains to be almost
Keplerian with a small drift velocity, i.e. $v_{\varphi, d} \sim v_K$
and $|v_{r,d}| \ll |v_{\varphi, d}|$, where $v_K$ is the Keplerian
velocity.  In this situation in a steady state, the dust to a first
approximation comoves with the gas so that $v_{\varphi, d
}=v_{\varphi}.$ This means that accelerations due to radial pressure
gradients acting on the gas have to be balanced by the accelerations
due to drag acting on the dust which occur because of the radial drift
velocity which will be given by:
$$\frac{v_{r,d}}{\tau_e} =\frac{1}{r}(v_{\varphi}^2 - v_K^2) =
\frac{1}{\rho} \frac{\partial P}{\partial r}. \label{drag0}$$ As
expected, if the pressure increases inwards, $v_{r,d}<0$.  Also
$v_{r,d} \rightarrow 0$ as $r_d \rightarrow 0$, which means that very
small grains are ``carried'' by the gas.  With $|\partial P/\partial
r| \sim P/r$, $c_s^2 \sim P/\rho$ and $H/r \sim c_s/v_K$, where $H$ is
the disk semi--thickness, we get $v_{r,d} \sim - c_s (H/r) \tau_e
\Omega_K$, where $\Omega_K$ is the Keplerian angular velocity.  With
the values of the parameters used above, this gives $v_{r,d}$ on the
order of a few $10^{-2}$~cm~s$^{-1}$ at $r=1$~AU.

If the grains were not subject to the gas drag force, they would
oscillate around the disk midplane with the frequency $\Omega_K$.
Since $2\pi/\Omega_K \gg \tau_e$, the oscillations are suppressed and
the grains sediment towards the disk midplane.  As for a highly damped
harmonic oscillator, the characteristic timescale for sedimentation is
$$\tau_s = {1\over \Omega_K^2 \tau_e} = \frac{3\rho c_s }{4\Omega_K^2
\rho_d r_d}.$$ Here $\rho_d$ is the mass density of the grains.  For
the values of the parameters used above, $\tau_s \simeq 10^5$~years at
1~AU.  This is small compared to the planet formation timescale, as
will be seen below.  We have supposed that $m_d$ and $r_d$ are
constant, so that $\tau_s$ given above is actually a maximum.  Indeed,
if the grains grow through collisions as they fall toward the
midplane, the drag force due to the gas decreases and the
sedimentation is accelerated.  Note that during $\tau_s$, the grains
drift radially over a distance $|v_{r,d}| \tau_s$, which is on the
order of $10^{-2}$~AU at 1~AU.  This is very small compared to the
disk semithickness ($\sim$~0.1~AU), which justifies the neglect of
$v_{r,d}$ in the discussion of the vertical motion. However, grains
grow through collisions during their descent.  Let us suppose that
they ``absorb'' all other grains on their way to the midplane.  When a
grain moves a distance $|dz|$, its mass then increases by $dm_d=\pi
r_d^2 |dz| \rho_s$, where $\rho_s$ if the mass density of solids
(dust) in the disk.  This corresponds to an increase of its radius of
$dr_d=\rho_s |dz| /(4 \rho_d)$.  Therefore, for the values of the
parameters used above, a grain initially at the altitude $H$ would
have a size $\sim$~mm when it settles toward the midplane at a
few~AU. Its radial drift velocity then cannot be neglected.

Numerical simulations in laminar disks (Weidenschilling 1980, Nakagawa
\etal 1981) show that grains in the process of settling toward the
midplane undergo a significant radial drift which promotes collisions.
This enables them to grow further up to 0.1--1~m, and sedimentation is
then found to occur in a few $10^3$~years (sedimentation being faster
for larger grains, as $\tau_s \propto 1/r_d$).  Note however that
these simulations assume that the grains stick perfectly to each
other, so that they agglomerate when they collide.

Even though the growth of the grains during their sedimentation is not
yet fully modeled by numerical simulations, it is generally accepted
that this process leads to cm/m sized particles around the (laminar)
disk midplane in less than $10^4$~years.

Calculations including turbulence have been performed by
Weidenschilling (1984).  He found that the coagulation rate is
initially much greater than in a non turbulent disk.  Aggregates
quickly reach sizes $\sim 0.1$--1~cm, but erosion and breakup in
collisions prevent growth of larger bodies.  These aggregates are too
small to settle to the plane of the disk in the presence of
turbulence.

\subsubsection{Growth from cm/m sized particles 
to 0.1/1~km sized planetesimals: \\ \\}
\label{sec:planetesimals}

Before examining the processes that could lead to further growth of
the solid particles, let us discuss the timescale on which such
processes  may have to operate.

{\em Maximum radial drift:} When the gas mean free path $l$ is no
longer large compared to the size of the grains $r_d$, the Epstein law
cannot be used to calculate the drag force between the gas and the
grains.  In general, the drag force acting on a single grain is given
by $-f \left( {\bf v}_d - {\bf v} \right)$, where $f$ is a coefficient
that depends on the characteristics of the grains and the gas.  In
steady state, the horizontal equations of motion for the grains can
therefore be written:

\begin{eqnarray}
\frac{2 v_\varphi \Delta v_\varphi}{r} + \frac{1}{\rho} \frac{\partial
P}{\partial r} & = & \frac{f}{m_d} v_{r,d} ,  \\ 
\left( \frac{v_\varphi}{r} + \frac{\partial v_\varphi
}{\partial r} \right) v_{r,d} & = & - \frac{f}{m_d} \Delta
v_\varphi ,
\label{motiongrain}
\end{eqnarray}

\noindent where $\Delta v_\varphi = v_{\varphi,d} - v_\varphi$.  
Note that:

\begin{equation}
v_\varphi^2 = \frac{r}{\rho} \frac{\partial P}{\partial r} + v_K^2 .
\label{vgas}
\end{equation} 

\noindent In equations~(\ref{motiongrain}) we have used $|v_{r,d}| \ll
|v_{\varphi,d}|$ and $|\Delta v_\varphi| \ll v_\varphi$.  We can
further eliminate $f/m_d$ to get a quadratic equation for $\Delta
v_\varphi$.  It is straightforward to show that this equation has a
solution only if

\begin{equation}
\left( \frac{1}{\rho} \frac{\partial P}{\partial r} \right)^2
-8 \left[ \frac{2}{\rho r} \frac{\partial P}{\partial r} +
\frac{2}{r} \frac{\partial}{\partial r} \left( \frac{1}{\rho r}
\frac{\partial P}{\partial r} \right) + \frac{\Omega_K^2}{2} \right]
v_{r,d}^2 \ge 0 .
\end{equation}

\noindent The first two terms in bracket are negligible compared to
the last one since $c_s/(r \Omega_K) \sim H/r \ll 1$.  This condition
can therefore be written $\left| v_{r,d} \right| \le \left| v_\varphi
- v_K \right| \sim c_s^2/(r \Omega_K)$, where again we have used
$v_\varphi / v_K \simeq 1$.  At 1~AU, this gives a maximum value of
$|v_{r,d}|$ on the order of $10^4$~cm~s$^{-1}$, independent of
$r_d$. From subsection~\ref{drag0}, $|v_{r,d}| \sim
10^{-2}$~cm~s$^{-1}$ when $r_d \sim 1 \mu {\rm m} \ll l$, i.e. when the Epstein law for
drag still holds, much smaller than the maximum speed derived here.
As $|v_{r,d}| \propto r_d,$ 
the maximum speed is reached when $r_d$ is at least as
large as $l \sim 1$~m.  The inflow time corresponding to
the maximum speed is very short, being less than $100$~years at 1~AU.

{\em Radial drift when $r_d \gg l$:} In the regime $r_d \gg l$, the
drag force acting on a single grain is $\simeq -C_D \pi r_d^2 \rho
\left| {\bf v}_d - {\bf v} \right| \left( {\bf v}_d - {\bf v}
\right),$ where $C_D$ is the drag coefficient, being a number of order
unity (Stokes law; e.g., Weidenschilling 1977).  In steady state, the
radial equation of motion for the solid particles leads to
$v_{\varphi,d} = v_K$ if we assume $|v_{r,d}| \ll |v_{\varphi,d}|$.
This means that the angular velocity of large particles around the
star is not affected by the gas.  The azimuthal equation of motion
(\ref{motiongrain}), after using equation~(\ref{vgas}), then gives:

\begin{equation}
v_{r,d} = \frac{C_D r_d^2}{2 m_d \rho \Omega_K^3} \left( \frac{\partial P}
{\partial r} \right)\left| \frac{\partial P}
{\partial r} \right| .
\label{Pgrdmig}
\end{equation}

\noindent Important results follow from the dependence of the
migration speed on the pressure gradient. In particular, if the
pressure increases inwards, inward drift occurs.  Similarly, an
outwardly increasing pressure results in outward drift. Thus,
particles will accumulate at pressure maxima and be depleted at
pressure minima.
We comment that this is a general feature 
of a non magnetic gas and dust mixture 
that arises because the forces acting on the two components are the same
at pressure extrema so that they 
can comove there without drag.
 To estimate the typical drift speed magnitude in the general case, we set
$ \partial P/\partial r = nP/r.$ Then equation~(\ref{Pgrdmig}) gives:

\begin{equation}
v_{r,d} = c_s \frac{3 C_Dn|n| \Sigma}{16\pi \rho_d r_d} 
\frac{c_s^2}{r^2 \Omega_K^2}  ,\label{Pgrdmig1}
\end{equation}

\noindent where $\Sigma$ is the surface density of the gas disk.
 
\noindent For $C_D =0.2,$ $|n|=3,$ $\rho_d = 1$~g~cm$^{-3},$ $r_d =
10$~m and $\Sigma = 10^3$~g~cm$^{-2},$ we get $\left| v_{r,d} \right|=
0.1 c_s [c_s/(r\Omega_K)]^2.$ The radial flow time scale is thus
$(5/\pi) (r\Omega_K/c_s)^3 P_{orb},$ with $P_{orb}$ being the local
orbital period.  For $c_s/(r\Omega_K) =0.07,$ this is
characteristically $\sim 5000$~ orbits, being more than  an order of magnitude
longer than the timescale corresponding to the maximum drift speed.
Therefore, the maximum speed is reached when $r_d \sim l \sim 1$~m.

\noindent Although there is a dependence on the local state variables,
the drift timescale for meter sized bodies at 1~AU is potentially as
short as $\sim$~100~years.  Therefore, either the process by which
sub--meter sized bodies grow to become planetesimals has to be very
fast, or possibly special flow features, like pressure extrema or
vortices, play an important role (see below).

Note however that the calculation presented above is for a particle
which is isolated.  In a laminar disk in which a dust layer forms
around the midplane, the drift velocity cannot be calculated by
assuming that the gas velocity is given by equation~(\ref{vgas}).  In
the dust layer, the dust density is probably large enough that the gas
is carried around at a velocity closer to the Keplerian value.  The
relative velocity of the solid particles with respect to the gas is
therefore smaller than what we have assumed.  It is actually the {\em
collective drag} on the surface of the layer which is responsible for
orbital decay.  The lifetime of the dust layer against orbital decay
at 1~AU is estimated to be $\sim 10^3$~years (Goldreich and Ward
1973) so that the problem still remains.

The process by which 0.1/1~km sized bodies form in the dust layer in a
laminar disk is still unclear.  Safronov (1969) and Goldreich and Ward
(1973) independently pointed out that the dust may concentrate
sufficiently in the layer to undergo gravitational instability.  This
occurs when the Toomre parameter $Q = \Omega c_d / (\pi G \Sigma_d)$
(with $c_d$ and $\Sigma_d$ being the velocity dispersion and surface
mass density, respectively, of the dust) approaches unity.  In a
typical disk, this requires the mass density of dust to be $\sim
10^{-7}$~g~cm$^{-3}$.  Fragmentation of this layer then produces
planetesimals which size is on the order of 1~km.  The advantage of
this scenario is that it operates very fast, on a timescale on the
order of the orbital timescale.

However, it was pointed out by Weidenschilling (1980) that, if the
dust density reaches such high values, then it dominates the disk in
the midplane, i.e. it carries the gas at a Keplerian velocity around
the star.  As mentionned above, because there is no pressure
associated with the dust, it rotates at a different rate than the gas
around the layer. In these conditions, the velocity shear between the
dust layer and the gas around it gives rise to Kelvin--Helmholtz
instabilities that result in turbulence, even in a disk which is
laminar to begin with.  This leads to mixing of the dust so that the
density in the layer is decreased and gravitational instabilities are
prevented.  This result was confirmed by detailed calculations by
Cuzzi \etal (1993) who showed that rapid accretion of planetesimals by
gravitationally unstable fragmentation on an orbital timescale is
unlikely to occur until objects have already accreted by some other
process to the mass of the largest known meteorite samples, if at all.
Note that if the disk is turbulent to begin with, settling is
inhibited for particles below a certain size $\sim 1$~cm, as already
mentioned above.

The instability scenario has recently been revisited.  Sekiya (1998)
showed that gravitational instabilities can occur if the dust to gas
ratio is much larger than the cosmic value.  Youdin and Shu (2002)
have discussed mechanisms for increasing this ratio.  In particular,
they showed that gas drag in a laminar disk causes global
redistribution and concentration of small solids as they drift
inwards.  This is because the drift velocity of these particles
decreases as they move in to smaller radii.  Youdin and Shu (2002)
conclude that the dust to gas ratio can become high enough for
gravitational instabilities to develop in less than a few million
years.  However, Weidenschilling (2003) pointed out their arguements
are based on motions obtained for independent particles and claimed
that collective motion due to turbulent stress on the particle layer
acts to inhibit concentration of particles and may prevent
gravitational instability.

Recently, Goodman and Pindor (2000) discovered a secular instability
that may lead to rapid formation of planetesimals in the dust
layer. This instability is not driven by self--gravity but by drag.
If the drag exerted by the (turbulent) gas on the surface layer is
collective, then the drift speed of the dust layer varies inversely
with its surface density.  A slight increase of the mass of some
annulus would therefore result in a decrease of its drift velocity,
leading to the accumulation of the dust inflowing from adjacent larger
radii.  The instability, which grows exponentially in a frame moving
at the unperturbed drift speed, leads to the formation of overdense
rings on an orbital timescale.  These rings further fragment to form
planetesimals of size $\sim 10$~km at 1~AU.  However, turbulent
stresses acting on the particle layer may act to inhibit concentration
of particles in this case also.

If none of the mechanisms described above operate, one has to rely on
sticky particle collisions (coagulation) to obtain planetesimals.
Cuzzi \etal (1993) argue that 10--100~km sized objects can be formed
through this process in the layer in about a million years (also see
below).  Note that coagulation and growth can be made more efficient
if particles can be trapped in vortices or overdense regions of the
disk.  Planetesimal formation in vortices has been studied by, e.g.,
Barge and Sommeria (1995), Tanga \etal (1996) and Klahr and Henning
(1997) who found that growth was indeed much faster.  But note
that Cuzzi et al. (2001) make the point that dust density enhancement
may only be significant for a narrow particle size range.
Furthermore, it is not clear that vortices can live long enough in
disks for any dust densification process to be efficient.  More
recently, trapping of particles in the spiral arms of a
gravitationally unstable disk has been examined.  Numerical
simulations of 1--10~m sized objects in marginally stable,
self--gravitating disks by Rice \etal (2004) show that the drag force
causes the solid objects to drift towards the peaks of the spiral arms
where the density and pressure are highest, in accord with the
behavior predicted by using equation~(\ref{Pgrdmig}) (self--gravity
was not taken into account in this equation, but it would not be
expected to modify the situation significantly as it acts equally on
dust and gas). It is further speculated that the density enhancements
may be sufficient for the growth of planetesimals through direct
gravitational collapse.  Similarly, Durisen \etal (2005) suggest that
the dense rings that appear in their simulations of gravitationally
unstable disks are conducive to accelerated growth of planetesimals.

However, there remains an issue as to the lifetime of these features
in an unstable self--gravitating disk.  In future years, more may be
learnt about the feasibility of the processes discussed here through
more extensive three dimensional numerical simulations.

\subsubsection{Accretion of a planetary core from planetesimals: \\ \\}

Once planetesimals of radius $\sim 10$~km have formed, they proceed to
evolve by processes of velocity dispersion growth due to mutual
gravitational interaction, gas drag, disk planet interaction and
collisional accretion of mass which results in the accumulation of
massive solid cores.

It is believed that the process starts with some planetesimals
entering a regime of runaway growth and forming a population of embryo
cores.  Once these are massive enough to significantly increase the
velocity dispersion equilibrium of the smaller planetesimals, runaway
growth ceases and the system enters an oligarchic growth phase in
which cores separated by 15 or so Hill radii accrete at comparable
rates. Throughout these phases, gravitational focusing plays an
important part in making the accretion rates rapid enough that cores
can enter the Earth mass range within the lifetime of the disk on a
scale of 5~AU.

After this, the cores can become isolated and processes like gap
formation may come into play to inhibit further accretion. The
mobility of the cores as a result of disk protoplanet interaction in
the presence of turbulence may be important.  Note that, if the disk
is laminar, rapid inward type~I migration due to disk protoplanet
interaction poses a serious threat to core survival.  Thus the final
accumulation of cores up to the 5--15~M$_{\oplus}$ range is still
poorly understood. Nonetheless, if favourable circumstances exist
(see discusion and references in sections~\ref{sec:timescales}
and~\ref{sec:turbulence} below), this may be possible within the
lifetime of the gas disk and thus giant planets can form.

A population of cores may be left behind to undergo gas free
accumulation when the gas disk has disappeared. The formation time of
the Earth through this process is expected to be $\sim 2\times
10^8$~yr.  In the outer solar system, gravitational scattering of the
cores can produce escapers leaving a residue to accumulate the
remaining small planetesimals and form Uranus and Neptune in a few
billion years.

We now go on to discuss these processes in more detail.  In doing so,
we follow many authors and adopt a so--called standard, or minimum
mass, solar nebula model. This has a gas surface density profile
$\Sigma \propto r^{-3/2}$ and 2~Jupiter masses within 5.2~AU (Hayashi
1981). This is an estimate of the spread out minimum mass required to
form the planets in the solar system. We further adopt a solid or
condensate surface density of 1\% of the gas surface density.  The
characteristic lifetime of the gas disk is between one and ten million
years and giant planets have to be formed on this timescale in the
solid core followed by gas accretion scenario.

The gas disk is also believed to undergo outward angular momentum
transport allowing accretion onto the central star in the manner of an
accretion disk.  Associated with this is an effective kinematic
viscosity, $\nu$, the origin of which is most likely as\-socia\-ted with
turbulence driven by the magnetorotational instability (Balbus and
Hawley 1991, 1998).  In the standard model, we adopt $\nu/(r^2\Omega) =
\alpha (H/r)^2\Omega = 10^{-5},$ with $\alpha$ being the well known
Shakura \& Sunyaev (1973) dimensionless viscosity parameter. \\

\noindent $\bullet$  {\em Gravitational focusing and runaway accretion: \\ }

\noindent We first show that {\em direct accumulation} of
planetesimals cannot lead to the formation of Earth sized objects
within the disk lifetime.  Consider the simplest possible accumulation
scenario where a core of mass $m_{\alpha}$ grows through the accretion
of sticky solid particles of mass $m_p$ which have number density
$n_p,$ and thus mass density $\rho_{p} =n_p m_p$, through direct
impacts. We suppose that the accreting particles are decoupled from
any gas present, so that their local root mean square velocity
dispersion, $v_p,$ is related to their estimated vertical
semithickness, $H_p$, through $H_p = v_p/\Omega$ and their surface
density is $\Sigma_p = 2\rho_p H_p.$

\noindent Neglecting any effects due to gravitational focusing of the
colliding masses, $m_{\alpha}$ increases at a rate given by:

\begin{equation} {{\rm d}{m_{\alpha}}\over{\rm d}t} = \pi a^2 \rho_{p}v_p,
\end{equation}

\noindent where $a$ is the core radius.  Thus the mass accumulation
timescale is given by:

\begin{equation} {1\over t_{acc}} = {{\rm d}{m_{\alpha}}\over{\rm d}t}{1\over
m_{\alpha}} = {3 \Sigma_p \Omega \over 8 a
\rho_{\alpha}}, \label{Acc}\end{equation}

\noindent where the density of the accumulating core is
$\rho_{\alpha}.$  This gives:

\begin{equation} t_{acc} = {8 a \rho_{\alpha} \over 6\pi \Sigma_p }
\left({r \over
1 \; {\rm AU}}\right)^{3/2} {\rm yr}.\end{equation}

\noindent Adopting $\rho_{\alpha} =1$~gm cm$^{-3}$ 
and $\Sigma_p/\Sigma =0.01,$ we
find:

\begin{equation} t_{acc} = 40  
\left( {\Sigma \over 1 \; {\rm g \; cm^{-2}}} \right)^{-1} {a \over 1
\; {\rm cm}} \left({r \over 1 \; {\rm AU}}\right)^{3/2} {\rm
yr}. 
\label{actsc}
\end{equation}

\noindent This indicates that, for disk models like the minimum mass
solar nebula model with $\Sigma \sim 200$~g~cm$^{-2}$ at 5~AU, 
straightforward accumulation processes operating with ef\-ficient
sticking could conceivably form objects of up to 10--100~km in size
at 5~AU, within the lifetime of protoplanetary disks of a
few million years.  However, to produce objects of significantly
larger size, an increase of the accretion cross section resulting from
gravi\-ta\-tional focusing, which can occur at low relative impact
velocities, needs to be invoked. \\

To take into account {\em gravitational focusing}, we have to include
the gravitational attraction of the core on the colliding particles.
This increases the impact parameter for a collision from the radius
$a$ to $a_{\rm eff} = a \left[ 1+2Gm_{\alpha}/(av_p^2) \right]^{1/2}.$ The
effect of this gravitational focusing is to increase the effective
cross section for impacting collisions accordingly.  Thus
equation~(\ref{Acc}) is modified to become:

\begin{equation} {1\over t_{acc}} = {{\rm d}{m_{\alpha}}\over{\rm d}t}{1\over
m_{\alpha}} = {3 \Sigma_p \Omega \over 8 a \rho_{\alpha}} \left( 1+ {2
Gm_{\alpha} \over v_p^2 a }\right). \label{focus1}\end{equation}

\noindent The enhancement factor given by the term in brackets on the
right hand side of equation~(\ref{focus1}) is large when $v_p$ is very
much smaller than the escape velocity from the core.  For one Earth
mass with a mean density of unity at 5~AU immersed in a planetesimal
swarm with $v_p = 0.01 r\Omega,$ the enhancement factor is close to
4250.

\noindent Thus, under these conditions, forming cores in the Earth
mass range in disks with pro\-per\-ties similar to those of the
minimum mass solar nebula becomes a possibility. Whether the relative
impact speeds are at an appropriate level depends on the velocity
dispersion equilibrium attained by the distribution of cores and
impacting particles or planetesimals.  This we shall consider
below. \\

At this point, we comment that when gravitational focusing dominates,
a phe\-nomenon known as {\em runaway accretion} is possible (e.g.,
Safronov 1969; Wetherill \& Stewart 1989).  This occurs when the
accretion rate increases rapidly enough with mass such that, if a
particular core has a mass slightly exceeding that of other cores, it
will grow increasingly faster and so run away from its neighbors in
mass.

\noindent For example, consider the situation when $v_p$ is
fixed. Then, given $a \propto m_{\alpha}^{1/3},$ from
equation~(\ref{focus1}) we obtain ${\rm d}m_{\alpha} / {\rm d}t
\propto m_{\alpha}^{k} ,$ with $k = 4/3$.  When, as in this case, $k >
1,$ $m_{\alpha} \rightarrow \infty$ in a finite time, such that most
of the evolution time is spent with $m_{\alpha}$ in the neighborhood
of the smallest mass followed by rapid growth to large values.  Hence
the term 'runaway'. In this situation, ultimately a single mass
dominates. More generally, runaway accretion requires $v_p \propto
m_{\alpha}^{k'}$ with $k' < 1/6.$ However, note that accumulation
with strong gravitational focusing may occur without runaway. Then
most time is not necessarily spent with $m_{\alpha}$ close to the
smallest mass. \\

Determination of the root mean square velocity dispersion $v_p$ for
the impacting planetesimals requires consideration of the balance
between growth due to gravitational scattering and damping due to
interaction with the gas which we now consider. \\

\noindent $\bullet$ {\em Analysis of Gravitational Scattering: \\ }
\label{sec:VGR}

\noindent When the planetesimals which form the building blocks from
which the core must accrete have a single characteristic mass
$m_{\alpha} \sim 10^{18}$~g, we require there to be $\sim 10^{10}$
such objects that can be accumulated to form a body in the Earth mass
range.  Such a large number is best treated by statistical means.  A
system of many bodies interacting under their mutual gravitation can
be described by the Fokker--Planck equation which we write in the
form:

\begin{equation}
{{\rm D}f_{\alpha}\over {\rm D}t}=
\Gamma_{\rm coll}(f_{\alpha}) +\Gamma_{\rm gas}(f_{\alpha})
\label{FPEQ}.
\end{equation}

\noindent Here $f_{\alpha}$ denotes the phase space number density of
bodies with mass $m_{\alpha}.$ The operator giving rise to evolution
due to mutual interactions among the planetesimals which lead to
gravitational scattering is $\Gamma_{\rm coll}.$ The operator causing
evolution due to interaction with the gaseous disk is $\Gamma_{\rm
gas}.$ The latter combines the effects of gas drag, eccentricity and
inclination damping as well as orbital migration.  The derivative
operator is taken following a particle orbit under the gravitational
force due to the central mass. Thus:

\begin{equation}
{{\rm D}\over {\rm D}t} \equiv {\partial \over \partial t} +
{\bf v}\cdot {\partial \over \partial {\bf r}}
-{\nab \Phi}\cdot{\partial \over \partial {\bf v} } ,
\end{equation}

\noindent where the central potential at a distance $r$ from the
central mass $M_\star$ is $\Phi = -GM_\star/r.$ The particle position
and velocity vectors measured with respect to an arbitrary coordinate
system are denoted by ${\bf r}$ and ${\bf v}$, respectively.

\noindent We consider the situation when the planetesimals have a
dispersion velocity that is very small compared to the orbital
velocity.  In this case, the position of a particular plane\-tesi\-mal
does not deviate much from a particular radius on orbital timescales,
implying that a local description of the interactions is possible.
Such a description can be given in the context of a local shearing box
(Goldreich \& Lynden--Bell 1965). \\

\noindent {\em --- Local shearing sheet approximation: Distribution
function for the planetesimal swarm: \/} In the shearing box,  we
consider a uniformly rotating local Cartesian coordinate system for
which the origin, located at some point of interest in circular orbit
with radius/semimajor axis $r$, corotates with the Keplerian angular
velocity $\Omega$.  The $x$--axis points radially outwards, the $y$--axis
points in the azimuthal direction in the direction of rotation while
the $z$--axis points in the vertical direction.  A linear expansion for
the combined acceleration due to gravity and the centrifugal force
is used such that:

\begin{equation} -\nab \left( \Phi -{1\over 2}\Omega^2r^2 \right) = 
\left( 3\Omega^2 x, 0, -\Omega^2 z \right).
\end{equation}

\noindent Then for an axisymmetric disk with no dependence on $y$:

\begin{eqnarray} 
{{\rm D} f_{\alpha} \over {\rm D}t} \equiv & & {\partial f_{\alpha}
\over \partial t} + v_x {\partial f_{\alpha}\over \partial x} + v_z
{\partial f_{\alpha} \over \partial z} \nonumber \\
& & - 2\Omega v_x{ \partial
f_{\alpha} \over \partial v_y } + (3\Omega x+2v_y)\Omega { \partial
f_{\alpha} \over \partial v_x } -\Omega^2 z{ \partial f_{\alpha}\over
\partial v_z }.
\label{CBE}
\end{eqnarray}

\noindent 
The Liouville equation ${\rm D} f_{\alpha} / {\rm D}t =0$ does not
have a solution corresponding to an isotropic Gaussian but does have
one for which $f_{\alpha}$ takes the form of an anisotropic Gaussian
given by:

\begin{equation} f_{\alpha} = C_{\alpha} \exp\left( -{v_x^2\over
2\sigma_x^2} -{u_y^2\over 2\sigma_y^2} -\frac{v_z^2 +\Omega^2 z^2}{2
\sigma_z^2}\right)
\label{VelI}.
\end{equation}

\noindent Here the root mean square velocity dispersions
$(\sigma_x,\sigma_y,\sigma_z)$ are constant and such that $\sigma_y
=\sigma_x /2.$ There is no constraint on $\sigma_z$.  The velocity
$u_y =v_y +3\Omega x/2$ is measured relative to the local circular
velocity, $v_y = -3\Omega x/2$, and we shall adopt local relative
velocity vectors ${\bf v} = (v_x, u_y, v_z).$ The constant
$C_{\alpha}$ is related to the spatial number density in the midplane,
$n_{\alpha},$ through: 

\begin{equation} C_{\alpha} = { n_{\alpha} \over (2 \pi)
^{3/2} \sigma_x \sigma_y \sigma_z }. 
\end{equation}

\noindent {\em --- Evolution of the cores and planetesimal swarm: \/}
In general, a growing core is expected to accrete from a planetesimal
swarm with a distribution of masses that itself evolves under
gravitational scattering and accretion. However, following Wetherill
\& Stewart (1989) and Ida \& Makino (1993), we simplify matters by
assuming that there are only two populations present. One corresponds
to the growing cores and the other to the planetesimal swarm from
which they accrete.  The latter distribution consists of a large
number of objects with a small fixed mass, $m_{\alpha}.$ The former
consists of a much smaller number of cores of larger mass,
$m_{\beta},$ which increases as the system evolves.  We adopt
$m_{\alpha} = 10^{18}$~g.  This is the mass that can be accumulated
within the disk lifetime without the need for gravitational focusing
as indicated above.  This is also the mass that emerges from the
assumption that planetesimals are able to form through the
gravitational instability in a dust layer (Goldreich and Ward 1973).

\noindent The evolution of each component may be described by a
Gaussian such as that given by equation~(\ref{VelI}) with velocity
dispersions and number density that slowly evolve under the action of
collisions. \\

\noindent {\em --- System evolution through gravitational scattering:
\/} Encounters between planetesimals that occur without direct
physical impacts tend to convert kinetic energy from shear into that
of random motions, in the same way as the action of viscosity converts
energy from shear into thermal motions in a gaseous disk. For the
planetesimal swarm, effects due to evolution resulting from
gravitational scattering occur on a timescale that is much longer
than orbital.

\noindent To model this, we use the form of $\Gamma_{\rm coll}$ given
by Binney \& Tremaine (1987).  This neglects rotation about the
central mass, which should be a reasonable approximation as long as
the timescale associated with a binary encounter is short compared to
$\Omega^{-1}.$ This in turn requires that $\sigma_x/\Omega > r_H,$
where $r_H = [ m_{\gamma}/ (3M_\star)]^{1/3} r$ is the Hill radius
appropriate to the largest characteristic mass, $m_{\gamma},$ involved
in the encounter.

\noindent Following Binney \& Tremaine (1987), we write using the
summation convention for repeated indices:

\begin{equation} 
\Gamma_{\rm coll}(f_{\alpha}) = -\frac{\partial}{\partial v_i}(A_i
f_{\alpha}) +\frac{1}{2}\frac{\partial}{\partial v_i} \left( D_{ij}
\frac{\partial f_{\alpha}}{\partial v_j} \right)
\label {COP}.
\end{equation}

\noindent We consider two kinds of interaction. The first, which is
important during the early stages of the core mass build up, arises
when objects in the distribution with small mass, $m_{\alpha},$
interact with each other. Then

\begin{equation} A_i = 4\pi G^2 \ln(\Lambda_{\alpha}) 
m_{\alpha}^2{\partial h \over
\partial v_i} \label{light0}, \end{equation}

\noindent and

\begin{equation}
D_{ij}=  4\pi G^2 \ln(\Lambda_{\alpha}) m_{\alpha}^2{\partial^2 g
 \over \partial v_i \partial v_j},
\label{light}
\end{equation}

\noindent with

\begin{equation}
g = 
\int f_{\alpha}({\bf v'})|{\bf v}-{\bf v'}|
\,{\rm d}^3{\bf v}' \; \; \; {\rm and} \; \; \;
h = \int {f_{\alpha}({\bf v'})\over |{\bf v}-{\bf v'}|}
\,{\rm d}^3{\bf v}'.
\end{equation}

\noindent We take $\Lambda_{\alpha} = 3\sigma_x^2 H_\alpha /(4Gm_{\alpha})$,
giving the ratio of maximum to minimum impact parameters as disc
semithickness to impact parameter for a typical deflection
(e.g., Binney \& Tremaine 1987).

\noindent The second type of interaction we consider, which is
important during the later stages of core mass build up, is between an
object from the first distribution with small mass, $m_{\alpha},$ and
an object from the second distribution with large mass, $m_{\beta},$
the number density of these being $n_{\beta}.$ Because of their high
inertia, these can be assumed to be in circular orbit and have zero
velocity dispersion.  In this limit: 

\begin{equation} A_i = 0, \label{heavy0}\end{equation} 

\noindent and 

\begin{equation} D_{ij}= 4\pi G^2 n_{\beta} \ln(\Lambda_{\beta}) m_{\beta}^2\left(
\delta_{ij}v^2 - v_i v_j\right)/v^3,\label{heavy} \end{equation} 

\noindent where $v = |{\bf v}|.$ Here $\Lambda_{\beta} = 3\sigma_x^2
H_\alpha /(4Gm_{\beta}).$ \\

\noindent {\em --- Growth of the velocity dispersion: \/} The effect
of gravitational scattering is to cause the velocity dispersion to
increase. This is most easily seen by formulating the Boltzmann ${\cal
H}$ theorem for the problems we are considering. This states that for
a single mass species interacting either with itself or another in
fixed circular orbits with no velocity dispersion,

\begin{equation}
{\cal H}\equiv -\int f_{\alpha}\ln(f_{\alpha})\,{\rm d}^3{\bf v}
\label{Hdef}
\end{equation}

\noindent increases monotonically with time.  ${\cal H}$, which can be
related to the entropy, can remain constant only for an isotropic
Gaussian which cannot be attained here because of the form of particle
orbits in the central potential (see eq.~[\ref{VelI}]). Using the
distribution function given by equation~(\ref{VelI}), and assuming
that $n_{\alpha}$ and the ratio of velocity dispersion components are
independent of time (see Papaloizou \& Larwood 2000), we obtain at the
midplane ($z=0$):

\begin{equation}
{{\rm d}{\cal H}\over{\rm d}t} =
{ 3 n_{\alpha} \over \sigma_x} {{\rm d}\sigma_x  \over {\rm d}t}.
\label{Hdot}
\end{equation}

\noindent Thus ${\rm d}{\cal H}/{\rm d}t > 0$ implies that the
velocity dispersion must increase with time. \\

We first evaluate ${\rm d}{\cal H}/{\rm d}t$ for a single mass species
interacting with itself.  Using the expression~(\ref{Hdef}) of ${\cal
H}$ and $\Gamma_{\rm coll} (f_{\alpha})$ given by equations~(\ref{COP}),
(\ref{light0}) and~(\ref{light}), we obtain:

\begin{eqnarray}
{{\rm d}{\cal H}\over {\rm d}t}  
& = & -4\pi G^2 m_{\alpha}^2\ln(\Lambda_{\alpha})
\int { 1 \over |{\bf v} - {\bf
v}'|} {\partial f_{\alpha}({\bf v}) \over \partial v_i} {\partial
f_{\alpha} ({\bf v'}) \over \partial v_i^\prime} 
\,{\rm d}^3{\bf v} \,{\rm d}^3{\bf v'} \nonumber \\
& + & 2\pi G^2
m_{\alpha}^2\ln(\Lambda_{\alpha}) \int {\partial f_{\alpha} ({\bf v})
\over \partial v_i} {\partial f_{\alpha} ({\bf v}) \over \partial v_j} { 
f_{\alpha} ({\bf v'})\over f_{\alpha} ({\bf v})} 
{\partial^2 |{\bf v} - {\bf v}'| \over \partial v_i \partial
v_j} \,{\rm d}^3{\bf v} \,{\rm d}^3{\bf v'}.
\label{CAP}
\end{eqnarray}

\noindent When $\sigma_y=\sigma_z$, the integral may be evaluated
analytically for an anisotropic Gaussian specified by
equation~(\ref{VelI}).  The relaxation time defined through $t_R =
\sigma_x/({\rm d} \sigma_x/ {\rm d}t)$ can then be calculated from
equation~(\ref{Hdot}) in the disk midplane:

\begin{equation}
{1\over t_R}=
 {8 \sqrt{\pi} G^2 m_{\alpha}^2\ln(\Lambda_{\alpha}) 
n_{\alpha}\over \sigma_x^3}
\left[{\sqrt{3}\over 4}\ln\left({2+\sqrt{3} \over
2-\sqrt{3}}\right)-1\right].
\label{tr}
\end{equation}

\noindent Here we have used the relation $\sigma_y =\sigma_x /2$
which is valid in the Keplerian case.

\noindent In terms of the semithickness of the planetesimal disk
$H_{\alpha}=\sigma_z/\Omega$ and the planetesimal surface density
$\Sigma_{\alpha} =\sqrt{2\pi}H_{\alpha} m_{\alpha}n_{\alpha},$ we have

\begin{equation}
\Lambda_{\alpha} = \left( \frac{3M_\star}{m_\alpha} \right)
\left( \frac{H_\alpha}{r} \right)^3 ,
\label{lambda}
\end{equation}
and
\begin{equation}
{1\over t_R} = 0.03\ln(\Lambda_{\alpha}){M_D m_{\alpha}\over M_\star^2}
\left({r\over H_{\alpha}}\right)^4\Omega,
\label{trr}
\end{equation}

\noindent where $M_D =\pi \Sigma_{\alpha}r^2$ gives an estimate of the
mass within the protoplanetary disk comprising the planetesimal swarm
that is contained within $r$.

\noindent A number of effects may act to oppose the growth of the
velocity dispersion induced by gravitational scattering and allow a
quasi steady state distribution to be reached.  These include (see
below for additional details) gas drag, tidal interactions with the
disk and physical collisions between planetesimals which, if very
inelastic, will prevent the dispersion velocity from significantly
exceeding the planetesimal escape velocity.

\noindent To estimate likely relaxation times, we adopt
$M_\star=1$~M$_\odot$, $m_{\alpha} =10^{18}$~g together with
$H_{\alpha}/r =3\times 10^{-4},$ which corresponds to a dispersion
velocity comparable to the escape velocity, and $M_D=10$~M$_{\oplus}$.
This gives $t_R \sim 5 \times 10^5$~yr at $r=1$~AU.

\noindent But note that the relaxation time is very sensitive to the
velocity dispersion, making the processes that determine it significant
for in turn determining the timescale for core accumulation. \\


We now evaluate ${\rm d}{\cal H}/{\rm d}t$ in the case when the
gravitational interactions of the planetesimal swarm, labelled by
$\alpha,$ with the system of cores with large masses in circular
orbits with zero velocity dispersion, labelled by $\beta$, become more
important than interactions between its own members. Using
equations~(\ref{COP}), (\ref{heavy0}) and~(\ref{heavy}), conjointly
with the expression~(\ref{Hdef}) of ${\cal H}$, we find:

\begin{equation} {{\rm d}{\cal H}\over {\rm d}t} = 2\pi G^2 m_{\beta}^2
n_{\beta}\ln(\Lambda_{\beta}) \int {\partial {\rm ln} f_{\alpha} ({\bf v})
\over \partial v_i} {\partial f_{\alpha} ({\bf v}) \over \partial v_j}
{\delta_{ij}v^2 - v_i v_j \over v^3 } \,{\rm d}^3{\bf v}
\label {CAP1}.
\end{equation} 

\noindent In this case, with identical assumptions about the ratios of
the velocity dispersion components as in the self--interaction case,
the integral is readily performed to give the relaxation time in the
midplane through: 

\begin{equation} {1\over t_{R,mid}}= {16 \sqrt{\pi} 
G^2 m_{\beta}^2\ln(\Lambda_{\beta})
n_{\beta}\over  \sqrt{2} \sigma_x^3} \left[{\sqrt{3}\over
2}\ln\left(2+\sqrt{3} \right)-1\right].
\label{Ogsctmid}
\end{equation}

\noindent Here $\Lambda_{\beta} = (3M_\star H_{\alpha}^{3})/(m_{\beta} r^3).$
When, as here, $m_{\beta} \gg m_{\alpha},$ the disk semithickness
associated with the heavy distribution, $H_{\beta} =
\Sigma_{\beta}/(\sqrt{2\pi} m_{\beta} n_{\beta})$, is much smaller
than that associated with the light distribution.  Accordingly, we take
an appropriate average of the relaxation rate by multiplying it by the
ratio $3H_{\beta}/(\sqrt{2}H_{\alpha}),$ this being a measure of the proportion
of time spent by a member of the light distribution interacting with
the heavy distribution.  Thus we obtain the final mean relaxation rate:

\begin{equation} {1\over t_{R}}= {24  G^2
m_{\beta}\ln(\Lambda_{\beta})\Sigma_{\beta}\over \sqrt{2} H_{\alpha}
\sigma_x^3} \left[{\sqrt{3}\over 2}\ln\left(2+\sqrt{3}
\right)-1\right]. \\
\label{Ogsct}
\end{equation}

\noindent {\em --- Comparison of scattering by small and large masses:
\/} As an ensemble of planetesimals evolves, the direction of
evolution is expected to be such that, because of runaway accretion,
the distribution becomes dominated by an increasing number of larger
masses.  Prior to this, when the smaller masses dominate the dynamical
relaxation, because their mass does not change, and their velocity
dispersion equilibrium is determined by balance between dynamical
relaxation and damping through interaction with the gas disk, the 
relaxation rate is independent of the magnitude of the larger mass
$m_{\beta}.$ Accordingly, runaway accretion by these may occur (see
above).

\noindent The condition for the larger masses to dominate the
relaxation is determined by equating the relaxation rates given by
equations~(\ref{tr}) and~(\ref{Ogsct}) and is, neglecting any
variation in $\ln(\Lambda)$: 

\begin{equation} 3 m_{\beta}\Sigma_{\beta} > m_{\alpha}\Sigma_{\alpha}
. \label{Olcon} \end{equation}

\noindent Simulations (see, e.g., Ida \& Makino 1993, Thommes \etal
2002) indicate that, once the larger masses start to dominate the
relaxation, the system enters the orderly mode of oligarchic accretion
in which the masses $m_{\beta}$ remain in near circular orbits
separated by $\sim 10f$ times the Hill radius $r_H =
[m_{\beta}/(3M_\star)]^{1/3} r,$ with $f$ being a number of order
unity, and grow by capturing the low mass material in their
neighborhood.  Then $\Sigma_{\beta} = m_{\beta}/(20 f \pi r r_H).$
Using this, the condition~(\ref{Olcon}) implies that for oligarchic
accretion we require:

\begin{equation} m_{\beta} > 2.2\times 10^{-7}f^{3/5} \left({r\over 1 \; {\rm
AU}}\right)^{6/5} \left({\Sigma_{\alpha} m_{\alpha} \over
10^{19}{\rm  gm}^2 {\rm cm}^{-2}}\right)^{3/5} \; {\rm M}_{\oplus} .\end{equation} 

\noindent This strongly suggests that the transition to oligarchic
accretion occurs at quite an early stage in the core accumulation
process. \\

\noindent {\em --- Damping processes: \/}

\noindent Gas Drag: We consider a planetesimal of mass $m_{\alpha},$
density $\rho_{\alpha}$ and radius $a$ moving relative to the disk gas
with speed $v.$ The equation of motion for $v$ is:

\begin{equation} m_{\alpha}{{\rm d}{v}\over{\rm d}t} = 
-\pi a^2 C_D \rho v^2, \end{equation}

\noindent where $C_D,$ being a dimensionless number normally in the
range 0.1--1, is the drag coefficient.  This leads to a relative
velocity damping time, $t_{gd} \equiv v/(dv/dt)$, equivalent to:

\begin{equation} {1\over \Omega t_{gd}}= {(36\pi)^{1/3} \Sigma C_D \over
8\rho_{\alpha}^{2/3} m_{\alpha}^{1/3}} {er\over H}, \label{GDRAG}\end{equation}

\noindent where we have set $v=er\Omega \sim \sigma_x $ with $e$
corresponding to an orbital eccentricity for a planetesimal.  In cgs
units, this gives: 

\begin{equation} t_{gd} =5.27\times 10^3 \left(m_{\alpha}\over
10^{18}{\rm gm}\right)^{1/3} {H\over er}{50 \rho_{\alpha}^{2/3} \over \Sigma
C_D} \left({r\over 1 \; {\rm AU} }\right)^{3/2} \; {\rm yr}. \label{GDRAG1}
\end{equation}

\noindent Note that this time can be small for small masses, which is
related to the need to evolve fast through the size range where
decoupling from the gas first occurs to objects of $\sim 10^{18}$~g
(see section~\ref{sec:planetesimals}). \\

\noindent Relative importance of gas drag and disk tides: Because of
the dependence on protoplanet mass, gas drag is more effective for
smaller masses while disk tides take over for larger masses.  The
cross over mass, which is also the mass for which the damping time is
a maximum, can be estimated by equating the gas drag time given by
equation (\ref{GDRAG1}) with the eccentricity decay time resulting from
disk tides given by equation (\ref{efit}) (see below).  This gives:

\begin{equation} { m_{\alpha} \over {\rm M}_{\oplus} } = 5.2\times
10^{-4} \; {f_{s}^{15/8} C_D^{3/4}\over \rho_{\alpha}^{1/2}}
\left({H/r\over 0.05}\right)^{3} \left({H/r\over e}\right)^{-3/4}
\left({r\over \; 5.2{\rm AU} }\right)^{-3/2}.
\label{efit1}
\end{equation}

\noindent Although there is significant sensitivity to 
protoplanet parameters, disk location, eccentricity and aspect ratio 
(going from $5.2$~AU to $1$~AU, other things being equal, increases it
by an order of magnitude) the crossover mass is characteristically
on the in the range $10^{-4} - 10^{-2}$ earth masses.  Thus gas drag
will be the dominant eccentricity damping process when the
planetesimals first form. However, if many larger mass cores form and
then interact during the later stages of core accumulation, disk tides
will dominate. \\

\noindent $\bullet$ {\em Velocity dispersion equilibrium: \\ }

\noindent From the above estimates we anticipate the following
evolutionary sequence: In the early phases, when the core masses are
comparable to the planetesimal mass, the ba\-lance is between relaxation
by self--interaction among the small masses against gas drag. In this
case the velocity dispersion depends only on the small unchanging mass
$m_{\alpha}.$ A few masses can then undergo runaway accretion growing
to large values $m_{\beta}.$ When this is high enough, the larger
masses dominate the gravitational scattering of the planetesimals
which is again balanced against gas drag. The velocity dispersion in
this phase increases with $m_{\beta}$ and the core accumulation moves
into the more orderly oligarchic regime (Ida \& Makino 1993). \\

\noindent {\em --- Scattering due to larger bodies balanced against
gas drag: \/}
In this case we have:

\begin{equation} 1/ t_R = 1 / t_{gd}, \label{VDSPEQ} \end{equation}

\noindent where we use equation~(\ref{Ogsct}) for $t_R.$ This gives

\begin{equation} {16 r m_{\beta}\ln(\Lambda_{\beta})\Sigma_{\beta}\over
M_\star^2}\left[{\sqrt{3}\over 2}\ln\left(2+\sqrt{3} \right)-1\right] =
\left(\sigma_x\over r\Omega\right)^5 {(36\pi)^{1/3} \Sigma C_D \over
16\rho_{\alpha}^{2/3} m_{\alpha}^{1/3} H}.\end{equation}

\noindent Numerically for $\ln(\Lambda_{\beta}) =1$: 

\begin{eqnarray}
\left({\sigma_x\over r\Omega}\right)^5 = & & 4 \times
10^{-6} \times \nonumber \\
& & \left({\Sigma_{\beta} \over C_D \Sigma}\right) \left({10^{18}
\rho_{\alpha}\over m_{\alpha} }\right)^{2/3} \left({r\over 5.2 \; {\rm
AU} }\right)^{2}\left( {20 H\over r}\right) \left({m_{\alpha}m_{\beta}
{\rm M}_{\odot}^2\over 10^{18} {\rm M}_{\oplus} M_\star^2} \right).
\end{eqnarray}

\noindent Thus, even when $m_{\beta}$ has grown to the order of an
Earth mass, the dispersion velocity $\sigma_x$ plausibly remains such
that $\sigma_x \ls H/r,$ for $r$ in the AU range, justifying the
application of gas drag and maintaining a strong degree of
gravitational focusing. For example, when $m_{\alpha} = 10^{18}$~g,
$m_{\beta}= 1$~M$_{\oplus}$ and $\Sigma_{\beta} = 2.5 \times 10^{-3}
C_D \Sigma,$ we find $H_{\beta}/r \sim 0.025$, which gives an
enhancement factor due to gravitational focusing (see above) of $\sim
400$ at $5.2$~AU.  This order of enhancement is needed in order to
accumulate cores in the Earth mass range.

\noindent In addition, the velocity dispersion $\sigma_x$ increases as
$m_{\beta}^{1/5}$ and $k$, defined in the paragraph on runaway
accretion, is equal to $14/15$.  Since this is less than unity, the
accumulation does not occur in a runaway manner in which a single core
outgrows everything else. Instead, a number of cores of comparable
mass, or oligarchs, that are adequately separated, grow together. But,
nonetheless, gravitational focusing can and must remain important if
core accumulation to the Earth mass range is to occur within the gas
disk lifetime. \\

\noindent {\em --- Scattering due to isolated larger bodies balanced
against disk tides when gas is present: \/} When all or most of the
small mass planetesimals are accumulated, the likely situation that
develops is one for which large mass cores, or oligarchs, of
comparable mass $m_{\beta}$ are in near circular orbits separated by
$10f$ Hill radii, in isolation (Chambers \etal 1996). The isolation
mass may be estimated from the mean surface density of the accumulated
solid material as:

\begin{equation} m_{\beta} =\left[{20 f \pi r^2\Sigma_{\beta}\over
(3 {\rm M}_\odot)^{1/3}} \right]^{3/2} = 0.46
f^{3/2}\left({\Sigma_{\beta}\over 1 \; {\rm g \;
cm}^{-2}}\right)^{3/2} \left({r \over 5.2 \; {\rm AU}}\right)^{3} {\rm
M}_{\oplus},\end{equation}

\noindent giving objects characteristically in the Earth mass range
for a minimum mass solar nebula at 5.2~AU. But note that, for such a
nebula, the isolation mass $\propto r^{3/4},$ being $\sim
0.11f^{3/2} {\rm M}_{\oplus}$ at 1~AU.

\noindent Such masses may slowly interact building up velocity
dispersion and resulting in further accumulation (Papaloizou \&
Larwood 2000).  For such large planetesimal masses $m_{\beta}$, tidal
interaction with the disk, should there be gas present, becomes more
important than gas drag. Then we should use the timescale $t_e$ as
given by equation~(\ref{efit}) (see below) rather than $t_{gd}$ in
equation~(\ref{VDSPEQ}) which determines the velocity dispersion
equilibrium.  The timescale $t_e$ is obtained from the
protoplanet--disk interaction theory described below.  As this
timescale is significantly shorter than the associated possible type~I
migration timescale $t_I$, we expect that a quasi--equilibrium is set
up before significant migration occurs and that this then drives the
evolution of the system, which will proceed under conditions of
quasi--equilibrium.  We recall the relaxation timescale for a central
point solar mass potential in the form:

\begin{equation}
t_R =
\frac{5}{\ln(\Lambda)}
\frac{{\rm M}_{\odot}^2}{\pi \Sigma_{\beta} r^2  m_\beta}
\left(\frac{H_{\beta}}{r}\right)^4
\left(\frac{r}{1 \; {\rm AU}}\right)^{3/2} \; {\rm yr},
\end{equation}

\noindent where $\Lambda = \left( 3M_\star/m_\beta \right) \left(
H_\beta/r \right)^3.$ Equating $t_e$ from equation~(\ref{efit}) and
$t_R$, assuming $e<H/r$ and $f_s =1 $, gives the equilibrium
semithickness of the protoplanet distribution $H_\beta$ in terms of
the gas disk semithickness $H$:

\begin{equation}
\frac{H_{\beta}}{H} = 0.6 [\ln(\Lambda)]^{1/4}
\left({ \Sigma_{\beta} \over 4\Sigma }\right)^{1/4}
\left({r\over 1 \; {\rm AU}}\right)^{-1/8}.
\label{EQUIL}
\end{equation}

\noindent Since $H_{\beta}/H$ depends only very weakly on $\Lambda$,
it is essentially independent of the protoplanet mass $m_{\beta}$, and
it depends only weakly on other parameters. For the representative
values $m_{\beta} =0.1 {\rm M}_{\oplus},$ $\Sigma_{\beta} / (4 \Sigma)
= 10^{-3}$ and $H_{\beta}/r =0.01,$ we obtain $H_{\beta}= 0.15H$ at
1~AU.  Hence the protoplanet swarm is expected to remain thin and
confined within the gaseous nebula as is confirmed by numerical
simulations (Papaloizou \& Larwood 2000). Note too that this implies
that gravitational focusing will be important, enabling core mass
build up to the Earth mass regime.  However, disk tide induced type~I
orbital migration then becomes important on a $10^6$~yr timescale
(Ward 1997, Tanaka \etal 2002), threatening the survival of the cores
against falling into the central star (see below). \\

\noindent {\em --- The gas free case: \/ } When gas is absent so is
the problem of inward core migration due to type~I migration. However,
should all the mass of solids be within the massive cores, there is
nothing to balance velocity dispersion growth other than the effects
of inter--core collisions. These are only likely to be limited when
the relative velocity is comparable to the escape velocity (Safronov
1969). If the final phase of the accumulation of the terrestrial
planets occurs in this way (Chambers \& Wetherill 1998), we may
estimate the evolution timescale for attaining 1~M$_{\oplus}$ at 1~AU
using equation~(\ref{actsc}), as gravitational focusing is of marginal
importance, to be $\sim 2 \times 10^{8}$~yr.  This timescale being
$\propto r^{3}$ implies an unacceptably long time to form Uranus and
Neptune for minimum mass solar nebula models.  In addition, the escape
velocity from a 1~M$_{\oplus}$ core exceeds the escape velocity from
the sun if $ r > 12$~AU, implying that escapers will be generated in
this region.  Goldreich \etal (2004) suggest that a residue of small
particles derived from collisions amongst the original planetesimals,
such as has been produced in calculations directed towards
explaining the distribution of Kuiper belt objects, by
Kenyon  \& Bromley  (2004), 
may remain and damp the velocity dispersions, reducing both evolution
times and the orbital eccentricities of the remaining growing cores,
once their number has been adequately reduced.

\subsection{Formation of giant planets: the core accretion model}

In order to produce a Jovian mass giant planet, the solid cores
discussed above must accrete gas.  According to the {\em critical core
mass} model first developed by Perri and Cameron~(1974) and Mizuno (1980;
see also Stevenson 1982, Wuchterl 1995, Papaloizou \& Terquem 1999),
the solid core grows in mass along with the atmosphere in
quasi--static and thermal equilibrium until it reaches the critical
mass, $M_{crit},$ above which no equilibrium solution can be found for
the atmosphere.  As long as the core mass is smaller than $M_{crit}$ ,
the energy radiated from the envelope into the surrounding nebula is
compensated for by the gravitational energy which the planetesimals
entering the atmosphere release when they collide with the surface of
the core.  During this phase of the evolution, both the core and the
atmosphere grow in mass relatively slowly.  By the time the core mass
reaches $M_{crit}$, the atmosphere has grown massive enough so that
its energy losses can no longer be compensated for by the accretion of
planetesimals alone.  At that point, the envelope has to contract
gravitationally to supply more energy.  This is a runaway process,
leading to the very rapid accretion of gas onto the protoplanet and to
the formation of giant planets such as Jupiter.  In earlier studies, it
was assumed that this rapid evolution was a dynamical collapse, hence
the designation 'core instability' for this model.  The critical mass
can be estimated to be in the range 5--15~M$_{\oplus}$ depending on
physical conditions and assumptions about grain opacity etc.  (see,
e.g., Pollack \etal 1996, Ikoma \etal 2000, Papaloizou \& Nelson
2005).  There is thus a need to evolve beyond a typical isolation mass
appropriate to a minimum solar nebula model and some authors have
suggested that a nebula up to ten times more massive than the minimum
mass solar nebula is needed (e.g., Pollack \etal 1996).

However, this late evolutionary phase is not yet well studied because
it depends on uncertain global phenomena such as inhibition of
planetesimal accretion due to isolation, competition with other cores,
gap formation (Thommes \etal 2002) and how mobile the cores are under
a modified type~I migration which may be stochastic in the presence of
turbulence (Nelson \& Papaloizou 2004) or be reversed by large scale
toroidal magnetic fields (Terquem 2003).  Indeed, it is believed that
in order for giant planet formation to occur and to be able to fit the
observed distribution of extrasolar planets, type~I migration rates
obtained for a laminar disk have to be reduced by such processes
(e.g., Alibert \etal 2004).  However, this has yet to be explored fully.

Time--dependent numerical calculations of protoplanetary evolution by
Bodenheimer~\& Pollack (1986) and Pollack \etal (1996) support the
critical core mass model, although they show that the core mass beyond
which runaway gas accretion occurs, which is referred to as the
'crossover mass', is slightly larger than $M_{crit}$, and that the
very rapid gravitational contraction of the envelope is not a
dynamical collapse.  The designation 'crossover mass' comes from the
fact that rapid contraction of the atmosphere occurs when the mass of
the atmosphere is comparable to that of the core.  Once the crossover
mass is reached, the core no longer grows significantly through
accretion of planetesimals.

The evolutionary calculations by Pollack \etal (1996) show that the
evolution of a protoplanet with a gaseous envelope is governed by
three distinct phases. During phase~1, runaway planetesimal accretion,
as described above, occurs, bringing about the depletion of the
feeding zone of the protoplanet leading to isolation. At this point,
when phase~2 begins, the atmosphere is massive enough that the
location of its outer boundary is determined by both the mass of gas
and the mass of the core. As more gas is accreted, this outer radius
moves out, so that more planetesimals can be captured enabling yet
more gas to enter the atmosphere.  The protoplanet grows in this way
until the core reaches the crossover mass, which should be identified
with the Mizuno (1980) critical core mass, at which point runaway gas
accretion occurs and phase~3 begins.  The timescale for planet
formation is determined almost entirely by phase~2, i.e. by the rate
at which the core can accrete planetesimals, and is found to be a few
million years at 5~AU, being comparable to the gas disk lifetime.
Note that although phase~3 is relatively rapid compared to phase~2 for
cores as large as 15~M$_{\oplus}$ , it may become much longer for
smaller mass cores which become critical for reduced planetesimal
accretion rates (see Pollack \etal 1996, Papaloizou \& Terquem 1999
and below).

Note that isolation of the protoplanetary core, as it occurs at the
end of phase~1, may be prevented under some circumstances by tidal
interaction with the surrounding gaseous disk (Ward~\& Hahn 1995) or
by interactions with other cores (Chambers \etal 1996).

Conditions appropriate to Jupiter's present orbital radius are
normally considered and then for standard 'interstellar' dust
opacities the critical or crossover mass is found to be around
15~M$_{\oplus}$.  This is rather large as recent observations indicate
that Jupiter has a solid core not exceeding $\sim 5$~M$_{\oplus}.$

The similarity between the critical and crossover masses is due to the
fact that, although there is some liberation of gravitational energy
as the atmosphere grows in mass together with the core, the effect is
small as long as the atmospheric mass is small compared to that of the
core.  Consequently, the hydrostatic and thermal equilibrium
approximation for the atmosphere is a good one for core masses smaller
than the critical value.

\subsubsection{Equations governing an envelope at equilibrium:\\ \\} 

We denote $R$ the spherical polar radius in a frame with origin at the
center of the protoplanet.  We assume that we can model the
protoplanet as a spherically symmetric nonrotating object.  We
consider the envelope at hydrostatic and thermal equilibrium.  The
equation of hydrostatic equilibrium is:

\begin{equation}
\frac{dP}{d R} = - g \rho ,
\label{dpdvarpi}
\end{equation}

\noindent where $P$ is the pressure, $\rho$ is the mass density per
unit volume and $g=G M(R) / R^2$ is the acceleration due to gravity,
$M(R)$ being the mass contained in the sphere of radius $R$ (this
includes the core mass if $R$ is larger than the core radius).  We
also have the definition of density:

\begin{equation}
\frac{dM}{d R} = 4 \pi R^2 \rho.
\label{dmdvarpi}
\end{equation}

At the base of a protoplanetary envelope, the densities are so high
that the gas cannot be considered to be ideal.  Thus we adopt the
equation of state for a hydrogen and helium mixture given by Chabrier
\etal (1992).  We take the mass fractions of hydrogen and helium to
be 0.7 and 0.28, respectively.  We also use the standard equation of
radiative transport in the form:

\begin{equation}
\frac{dT}{d R} = \frac{-3 \kappa \rho}{16 \sigma
T^3} \frac{L}{4 \pi R^2} .
\label{dtdvarpi}
\end{equation}

\noindent Here $T$ is the temperature, $\kappa$ is the opacity, which
in general depends on both $\rho$ and $T$, $\sigma$ is the
Stefan--Boltzmann constant and $L$ is the radiative luminosity.
Denoting the radiative and adiabatic temperature gradients by
$\nabla_{rad}$ and $\nabla_{ad}$, respectively, we have:

\begin{equation}
\nabla_{rad} = \left( \frac{\partial \ln T}{\partial \ln P}
\right)_{rad} = \frac{3 \kappa L_{core} P}{64 \pi
\sigma G M T^4} ,
\label{dTdr_rad}
\end{equation}

\noindent and

\begin{equation}
\nabla_{ad} = \left( \frac{\partial \ln T}{\partial \ln P} \right)_s ,
\end{equation}

\noindent with the subscript $s$ denoting evaluation at constant
entropy.  In writing equation~(\ref{dTdr_rad}), we have assumed that
the only energy source comes from the accretion of planetesimals onto
the core.  The gravitational energy of the planetesimals going through
the planet atmosphere is released near the surface of the core which,
as a result, outputs a total luminosity $L_{core}$ given by (see,
e.g., Mizuno 1980; Bodenheimer \& Pollack 1986):

\begin{equation}
L_{core} = \frac{ G M_{core} \dot{M}_{core}}{ r_{core}} .
\end{equation}

\noindent 
Here $M_{core}$ and $r_{core}$ are respectively the mass and the
radius of the core, and $\dot{M}_{core}$ is the planetesimal accretion
rate onto the core (note that some of the planetesimals entering the
atmosphere may not end up colliding with the core).

Note however that, as the the interiors of these protoplanet models
are convective and approximately isentropic, the above formalism can be
used when the energy source corresponds to settling of the accreting
gaseous atmosphere.  One can use the conservation of mass and energy to
determine the evolution of the protoplanet (see Papaloizou \&
Nelson~2005, and sections~\ref{typeAev} and~\ref{typeBev} below).

\noindent When $\nabla_{rad} < \nabla_{ad}$, the atmosphere is stable to
convection and thus all the energy is transported by radiation,
i.e. $L=L_{core}$.  When $\nabla_{rad} > \nabla_{ad}$, there is
instability to convection. Then, part of the energy is transported by
convection, and $L_{core}= L+L_{conv}$, where $L_{conv}$ is the
luminosity associated with convection.  We use mixing length theory to
evaluate $L_{conv}$ (Cox \& Giuli~1968).  This gives:

\begin{equation}
L_{conv} = \pi R^2 C_p \Lambda_{ml}^2 \left[ \left( \frac{\partial
T}{\partial R} \right)_s - \left( \frac{\partial T}{\partial R}
\right) \right]^{3/2} \times \sqrt{ \frac{1}{2} \rho g \left| \left(
\frac{\partial \rho}{\partial T} \right)_P \right| } ,
\end{equation}

\noindent where $\Lambda_{ml}=|\alpha_{ml}P/(dP/d R)|$ is the mixing
length, $\alpha_{ml}$ being a constant of order unity, $\left(
\partial T/\partial R \right)_s = \nabla_{ad} T \left( d \ln P / d R
\right)$, and the subscript $P$ means that the derivative has to be
evaluated for a constant pressure.  All the required thermodynamic
parameters are given by Chabrier \etal (1992).  In the numerical
calculations presented below we take $\alpha_{ml}=1$. \\

\noindent {\em --- Inner Boundary: \/ } We suppose that the planet
core has a uniform mass density $\rho_{core}$.  The composition of the
planetesimals and the high temperatures and pressures at the surface
of the core suggest $\rho_{core}=3.2$~g~cm$^{-3}$ (e.g., Bodenheimer
\& Pollack~1986 and Pollack \etal 1996), which is the value we will
adopt throughout.  The core radius, which is the inner boundary of the
atmosphere, is then given by:

\begin{equation}
r_{core} = \left( \frac{3 M_{core}}{4 \pi \rho_{core}} \right)^{1/3}.
\end{equation}
At $R=r_{core}$, the total mass is equal to $M_{core}.$ \\

\noindent {\em --- Outer Boundary: \/} In the models that will be
discussed below, the outer boundary of the atmosphere will be taken to
be either free or at the Roche lobe radius $r_L$ of the protoplanet
given by:

\begin{equation}
r_L = \frac{2}{3} \left( \frac{M_{pl}}{3 M_{\star}} \right)^{1/3} r ,
\end{equation}

\noindent where $M_{pl} = M_{core} + M_{atm}$ is the planet mass,
$M_{atm}$ being the mass of the atmosphere, and $r$ is the orbital
radius of the protoplanet in the disk.\\

\subsubsection{Models enveloped by the disk
with luminosity derived from planetesimal accretion: \\ \\}

\label{sec:formgiant}

These models must match to the state variables of the accretion disk
at the surface of the Roche lobe.  We denote the disk midplane
temperature, pressure and mass density at the distance $r$ from the
central star by $T_m,$ $P_m$ and $\rho_m$, respectively.  To obtain
the needed matching, we require that at $R=r_L,$ the mass is equal to
$M_{pl}$, the pressure is equal to $P_m$ and the temperature is given
by:

\begin{equation}
T = \left( T_m^4 + \frac{ 3\tau_L L_{core}}{16 \pi \sigma r_L^2}
\right)^{1/4},
\end{equation}

\noindent where we approximate the additional optical depth above the
protoplanet atmosphere, through which radiation passes, by:

\begin{equation}
\tau_L = \kappa \left( \rho_m, T_m \right) \rho_m r_L .
\end{equation}

For a particular disk model, at a chosen radius $r,$ for a given core
mass $M_{core}$ and planetesimal accretion rate $\dot{M}_{core},$
Papaloizou \& Terquem~(1999) solve equations~(\ref{dpdvarpi}),
(\ref{dmdvarpi}) and~(\ref{dtdvarpi}) with the boundary conditions
described above. The opacity, taken from Bell \& Lin~(1994), has
contributions from dust grains assumed to be 'interstellar',
molecules, atoms and ions.

For a fixed $\dot{M}_{core}$ at a given radius, there is a critical
core mass $M_{crit}$ above which no solution can be found, i.e. there
can be no atmosphere in hydrostatic and thermal equilibrium confined
between the radii $r_{core}$ and $r_L$ around cores with mass larger
than $M_{crit}$, as explained above.  For masses below $M_{crit}$,
there are (at least) two solutions, corresponding to a low--mass and a
high--mass envelope, respectively.

 To estimate how large $M_{core}$ should be before a significant
atmosphere can start to form, we suppose it to be isothermal, which is
the most favourable situation for retaining a massive atmosphere.
When the atmosphere starts to form, we may neglect its mass, so the
gas density will be given by:

\begin{equation} 
\rho_b = \rho_m\exp\left(GM_{core}\over H^2\Omega^2 r_{core}\right). 
\end{equation}

\noindent We estimate that the atmosphere can start to be significant
when $4\pi r_{core}^3\rho_b/3 = f M_{core},$ where $f < 1$ is an
estimated mass fraction contained in the atmosphere.  This gives:
 
\begin{equation} M_{core} =\left( {H^2\Omega^2\over G}\right)^{3/2}
\left({3\over 4\pi\rho_{core}}\right)^{1/2}\left( \ln
 \left({2fH\rho_{core}\over \Sigma}\right)\right)^{3/2}.
\end{equation}
 
\noindent Taking $\Sigma = 200$~g~cm$^{2}$, $H/r = 0.05,$ $r=5.2$~AU
 and $f=0.01,$ we obtain $M_{core} = 0.064$~M$_{\oplus}.$ 

\begin{figure*}
\centerline{\epsfig{file=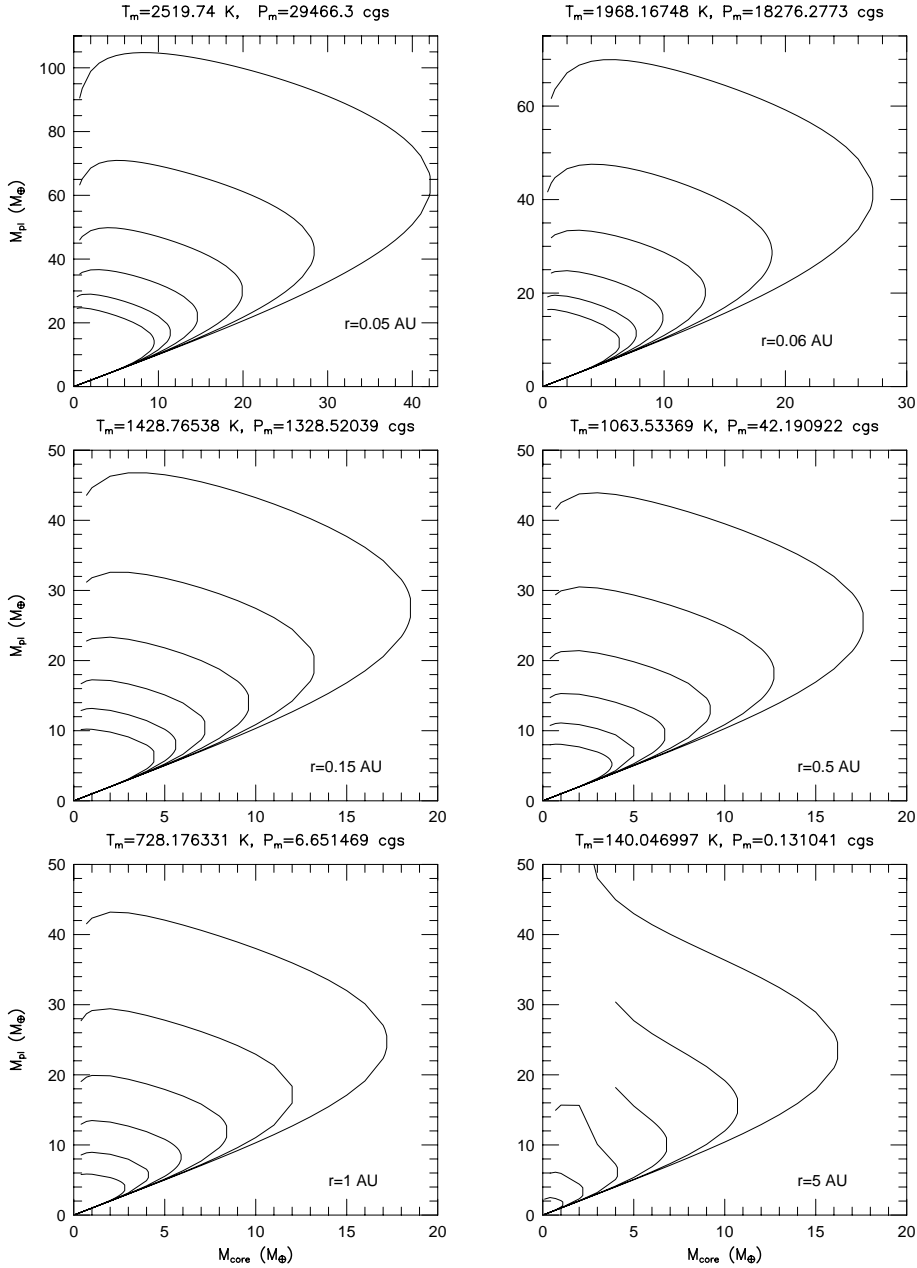, width=\textwidth, angle=270}}
\vspace{25mm}
\caption{\label{PT99_fig} Plots of total mass $M_{pl}$, in units
M$_{\oplus}$, as a function of core mass $M_{core}$, in units
M$_{\oplus}$, at different locations $r$ in a steady state disk model
with $\alpha=10^{-2}$ and gas accretion rate ${\dot M} =
10^{-7}$~M$_{\odot}$~yr$^{-1}$.  From left to right and top to bottom,
the frames correspond to $r=0.05$, 0.06, 0.15, 0.5, 1 and 5~AU,
respectively.  The midplane temperature and pressure at these
locations are indicated above each frame. Each frame contains six
curves which, moving from left to right, correspond to core
luminosities derived from planetesimal accretion rates of
$\dot{M}_{core}=10^{-11}$, $10^{-10}$, $10^{-9}$, $10^{-8}$, $10^{-7}$
and $10^{-6}$~M$_{\oplus}$~yr$^{-1}$, respectively.  The critical core
mass is attained when the curves first begin to loop backwards when
moving from left to right. These plots are taken from Papaloizou \&
Terquem (1999).}
\end{figure*}

Figure~\ref{PT99_fig} shows the total protoplanet mass $M_{pl}$ as a
function of core mass $M_{core}$ at different radii in a steady state
disk with $\alpha= 10^{-2}$ (Shakura \& Sunyaev~1973) and ${\dot
M}=10^{-7}$~M$_{\odot}$~yr$^{-1}$ (see Papaloizou \&
Terquem~1999). In each frame, the different curves correspond to
planetesimal accretion rates in the range
$10^{-11}$--$10^{-6}$~M$_{\oplus}$~yr$^{-1}$.  The critical core mass
is attained at the point where the curves start to loop backwards.

\noindent  It was demonstrated by Stevenson (1982) that the
qualitative behaviour of the curves shown in Figure~\ref{PT99_fig} can
be produced using a simplified model with only radiative energy
transport, ideal gas equation of state and constant opacity. It thus
appears to result fom the core plus envelope nature of the models
rather than details of the physics.

When the core first begins to gravitationally bind some gas, the
protoplanet is on the left on the lower branch of these curves.
Assuming $\dot{M}_{core}$ to be constant, as the core and the
atmosphere grow in mass, the protoplanet moves along the lower branch
up to the right, until the core reaches $M_{crit}$.  At that point,
the hydrostatic and thermal equilibrium approximation can no longer be
used for the atmosphere, which begins to undergo very rapid
contraction.  Figure~\ref{PT99_fig} indicates that when the core mass
reaches $M_{crit}$, the mass of the atmosphere is comparable to that
of the core, in agreement with Bodenheimer \& Pollack~(1986). Since
the atmosphere in complete equilibrium is supported by the energy
released by the planetesimals accreted onto the protoplanet, the
critical core mass decreases as $\dot{M}_{core}$ is reduced.

For $\alpha=10^{-2}$ and $\dot{M}=10^{-7}$~M$_{\odot}$~yr$^{-1}$, the
critical core mass at 5~AU varies between 16.2 and 1~M$_{\oplus}$ as
the planetesimal accretion rate varies between the largest and
smallest value.  The former result is in good agreement with that of
Bodenheimer \& Pollack~(1986).  Note that there is a tendency for the
critical core masses to increase as the radial location moves inwards,
the effect being most marked at small radii.  At 1~AU, the critical
mass varies from 17.5 to 3~M$_{\oplus}$ as the accretion rate varies
between the largest and smallest value, while at 0.05~AU these values
increase still further to 42 and 9~M$_{\oplus}$, respectively.

\begin{figure*}
\centerline{\epsfig{file=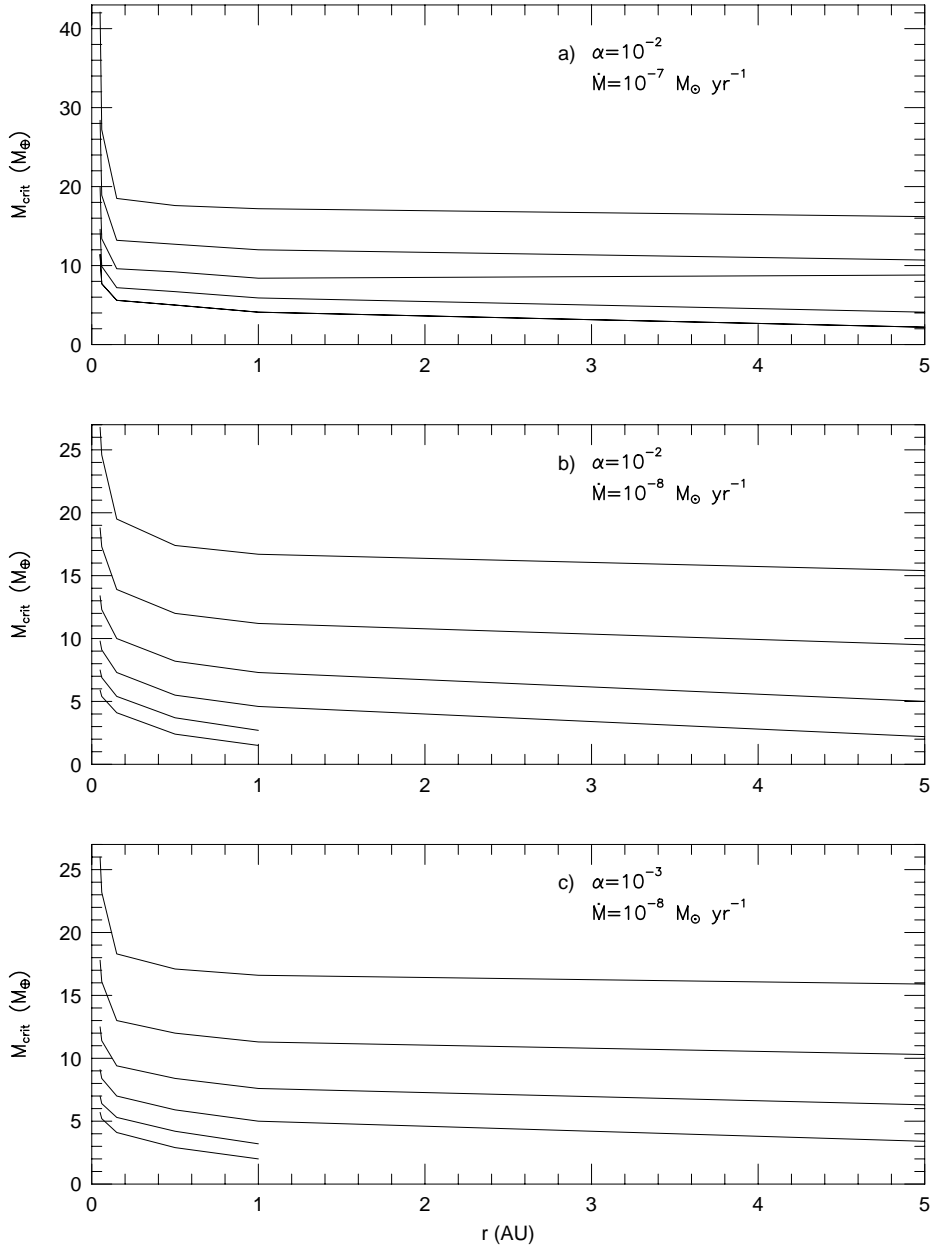, width=\textwidth, angle=270}}
\vspace{25mm}
\caption{\label{PT99_fig2} Critical core mass $M_{crit}$, in units
M$_{\oplus}$, {\em vs.}  location $r$, in units AU, in a steady state
disk model with viscous parameter $\alpha$ and gas accretion rate
$\dot{M}$.  The different plots correspond to $\alpha=10^{-2}$ and
$\dot{M} = 10^{-7}$~M$_{\odot}$~yr$^{-1}$ (a), $\alpha=10^{-2}$ and
$\dot{M} = 10^{-8}$~M$_{\odot}$~yr$^{-1}$ (b), and $\alpha=10^{-3}$
and $\dot{M} = 10^{-8}$~M$_{\odot}$~yr$^{-1}$ (c).  Each plot contains
six curves which, moving from bottom to top, correspond to core
luminosities derived from planetesimal accretion rates of
$\dot{M}_{core}=10^{-11}$, $10^{-10}$, $10^{-9}$, $10^{-8}$, $10^{-7}$
and $10^{-6}$~M$_{\oplus}$~yr$^{-1}$, respectively. These plots are
taken from Papaloizou \& Terquem (1999).}
\end{figure*}

In Figure~\ref{PT99_fig2} we plot the critical core mass $M_{crit}$
versus the location $r$ for three different steady disk models.  These
models have $\alpha=10^{-2}$ and $\dot{M} =
10^{-7}$~M$_{\odot}$~yr$^{-1}$, $\alpha=10^{-2}$ and $\dot{M} =
10^{-8}$~M$_{\odot}$~yr$^{-1}$, and $\alpha=10^{-3}$ and $\dot{M} =
10^{-8}$~M$_{\odot}$~yr$^{-1}$, respectively.  Here again, in each
frame, the different curves correspond to planetesimal accretion rates
in the range $10^{-11}$--$10^{-6}$~M$_{\oplus}$~yr$^{-1}$.  Similar
qualitative behavior is found for the three disk models, but the
critical core masses are smaller for the models with $\dot{M} =
10^{-8}$~M$_{\odot}$~yr$^{-1}$, being reduced to 27 and 6~M$_{\oplus}$
at 0.05~AU for the highest and lowest accretion rate, respectively.

\subsubsection{Models enveloped by the disk with no planetesimal accretion: 
\\ \\}
\label{typeAev}

For these models, subsequently denoted as of type~A, just as for those
with plane\-tesimal accretion as energy source, we assume the structure
extends to the Roche lobe or boundary of the Hill sphere beyond which
material must be gravitationally unbound from the protoplanet.  There
it matches to the external disk. Thus the boundary conditions for
these models are the same as those with planetesimal accretion as
energy source.
 
\noindent Papaloizou \& Nelson~(2005) adopt disk parameters
appropriate to 5~AU from the disk model of Papaloizou \&
Terquem~(1999) with $\alpha = 10^{-3}$ (Shakura and Sunyaev~1973) and
steady sate accretion rate of
$10^{-7}$~M$_{\oplus}$~yr$^{-1}$. Accordingly, $T_m = 140.047$~K and
$P_m = 0.131$~dyn~cm$^{-2}$.

\noindent For models of type~A, specification of the core mass and the
total mass $M_{pl}$ is enough to enable it to be constructed. This is
because, although the luminosity is apparently an additional
parameter, the degree of freedom associated with it is specified by
the requirement that the radius of the planet atmosphere should be the
Hill radius which, for a fixed orbital radius, depends only on
$M_{pl}.$ Thus, the luminosity and the total energy of the protoplanet
may be written as $L = L(M_{pl})$ and $E = E(M_{pl}).$

\noindent For small changes in mass, the change in $E$ is $dE =
(dE/dM_{pl})dM_{pl}.$ If the change occurs over a time interval $dt,$
we must have: 

\begin{equation} 
\frac{dE}{dM_{pl}} \frac{dM_{pl}}{dt} = -L. \label{eveA}
\end{equation}

\noindent This determines how the mass and hence the luminosity of the
protoplanet vary with time and therefore the evolutionary track is
specified.

\subsubsection{Models with free boundary accreting from the protostellar 
disk: \\ \\ }
\label{typeBev}

In contrast to embedded models of type~A, we can consider models that
have bounda\-ries detached from and interior to the Roche lobe which
still accrete material from the external protoplanetary disk that
orbits the central star.  Detachment from the Roche lobe and the
development of a free boundary are expected once the planet goes into
a rapid accretion phase.  This is also expected because numerical
simulations of disk--planet interactions have shown that, once it
becomes massive enough, a protoplanet forms a gap in the disk but is
still able to accrete from it through a circumplanetary disk (see,
e.g., Bryden \etal 1999, Kley 1999, Lubow \etal 1999, R.~Nelson \etal
2000).  We thus consider models with free boundaries which are able to
increase their mass and liberate gravitational energy through its
settling.  These models are referred to as type~B.

\noindent In contrast to models of type~A, for an assumed externally
supplied accretion rate, models of type~B form a two parameter family
in that, without specification of the energy source, and given their
freedom to determine their own radius, they require specification of
both $M_{pl}$ and $L$ in order for a model to be constructed. Thus $E
= E(M_{pl}, L).$ Accordingly, for small changes in mass and luminosity,
the change in $E$ is $dE = (\partial E/\partial M_{pl})d M_{pl} +
(\partial E/\partial L)dL.$

\noindent Now, for these models, matter is presumed to join the
protoplanet on its equator after having accreted through a
circumplanetary disk.  In this case, we assume the accretion rate to
be prescribed by the dynamics of the disk--planet interaction while
gap formation is taking place. This is in agreement with simulations
of disk--planet interactions where it is found that an amount of
material comparable to that flowing through the disk may be supplied
to the protoplanet (Bryden \etal 1999, Kley 1999, Lubow \etal 1999,
R.~Nelson \etal 2000 and simulations presented in section~\ref{DPI}).  In
arriving at the planet equator, all available gravitational binding
energy of $-GM_{pl}/r_s$ per unit mass, $r_s$ being the surface
radius, has been liberated and so an amount of energy $-GM_{pl}
dM_{pl}/r_s$ must be subtracted from $dE$ in order to obtain the
energy available to replace radiation losses.  Therefore, if the
changes occur over an interval $dt,$ we must have:

\begin{equation} 
dE + \frac{GM_{pl} dM_{pl}}{r_s} = \frac{\partial
E}{\partial M_{pl}} dM_{pl} + \frac{\partial E}{\partial L} dL + \frac{GM_{pl}
dM_{pl}}{r_s} = -Ldt.
\end{equation}

\noindent Thus, total energy conservation for models of type~B enables
the calculation of evolutionary tracks through:

\begin{equation}
\left[{\partial E\over \partial M_{pl}} +  {GM_{pl}\over r_s} \right ]
{dM_{pl}\over dt} + {\partial E\over \partial L}{dL\over dt}  = -L.
\label{eveB}
\end{equation}

\noindent Note that, as we regard the accretion rate $dM_{pl}/ dt$ as
specified for these models, equation~(\ref{eveB}) enables the
evolution of $L$ to be calculated.

\noindent Thus, equations~(\ref{eveA}) and~(\ref{eveB}) constitute the
basic equations governing the evolution of models of type~A and
type~B, respectively.

\begin{figure*}
\centerline{\epsfig{file=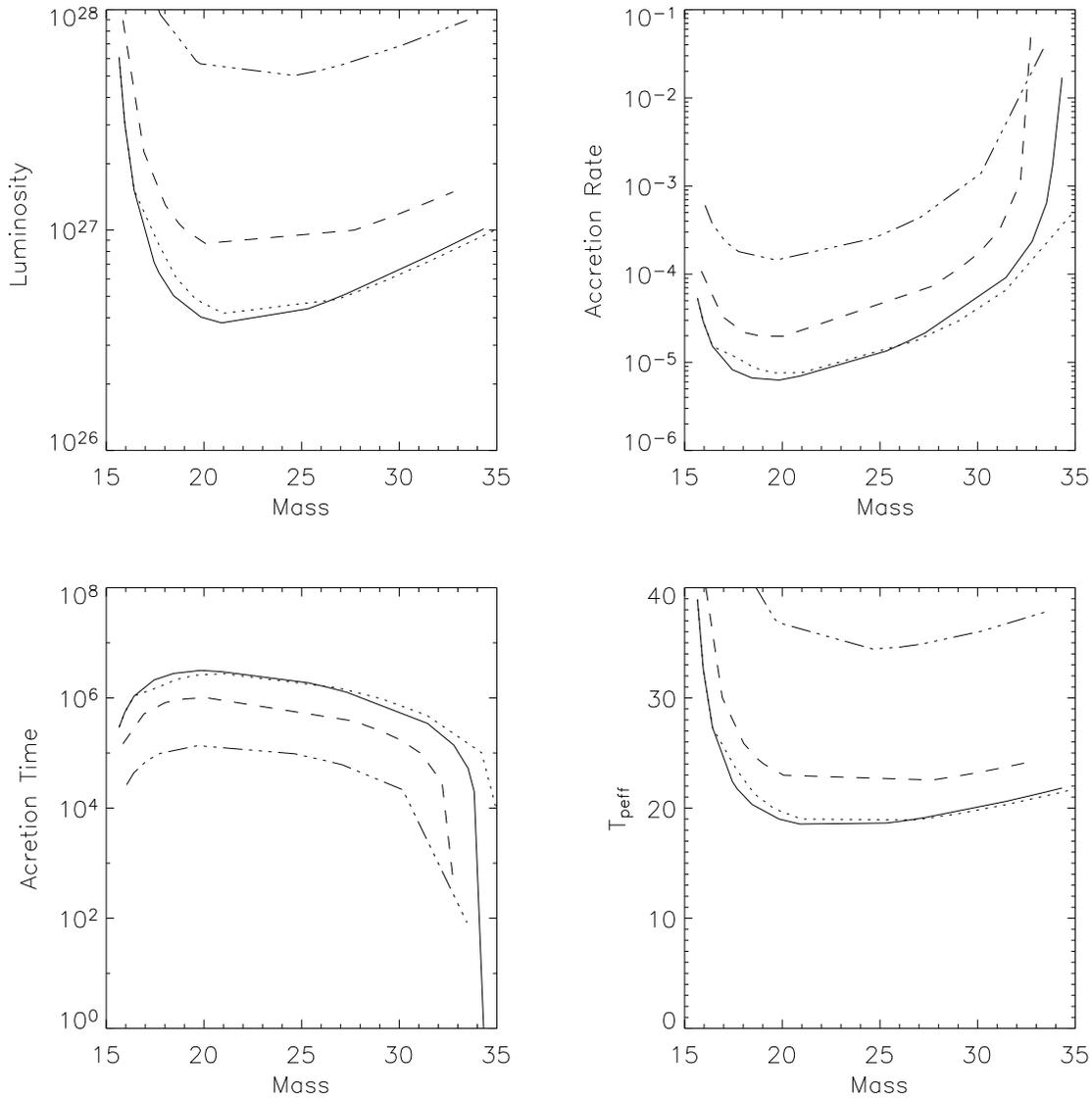,width=\textwidth, angle=0}}
\caption{\label{EV_fig6} This figure illustrates the evolution of
protoplanet models of {\bf type~A} which maintain contact with the
protoplanetary disk and fill their Roche lobes while they accrete from
it. They have fixed solid core masses of 15~M$_{\oplus} $ and are
situated at 5~AU. The upper left panel shows the luminosity, in cgs
units, as a function of their increasing mass $M,$ in cgs units.  The
upper right panel shows the gas accretion rate, in
M$_{\oplus}$~y$^{-1}$, as a function of mass, while the lower left
panel gives the accretion time $ M /{\dot M}$, in yr, as a function of
mass.  The lower right panel gives the temperature $T_{effp},$
being the radiation   temperature on the Roche lobe  required  account for the
 luminosity of the protoplanet, 
 as a function of mass.  The model
illustrated with the full curve is embedded in a standard disk while
the model illustrated with a dotted curve is embedded in a disk with
the same temperature but with a density ten times larger.  These two
protoplanet models with standard opacity show very similar behavior
indicating lack of sensitivity to the detailed boundary conditions.
In addition, we illustrate two models with this core mass embedded in
a standard disk but with opacities which have a reduction factors of
ten and one hundred (dashed curves and triple dot dashed curves,
respectively) that are constant for $T <1600$~K and which then
decrease linearly to unity at $T = 1700$~K. Note the rapid
evolutionary speed up that occurs when these models attain $\sim
35$~M$_{\oplus}.$ At this point, detachment from the Roche lobe and
transition to a model of type~B is expected. (These plots are taken
from Papaloizou \& Nelson 2005).  }
\end{figure*}

\noindent Note that any input from planetesimal accretion during and
after the phase when the core becomes critical is neglected. We recall
that after this point planetesimal accumulation proceeds by oligarchic
growth (Ida \& Makino~1993, Kokubo \& Ida 1998).  N--body simulations
of protoplanetary core formation indicate that obtaining cores of a
high enough mass $\sim 15$~M$_{\oplus}$ to obtain evolutionary
timescales within the gas disk lifetime is not an easy task to achieve
during the oligarchic growth phase. This is in part due to planetary
cores of a few M$_{\oplus}$ repelling the surrounding planetesimals
and opening gaps in the planetesimal disk (e.g., Thommes \etal 2002).
In addition, it is reasonable to suppose that the formation of a
series of neighboring cores may deplete planetesimals in the local
neighborhood.  We also note that if the slow down in evolution rate
due to the long thermal timescale of the atmosphere can be obviated,
a  depletion of planetesimals  brings about a favourable
situation for shortening the evolutionary timescale as a potential
energy source for sustaining the models without gas accretion is
removed.  The thermal timescale may be reduced by reducing the
contribution of grain opacity from the 'interstellar value' (see
figure~\ref{EV_fig6}, Ikoma \etal 2000 and Papaloizou \& Nelson~2005).
If the grain opacity in the outer layers of the atmosphere can be
reduced by a factor of $100,$ the evolutionary timescale for a
protoplanet with a 5~M$_{\oplus}$  core  can be reduced from
$3\times 10^8$~yr to a value within the expected gas disk lifetime.
Such a reduction may occur either because of grain depletion or
because the grains have been accumulated into particles of larger sizes
than the original (Ikoma \etal 2000).

\begin{figure*}
\centerline{\epsfig{file=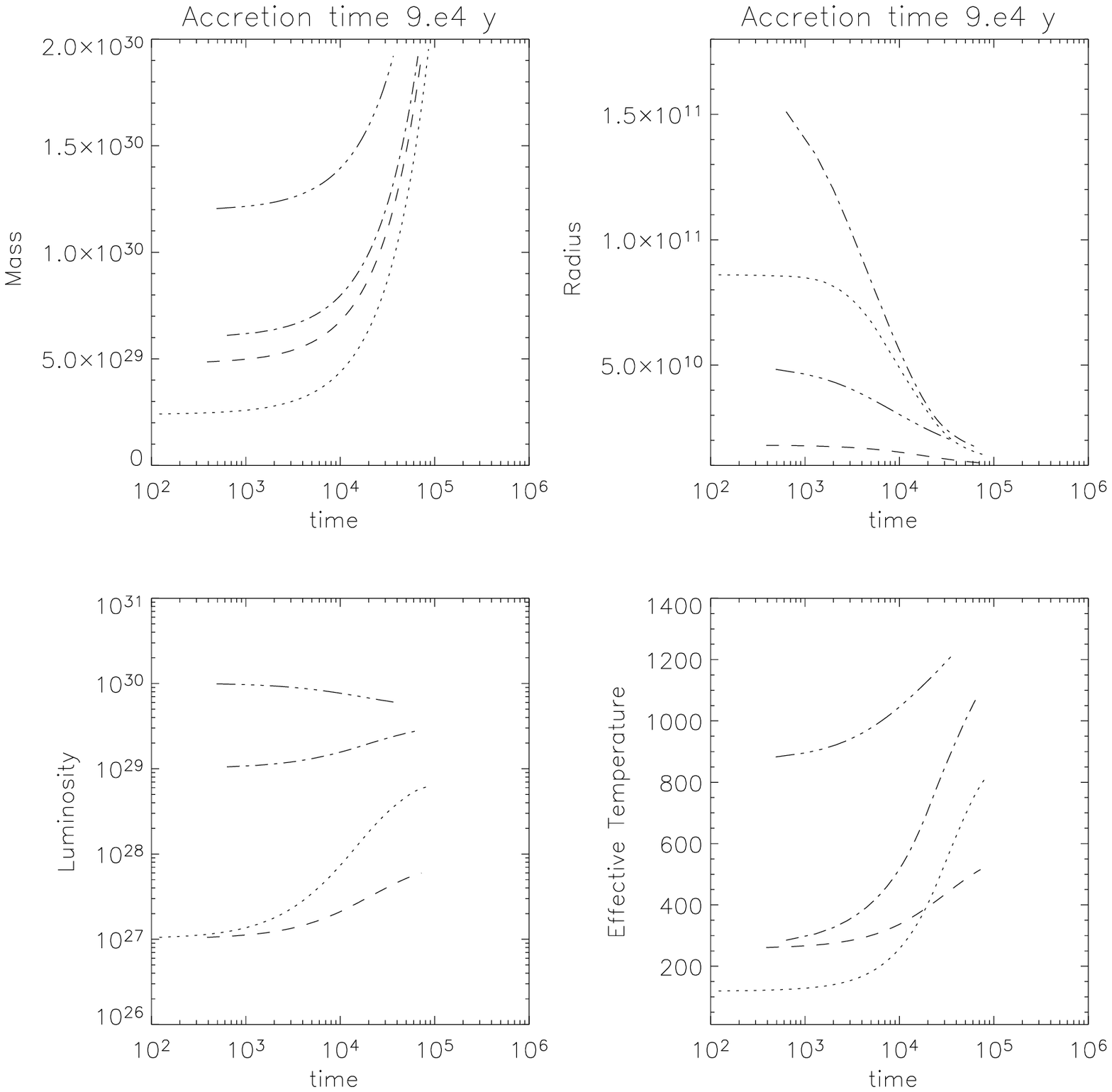,width=\textwidth, angle=0}}
\caption{\label{EV_fig7} This figure illustrates the evolution of
protoplanet models of {\bf type~B} which accrete from the
protoplanetary disk at an assumed rate of one Jupiter mass in $9\times
10^4$~yr but are detached from their Roche lobes.  They have fixed
solid core masses of 15~M$_{\oplus}$ and are situated at 5~AU. The
upper left panel shows the planet mass, in cgs units, as a function of
time, in yr.  The upper right panel shows the protoplanet radius, in
cgs units, as a function of time.  The lower left panel gives the
luminosity, in cgs units, as a function of time. The lower right panel
gives the effective temperature as a function of time.  The four
models shown correspond to differing initial conditions corresponding
to different starting masses and luminosities.  The resulting
evolutionary tracks tend to show convergence as time progresses.
(These plots are taken from Papaloizou \& Nelson 2005).}
\end{figure*}

However, regardless of opacity, it is found that the evolutionary
timescales for the type~A models decrease rapidly once the gas mass is
about twice the core mass (see figure \ref{EV_fig6}). At this point,
detachment from the Roche lobe and transition to a model of type B is
expected (Papaloizou \& Nelson~2005). Some of these are illustrated
in figure~\ref{EV_fig7}. It is found that these models rapidly attain
a radius $ \ls 2 \times 10^{10}$~cm for any reasonable accretion rate
from the disk.  Furthermore, these protoplanets may attain
luminosities of $\sim 10^{-3}$--$10^{-4}$~L$_{\odot}$ for a time $\sim
10^5$--$10^6$~yr.

\subsection{Formation of giant planets: gravitational instabilities}

Gravitational instabilities in protostellar disks have been considered
as a mechanism for forming giant planets since Kuiper (1949) proposed
that planets were formed through gravitational breakup of gaseous
rings.  The advantage of this mechanism is that it leads to giant
planets on a very short timescale, on the order of the dynamical
timescale.  This is an important argument, as we have seen above that
the build--up of a massive core may be prohibitively long in some
parts of the protostellar nebula.  Also, some of the extrasolar
planets discovered so far are very massive, with a mass several times
that of Jupiter.  It is not clear that these objects can form through
the core--accretion scenario, as gap formation inhibits gas accretion
above some critical mass (see below).

It was initially believed that axisymmetric instabilities in disks
would lead to the formation of rings that would further break--up into
condensations.  Using the theory of viscous accretion disks developed
by Lynden--Bell and Pringle (1974), Cameron (1978) calculated
evolution models of the primitive solar nebula which were unstable
against axisymmetric perturbations.  

The parameter governing the importance of disk self--gravity is the
Toomre $Q$~parameter which, in a Keplerian disk, can be written as $Q
\simeq \Omega c_s /(\pi G \Sigma) \sim M_{\star} H/(M_d r)$, with
$M_d$ being the disk mass contained within radius $r$, $H$ the disc
semithickness and $c_s$ the sound speed.  Typically $H/r \sim 0.1$ for
protostellar disks (Stapelfeldt \etal 1998) such that the condition
for the importance of self--gravity, $Q \sim 1$, gives $M_d \sim 0.1
M_{\star}$.  The characteristic scale associated with growing density
perturbations in a disk undergoing gravitational instability with $Q
\sim 1$ is on the order of $H$, and the corresponding mass scale is
$M_d (H/r)^2 \sim M_{\star} (H/r)^3$, which is on the order of one
Jupiter mass for $H/r \sim 0.1$ and $M_{\star}= 1$~M$_{\odot}$.

There are, however, problems with forming planets through
gravitational ins\-tabilities.  Fragmentation, which follows the
development of axisymmetric modes, requires $Q \le 1$, whereas non
axisymmetric modes develop for values of $Q$ slightly larger than
unity.  As noted by Laughlin \& Bodenheimer (1994), it is unlikely
that a disk would ever get massive enough for $Q$  decrease below unity.  The
disk builds up in mass as a result of the collapse of an envelope.
The $Q$ parameter therefore decreases with time from high values.
When it reaches values slightly larger than unity, non axisymmetric
(spiral) modes develop.  These lead to outward angular momentum
transport (Papaloizou \& Savonije 1991; Heemskerk \etal 1992;
Laughlin \& Bodenheimer 1994; Pickett \etal 1998) that results in
additional mass growth of the central star.  Therefore $Q$ is
decreased and gravitational stability is restored.  Since the
redistribution of mass may occur on the dynamical timescale (a few
orbits) of the outer part of the disk, it follows that $Q$ can never
decrease to below  unity unless the disk can be cooled down (so that $c_s$
decreases) or matter can be added (so that $\Sigma$ increases) on the
dynamical timescale, which is unlikely to happen in the regions of
planet formation (it could happen however in the optically thin parts
of the disk, typically beyond $\sim 40$~AU).

Nonetheless, condensations may form even if the disk is unstable only
against non--axisymmetric perturbations, providing the density gets
high enough in the spiral arms (Boss 1998, Mayer \etal 2004).  There
are indications, however, that the clumps that appear in the numerical
simulations of gravitationally unstable disks are not long--lived
(see, e.g., Durisen 2001 and references therein).  The outcome of
gravitational instabilities is very sensitive to the thermal physics
of the gas. Chemical composition dependence is expected but the correlation
of the existence of planets with metallicity has yet to be demonstrated.
 When the equation of state is adiabatic, heating of the
gas prevents the formation of clumps.  In contrast, condensations form
in an isothermal disk.  But if the outer radius of the disk is allowed
to expand under the effect of gravitational torques, these clumps tend
to disappear, dissolving back into the gas (Pickett \etal 1998, 2000a,
b; Boss 2000).  Simulations including heating via dynamical processes
and radiative cooling produce spiral structures that are less distinct
than in isothermal disks and that do not collapse into condensed
objects (A.~Nelson \etal 2000).  Although the question of whether the
clumps can survive or not is still being debated, it seems unlikely
that Jupiter mass planets can be formed through gravitational
instabilities in optically thick regions of disks.  However, as noted
in section~\ref{sec:planetesimals}, the accumulation of solid
particles into planetesimals may be assisted by gravitational
instabilities in the gas disk through particle accumulation at
pressure maxima in spiral structures (Rice \etal 2004, Durisen \etal
2005).

Note that fragmentation may also occur before a disk is completely
formed, during the initial collapse of the protostellar envelope.
Such opacity limited fragmentation has been estimated to produce
objects with a lower mass limit of 7~Jupiter masses (Low \&
Lynden--Bell 1976, Masunaga \& Inutsuka 1999), but there is no
definitive argument to rule out somewhat smaller masses (Bodenheimer
\etal 2000).  It is possible that both a disk and fragments may form
simultaneously out of the protostellar envelope, the relative
importance of the two processes depending for instance on the angular
momentum content of the envelope, on the strength of any magnetic
field (so far neglected in disk fragmentation calculations) and
possibly on the initial clumpiness.  Large scale observations of
class~0 envelopes so far do not rule out the presence of clumps with
masses smaller than about 10 Jupiter masses (Motte \& Andr\'e 2001).
To explain the large eccentricities of many of the extrasolar planets,
Papaloizou \& Terquem (2001) and Terquem \& Papaloizou (2002) have
considered a scenario in which a population of planetary mass objects
are assumed to form rapidly through a fragmentation process occurring
in a disk or protostellar envelope on a scale of 100~AU (see
section~\ref{multiplanet} below).

\section{Disk--planet interactions and orbital migration}
\label{DPI}

The discovery of the 'hot Jupiters' led to the realization of the
importance of orbital migration.  This comes about because of the
perceived difficulties in accumulating a core directly at small radii
interior to the so called 'snow line' (Bodenheimer \etal 2000).
Instead, it is possible the protoplanet formed at larger radii and
migrated inwards.

\noindent So far, three mechanisms have been proposed to explain the
location of planets at very short orbital distances.  One of them
relies on the gravitational interaction between two or more Jupiter
mass planets, which may lead to orbit crossing and to the ejection of
one planet while the other is left in a smaller orbit (Rasio \& Ford
1996, Weidenschilling \& Marzari 1996).  However, this mechanism
cannot account for the relatively large number (about 20\%) of
short--period planets observed.  Another mechanism is the so--called
'migration instability' (Murray \etal 1998, Malhotra 1993).  It
involves resonant interactions between the planet and planetesimals
located inside its orbit which lead to the ejection of a fraction of
them while simultaneously causing the planet to migrate inwards.  To
move a Jupiter mass planet from about 5~AU down to very small radii
through this process, a disk containing about 1 Jupiter mass of
planetesimals, and thus about 0.1~M$_{\odot}$ of gas, inside 5~AU is
required.  Such a massive disk is unlikely and furthermore it would be
only marginally gravitationally stable.  The third mechanism, that we
are going to focus on here, involves the tidal interaction between the
protoplanet and the gas in the surrounding protoplanetary nebula
(Goldreich \& Tremaine 1979, 1980; Lin \& Papaloizou 1979, 1993 and
references therein; Papaloizou \& Lin 1984; Ward 1986, 1997b).

\noindent Below we review the possible main types of orbital migration
associated with this tidal interaction.  These are (i) type~I
migration, which applies to an embedded protoplanet of small mass for
which the disk response can be considered using linear analysis (e.g.,
Ward 1997), (ii) type~II migration, which applies when the protoplanet
is massive enough to open a gap (Lin \& Papaloizou~1986) and (iii)
runaway or type~III migration (Masset \& Papaloizou~2003, Artymowicz~2004),
 which is a
new form of potentially fast migration, applicable to massive disks,
that could be driven by coorbital torques.

\noindent Orbital migration in each of these cases is induced by the
angular momentum exchange with disk material that occurs through
disk--planet interaction. This may be through wave excitation, as in
type~I migration, a combination of wave excitation and shock
dissipation, as occurs in type~II migration, or direct exchange with
disk material traversing the orbit, as has been suggested for type~III
migration. 

\noindent In the appendix we give a simple description of the global
angular momentum balance and conservation when a protoplanet in
circular orbit is embedded in a gaseous disk with which it interacts
gravitationally and consequently migrates radially. We show that the
angular momentum exchange with the protoplanet can be accounted for by
considering viscous and wave fluxes of angular momentum, that can be
measured a long way from the protoplanet, as well as the difference
between the rate of advection of angular momentum by gas flowing
through the orbit and the rate of increase in the angular momentum
content of regions in the neighborhood of the protoplanet because of
non steady conditions.  The angular momentum carried in wave/viscous
fluxes is associated with type I and type II migration, whereas the
advected flux is associated with the more recently proposed type III
migration.

\noindent We consider the timescales for the various migration
processes and the potential problems introduced for planet formation
theory.

\subsection{Angular momentum transfer due to scattering and type~II migration} 
\label{scatt}

\subsubsection{The impulse approximation applied to a protoplanet 
in a gap: \\ \\}

Here, we adopt a very simple picture of the angular momentum exchange
process between a protoplanet in circular orbit and the disk.  We work
in a frame in which the protoplanet appears at rest at the origin. For
a protoplanet in circular orbit, this frame would have to be in a
state of uniform rotation and we should include centrifugal and
Coriolis forces to describe the motion of the disk matter.  However,
we shall neglect these and make a local approximation, supposing that
a particle of disk matter approaches the protoplanet from large
distance moving in a sraight line with impact parameter $a$ which we
shall take to be much less than the orbital radius of the protoplanet
(Lin \& Papaloizou~1979).

\noindent On passing the protoplanet, the gravitational interaction
produces an angle of deflection $\delta$ between the asymptotic
tanjents to the trajectory of the passing element of disk matter.  In
the local approximation, the disk matter recedes from the protoplanet
along a straight line inclined to the line of approach by an angle
$\delta.$

\noindent We adopt local Cartesian coordinates with origin at the
center of mass of the protoplanet.  The $x$--axis points radially
outwards, the $y$--axis points in the azimuthal direction in the
direction of rotation while the $z$--axis points in the vertical
direction.  We may assume that the initial trajectory is along $x=a$
approaching from $y=-\infty$ with relative speed $u.$ This corresponds
to a disk with circular streamlines.

\noindent As a result of the interaction, a velocity component $v_x,$
parallel to the $x$--axis, is induced.  This can be found from the
$x$--component of the equation of motion in the form:
$$ {d v_x\over dt} ={-GM_p x/(x^2+y^2)^{3/2}},$$ where $M_p$ is
the planet mass.  Assuming the interaction to be weak, we set $x=a,$
$y=ut$ and integrate with respect to time between $-\infty$ and $+\infty.$
Hence, the final induced $x$--velocity is:
$$v_x =-G M_p a\int^{\infty}_{-\infty} {dt\over [a^2+(ut)^2]^{3/2}}
= -{2G M_p \over au}.$$ The angle of deflection is thus simply:
$$\delta = \left| \frac{v_x}{u} \right| = {2G M_p \over au^2}.$$

\noindent Associated with this interaction is a transfer of angular
momentum between the disk matter and protoplanet. To see this, we
first recall that, as the interaction between protoplanet and disk is
conservative, energy conservation implies that the ultimate speed of
recession is also $u.$ Accordingly, there must have been a parallel
momentum transfer per unit mass of magnitude $ \Delta u =
u(1-\cos\delta) \simeq u\delta^2/2.$ If the orbital radius of the
protoplanet is $R,$ the associated angular momentum transferred is:

\begin{equation} \Delta J = R u \delta =
{2RG^2M_p^2\over a^2 u^3} \label{DeltaJ}.\end{equation}

\noindent We should of course note that this transfer is directed
along the original direction of motion as seen by the protoplanet.
Thus, for disk matter on circular orbits interior to the protoplanet,
the transfer is positive, while for exterior disk matter, the transfer
is in the opposite sense and negative. The interaction is accordingly
frictional, with interior matter speeding up the protoplanet and
exterior matter slowing it down.

\noindent Note too that the trajectory prior to the local scattering
interaction is assumed linear (corresponding to circular orbit when
viewed globally).  Subsequent returns of disk matter on circular orbit
are required in order to make the frictional interaction persistent.
The scattering itself, of course, disturbs that situation, so that there
is an implicit assumption that dissipative or other processes work to
restore circular orbits for returning trajectories.  The indications
from many numerical simulations are that disk viscosity is able to
provide such an effect.

\noindent The simple calculation of $\Delta J$ given above neglects
the fact that the calculation should be done in a rotating frame.
However, this has been incorporated in calculations by Goldreich \&
Tremaine~(1980) and Lin \& Papaloizou~(1993).  As these calculations
add nothing new with regard to the essential physics, we refer the
reader to those works for the details. The end result is that the
expression given in equation~(\ref{DeltaJ}) is multiplied by a
correction factor
$$C_0 = {4\over 9} \left[ 2K_0(2/3) + K_1(2/3) \right]^2 ,$$ where
$K_0$ and $K_1$ denote modified Bessel functions.  Thus:

\begin{equation} \Delta J =
 C_0 {2RG^2M_p^2\over a^2 u^3} \label{DeltaJ1}.\end{equation}

\subsubsection{Angular momentum exchange rate with the disk: \\ \\}

We may use equation~(\ref{DeltaJ1}) to evaluate the total angular
momentum exchange rate with the disk. To do this, we replace the
relative speed $u$ by $u = R|\Omega - \omega|,$ where $\omega$ and
$\Omega$ are the angular rotation rates of the protoplanet and disk
matter, respectively. Given that we have implicitly assumed the
interaction occurs close to the protoplanet, we adopt the first order
expansion $|\Omega - \omega| = 3a\omega /(2R).$

\noindent Furthermore, each disk particle orbiting a distance $a$ from
the protoplanet suffers impulses separated by a time interval
$2\pi/|\Omega - \omega|.$ We can thus write down the rate of transfer
of angular momentum between the disk matter interior to the
protoplanet orbit and the protoplanet by integrating over the disk
mass as:

\begin{equation}
{d J\over dt} = \int^{\infty}_{a_{min}} R\Sigma \Delta J |\Omega -
\omega|da. \end{equation}

\noindent This gives:

\begin{equation}
{d J\over dt} = {8 G^2 M_p^2 R \Sigma C_0\over 27 \omega^2 a_{min}^3}
\label{Jdot}.\end{equation}

\noindent Here, $a_{min}$ is the minimum distance between disk matter
and the protoplanet orbit. Thus, there is an inherent assumption of a
gap in the disk, or lack of coorbital material, which would have to be
taken into account if the protoplanet is embedded in the disk and the
interaction is linear.  Therefore, the above analysis is applicable
only in a nonlinear regime when there is a significant gap,
corresponding to type~II migration. Note that the sign is positive for
the more rapidly moving inner disk material.  For the outer disk
material, one can see from symmetry considerations that the same
expression applies but with a sign reversal because it is more slowly
moving than the protoplanet and thus attempts to slow it down.

\subsubsection{Gap formation: \\ \\}

A gap forms because angular momentum exchange occurs locally in the
disk, on each side of the protoplanet, in such a way as to cause
the disk material to move away from the protoplanet. 


\noindent In order for a gap to be maintained, we should expect that
the angular momentum exchange rate given by equation~(\ref{Jdot}) be
able to balance the angular momentum flow rate due to viscosity. The
viscous flow rate is given by:

\begin{equation}
\left( {d J\over dt} \right)_{visc} = 3\pi\nu \Sigma R^2 \omega, \end
{equation}

\noindent where $\nu$ is the kinematic viscosity and $\Sigma$ is the
surface density of gas in the disk.  Of course, we recall that $\nu$
is not a real 'viscosity' but a way of representing the flow of
angular momentum resulting from turbulent stresses.  From what we have
indicated above, we expect the condition for gap formation to be:

\begin{equation}
{d J\over dt} > \left( {d J\over dt} \right)_{visc} . 
\end{equation}

\noindent This is equivalent to:
\begin{equation}
{8  M_p^2 R^3  C_0\over 81\pi M_{\star}^2  a_{min}^3} >
{\nu \over R^2 \omega},  \label{GAPvisc}\end {equation}

\noindent where $M_{\star}$ is the mass of the central star, around
which the planet orbits.  To work with equation~(\ref{GAPvisc}), we
need to know what value of $a_{min}$ to use.  Considerations of the
pressure response in type~I migration (Goldreich \& Tremaine 1980,
Artymowicz 1993) suggests that there should be a torque cut--off
leading to $a_{min} \sim H.$ This is also a statement that pressure
effects are able to smooth out smaller scale perturbations in the
vicinity of the protoplanet.  A similar conclusion was reached by
Papaloizou \& Lin~(1984) who considered the linear response of a
protoplanet orbiting in a disk already with a gap with lengthscale
$H,$ which was found to be the physical scale that led to the maximum
angular momentum exchange rate.

\noindent However, the condition for strong nonlinearity, which is
required for local gap formation, is that the   radial lengthscale of
the flow cannot  be less than the Hill radius, $r_H,$ which  being the largest
possible scale of signigicant influence by the protoplanet
gravitational field (Lin \& Papaloizou~1979), must exceed $H.$  Accordingly, we should
have $a_{min} \ge r_H.$ Taking $a_{min} =2r_H$ and absorbing
uncertainties in the constant coefficient $C_0,$
equation~(\ref{GAPvisc}) gives:

\begin{equation}
{3 M_p  C_0\over 81\pi M_{\star}} >
{\nu \over R^2 \omega}. \label{GAPvisc0}\end {equation}

\noindent Taking $C_0 =2/3 ,$ this leads to:
\begin{equation}
{ M_p \over M_{\star}} > { 40 \nu \over R^2 \omega}. \label{GAPvisc1}
\end{equation}

\noindent This condition is given by Lin \& Papaloizou~(1986, 1993).
But note that it is only expected to be accurate to within a constant
of order unity. For a typical disk model, $\nu/(R^2\Omega) \sim
10^{-5}$, and we expect gap formation for planets slightly above the
mass of Saturn.

\noindent  However, note that Rafikov (2002) argues that gap
formation may occur for planets of mass as low as 2~M$_{\oplus}.$ So
far, gap formation at such low masses has not been seen in simulations
(e.g., D'Angelo \etal 2003).  This may be because Rafikov (2002)
suggested that a gap can be produced as the density waves excited by a
planet become weak shock waves and dissipate.  However, this occurs
several scale heights away from the planet, beyond the region where a
gap would be expected.  Furthermore, the angular momentum deposition
from the waves also extends over a region of several scale heights,
being wider than an expected gap. In this context, note that an annulus
of width $5H$ at $5.2$~AU in a minimum mass solar nebula contains a
Jupiter mass, so that a planet of several earth masses could not
redistribute its angular momentum without significant migration.

\subsubsection{Type~II migration: \\ \\}

\begin{figure*}
\centerline{
\epsfig{file=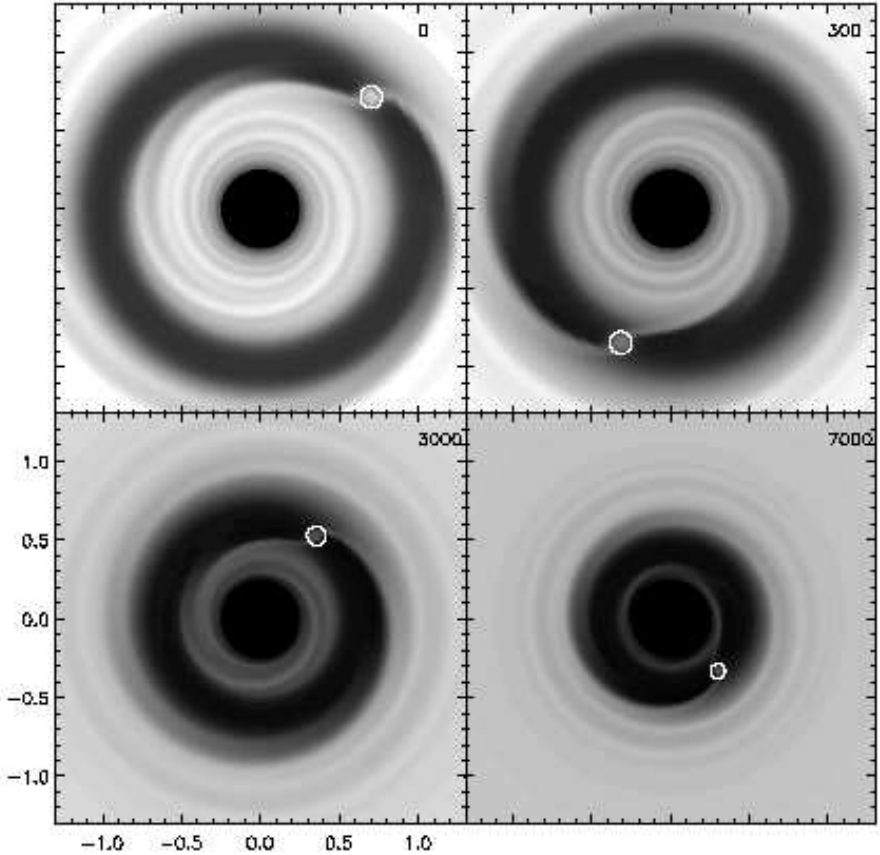,width=\textwidth} }
\vspace{5mm}
\caption[] {This figure shows the time evolution of the disk surface
density contours and orbital radius as a function of time measured in
orbital periods for a protoplanet with initial mass of 1~M$_{\rm J}$.
As the protoplanet undergoes type~II migration inwards, its mass grows
to 3.5~M$_{\rm J}$ under an assumption of maximal accretion.  (The
plots are taken from R. Nelson \etal 2000).  }
\label{type2_fig16}
\end{figure*}

When the conditions for gap formation are satisfied, the outward flux
of angular momentum due to viscous transport in the outer disk section
is supplied by the planet.  Similarly, the outward flux from the inner
section is taken up by the planet.  When these contributions do not
balance, the planetary orbit changes its angular momentum and thus
migrates. Normally, simulations indicate that the outer disk has the
dominant effect, causing inward migration.  This is illustrated in
figure~\ref{type2_fig16}, taken from R.~Nelson \etal~(2000), which
shows the evolution of a planet of initially 1~M$_{\rm J}$ in a
standard disk.  In this case, the protoplanet was allowed to accrete
as it migrated inwards on the viscous timescale of $\sim 10^5$~yr.
Inward migration on the viscous timescale $R^2/\nu$ is characteristic
of type~II migration driven by the evolution of the disk when the
protoplanet mass does not exceed the local disk mass.  In that case,
using the standard $\alpha$--prescription by Shakura and Sunyaev
(1973), the migration timescale is given by:

\begin{equation}
t_{\rm{II}} (\rm{yr}) = \frac{1}{ 3\alpha} \left( \frac{R}{H} \right)^2
\Omega^{-1} = \frac{0.05}{\alpha} \left( \frac{R}{H} \right)^2 \left(
  \frac{R}{1 \; \rm{AU}} \right)^{3/2} .
\label{tII}
\end{equation}

\noindent If we adopt $H/r=0.1$ and consider $\alpha$ in the range
$10^{-3}$--$10^{-2}$, we get $t_{\rm{II}} \sim 10^3$ or $10^4$~yr at
$R=1$ or 5~AU, respectively.  These timescales are much shorter than
the disk lifetime or estimated planetary formation timescales.  The
expression of $t_{\rm{II}}$ is independent of the protoplanet's mass
and the surface mass density of the disk, but it is implicitly assumed
here that the mass of gas within the characteristic orbital radius of
the planet is at least comparable to the mass of the planet itself.
If the disk is significantly less massive, the inertia of the planet
slows down the migration (Syer \& Clarke 1995, Ivanov \etal 1999).


\subsubsection{Stopping type~II migration: \\ \\}

\noindent {\em{--- Stopping type~II migration at small radii: \/}}
Planets subject to type~II migration would stop before falling onto
the star if they entered a magnetospheric cavity.  Tidal interaction
with a rapidly rotating star would also halts planet orbital decay at
a few stellar radii (where the interaction becomes significant; see
Lin \etal~2000 and references therein).  Both of these mechanisms have
been put forth to account for the present location of the planet
around 51~Pegasi (Lin \etal 1996), and to explain the location of hot
Jupiters more generally.

\noindent A planet overflowing its Roche lobe and losing part of its
mass to the central star would also halt at small radii.  This is
because during the transfer of mass the planet moves outward to
conserve the angular momentum of the system (Trilling \etal 1998).
The planet stops at the location where its physical radius is equal to
its Roche radius.  Recent observations of atomic hydrogen absorption
in the stellar Lyman~$\alpha$ line during three transits of the planet
HD209458b suggest that hydrogen atoms are escaping the planetary
atmosphere (Vidal--Madjar \etal 2003). \\

\noindent {\em{--- ``The last of the Mohicans''...: \/ }} The
mechanisms reviewed above cannot account for the presence of giant
planets orbiting their parent star at distances larger than about a
tenth of an AU.

\noindent It has been suggested that migration of a giant planet could
be stopped at any radius if migration and disk dissipation were
concurrent (Trilling \etal 1998, 2002).  If the disk dissipates while
migration is taking place, then the drift timescale may increase in
such a way that the planet stalls at some finite radius.  Note however
that this requires very fine tuning of the parameters (disk mass, disk
lifetime etc.), as for a given disk mass the migration (viscous)
timescale given by equation~(\ref{tII}) decreases as the orbital
radius decreases.  Also, a major problem with this mechanism is to
explain how the disk dissipates.  Within this scenario, there is
initially enough gas in the disk to push the planet down to some
orbital radius.  For typical disk parameters, only part of this gas
may be accreted by the planet or leak through the gap to be accreted
onto the star (Bryden \etal 1999, Kley~1999, Lubow \etal 1999,
R.~Nelson \etal 2000).  It is therefore not clear how the gas
disappears.

\noindent A giant planet could survive if after it formed there were
not enough material left in the disk for significant migration to
occur.  It has been suggested that a series of giant planets could
actually assemble in the disk and disappear onto the star (Gonzalez
1997, Laughlin \& Adams 1997).  Then at some point the disk mass may
be such that one more planet can be formed but not migrate (Lin 1997).
This survivor is sometimes refereed to as the last of the Mohicans.

\noindent Note that the prediction that the more massive giant planets
that are able to form gaps should be predominantly at larger radii is
supported by the observations. \\

\noindent {{\em ---Planets locked in resonances: \/ }} Orbital
migration occuring for different planets at different rates, dependent
on local disk parameters, can produce convergent migration and locking
on to mean motion commensurabilities in a similar manner as is
believed to occur as a result of tidally induced migration of
satellites in the solar system (Goldreich 1965).

\noindent Pairs of extrasolar planets have been discovered in such
commensurabilities.  Gliese~876 (Marcy \etal 2001) and HD~82943 (Mayor
\etal 2004) are in 2:1 commensurabilities while 55~Cancri (McArthur
\etal 2004) exhibits a 3:1 commensurability.  The origin of these
commensurabilities is satisfactorily accounted for by disk planet
interaction (e.g., Snellgrove \etal 2001, Lee \& Peale 2002,
Papaloizou 2003, Kley \etal 2004). However, there is no indication
that the existence of these commensurabilities slowed down or halted
inward migration. On the other hand, it is possible that for some
planet masses and disk conditions, resonant trapping of planets could
lead to a reversal of type~II migration.  This has been suggested to
occur when a giant planet (e.g., Jupiter) migrates inwards and
captures a lighter Saturn mass planet in 2:3 resonance (Masset \&
Snellgrove 2001).  However, it is unclear how often and for how long
such situations can be realized.
 
 \noindent We also comment that apsidal resonances in which the
apsidal lines of two planet orbits are aligned, but not commensurable,
could also be produced through disk planet interactions which are
dissipative and therefore tend to select a particular normal mode
involving non zero eccentricity of the planetary and disk orbits
jointly, while the latter is present (e.g., Papaloizou 2003).

\subsection{Linear torques and type~I migration }

\label{sec:typeI}

In this case, one is concerned with protoplanets that are not massive
enough to form a gap in the disk.  We are therefore interested in
calculating the linear response of the disk to an embedded orbiting
protoplanet and using that to calculate the angular momentum exchange
rate with the protoplanet.  

\noindent The perturbation exerted by the protoplanet on the disk
propagates as density waves outside the Lindblad resonances and is
evanescent inside these resonances, in the corotation region.  The
protoplanet exerts a torque on the density waves, which, together with
the torque exerted at corotation, is responsible for the exchange of
angular momentum between the disk rotation and the planet's orbital
motion.  Away from the Lindblad resonances, the waves have a small
wavelength, so that the contribution to the net torque is small.
Therefore, most of the torque is exerted in the vicinity of the
Lindblad resonances, where the perturbation has a large wavelength, or
in the vicinity of the corotation resonance (see Terquem \etal 2000
and references therein).

\noindent Thus, the torque exerted by the protoplanet on the disk can
be calculated either by performing a full linear analysis of the
excited waves in two or three dimensions (see, for example, Korycansky
\& Pollack~1993, Tanaka \etal 2002, Terquem~2003, Tanaka \&
Ward~2004) or by summing up the contribution of point like torques
exerted at Lindblad and corotation resonances (see, for example,
Goldreich \& Tremaine~1979, Artymowicz~1993, Ward~1997, Papaloizou
\& Larwood~2000).  The results of full numerical simulations in
general confirm the linear analysis (e.g., D'Angelo \etal 2003, Bate
\etal 2003).  Both analytical and numerical calculations find a
positive (negative) torque on the disk (protoplanet) when the disk is
laminar, circular and does not contain any magnetic field.  This
corresponds to inward migration of the protoplanet.

\begin{figure*}
\centerline{
\epsfig{file=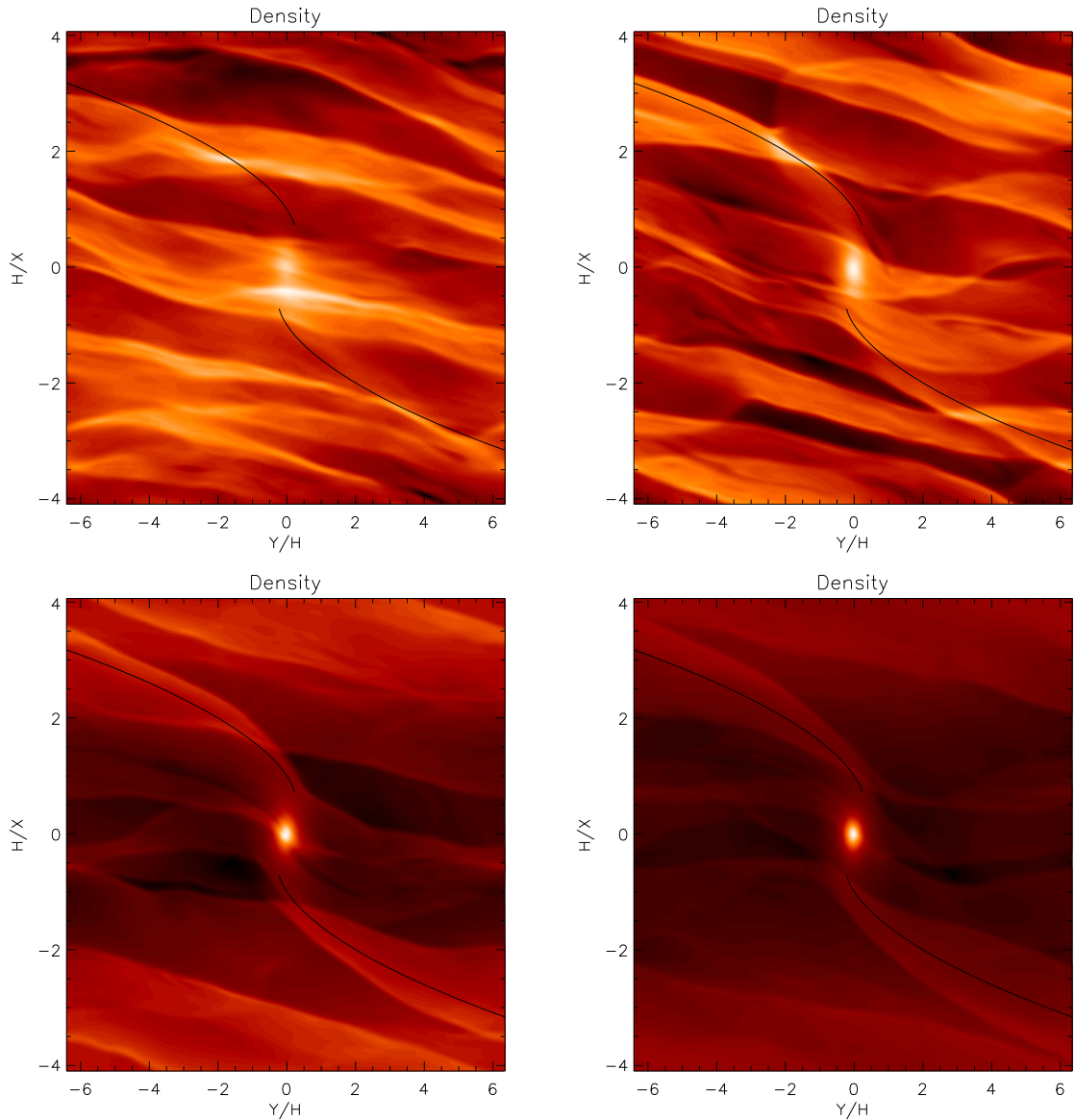,width=\textwidth,angle =0} }
\caption[]
{Density contour plots in a typical $(x,y)$ plane for simulations
with $M_p R^3/(M_{\star} H^3) = 0.1$ ({\em top left panel}), 0.3 ({\em
top right panel}), 1.0 ({\em bottom left panel}) and 2.0 ({\em bottom
right panel}).  As $M_p R^3/(M_{\star} H^3)$ increases, the wake
becomes more prominent, material is accreted by the protoplanet and is
pushed towards the radial boundaries as a gap is formed. The location
of the characteristic ray defining the wake is also plotted (see
Papaloizou \etal~2004).  The qualitative structure seen in these
simulations is maintained once a quasi--steady state has been
attained. (These plots are taken from Papaloizou \etal~2004).}
\label{SIM_fig3}
\end{figure*}

\noindent The conditions for non linearity, or gap formation, have been
explored by Papaloizou \etal (2004) using both local shearing box
simulations and fully global simulations.  A sequence of local
shearing box simulations, incorporating MHD turbulence with zero net
magnetic flux,  which have increasing $M_p R^3/(M_{\star} H^3),$
is illustrated in figure~\ref{SIM_fig3}.  The transition from being
embedded to opening a gap is seen to occur when this parameter is
around unity and it corresponds to the nonlinearity condition
mentioned above that the radius of the Hill sphere should exceed the
disk thickness.  When this condition is satisfied, the viscous
transport condition given by equation~(\ref{GAPvisc1}) is also
satisfied in these simulations.

\subsubsection{Timescales: \\ \\}

\label{sec:timescales}

The resulting dependence of the linear (inward) migration time $t_I$
and eccentricity damping time $t_e$, calculated for a laminar inviscid
disk as a sum of contributions from Lindblad resonances, on the gas
disk thickness is such that $t_{I} \equiv | J/{{\dot J}}| \propto
(H/R)^{2}$ and $t_e \equiv |e/{\dot e}| \propto (H/R)^{4}$.
  This
scaling results in $t_e$ being much shorter than $t_{I}$.
Accordingly, we expect circular orbits to be set up quickly and then
migrate.  But note that this may not be the case when the disk is
eccentric. Then the disk interaction is expected to lead to a finite
eccentricity in the protoplanet orbit (Papaloizou 2002). Furthermore,
the migration direction may be affected.

 Tanaka \etal (2002) performed a local three dimensional
calculation for a planet in circular orbit embedded in an isothermal
axisymmetric disk that summed contributions from both Lindblad and
corotation resonances introduced through the vertical dependence.
When $\Sigma \propto R^{-\gamma}$ and there is a central solar mass,
they obtained:

\begin {equation} 
t_I = 2(2.7+1.1\gamma)^{-1}\frac{M_{\odot}}{M_p} 
\frac{M_{\odot}}{\Sigma R^2} \left( \frac{H}{R} \right)^2
\Omega^{-1}.
\label{ttw}
\end{equation}

\noindent  Results obtained from three dimensional calculations
can be reproduced in two dimensional flat disk calculations if a
gravitational softening parameter is adopted.  When softening is
introduced, the $1/R$ potential for the protoplanet is changed to
$1/\left( R^2 + b^2 \right)^{1/2},$ the softening parameter being
$b$. This can be regarded as accounting for the vertical averaging of
the potential over the vertical thickness and accordingly $b$ should
be $\sim H.$

We note that for their two dimensional calculations, Papaloizou \&
Larwood~(2000) found that the torque results are dependent on the
gravitational softening parameter used.  Then $t_{I}$ and $t_e$ scale
as $b^{1.75}$ and $b^{2.5}$, respectively, for $b$ in the range
$0.4H$--$H$.  Papaloizou \& Larwood~(2000) find the approximate fits
to $t_{I}$ and $t_e$ {\bf when there is a central solar mass and
$\gamma = 3/2$ }to be:

\begin{eqnarray}
t_{I}({\rm yr}) = & & 3.5 \times 10^5 f_s^{1.75} \left[  1+ \left( {e R
\over 1.3H} \right)^5 \right] \left[ 1- \left( {e R \over 1.1H} \right)^4
 \right]^{-1}
\times \nonumber \\ 
& & 
\left({H/R
\over 0.07}\right)^2 \left({ 2 {\rm M}_{\rm J}\over M_{GD}}\right)
\left({ {\rm M}_{\oplus}\over M_p }\right) \left({R \over 1 \;
{\rm AU}}\right) ,
\label{mfit}
\end{eqnarray}

\noindent and:

\begin{eqnarray}
t_e({\rm yr}) = & & 2.5 \times 10^3 f_{s}^{2.5}
\left[1+\frac{1}{4}
\left({e \over H/R }\right)^3\right] \times \nonumber \\ 
& & 
\left({H/R \over 0.07}\right)^4
\left({ 2 {\rm M}_{\rm J}\over M_{GD}}\right)
\left({ {\rm M}_{\oplus}\over M_p }\right)
\left({R \over 1 \; {\rm AU}}\right) .
\label{efit}
\end{eqnarray}

\noindent Here the gas disk has a mass $M_{GD}$ contained within 5~AU
and $f_s \equiv 2.5b/H,$ where $b$ is the gravitational softening
length. {\bf  Equations~(\ref{ttw})~and~(\ref{mfit}) agree when 
$f_s = 2.3,$  and give }the characteristic timescale for
migration of an Earth mass core in the minimum mass solar nebula as
about $10^6$~yr.  This timescale decreases linearly with mass,
resulting in a survival problem for proto--Jovian mass planet cores.
In this context, even the smallest cores considered of 5~M$_{\oplus}$
have an infall timescale of only $2\times 10^5$~yr.

\noindent The resolution of this problem may be provided by magnetic
fields.  When they are present and produce turbulence with density
fluctuations that can be of order unity, type~I migration may be
stochastic (e.g., Nelson \& Papaloizou 2004, Laughlin \etal 2004). In
this case, the migration direction of small mass cores becomes
indeterminate over the timescales simulations can be performed and the
final outcome may depend on long term global fluctuations in the MHD
turbulence which in turn may depend on the global disk environment.
We comment here that it is not clear that the disk behaves as a
laminar disk with the time averaged density, upon which noisy
fluctuations, with a characteristic time comparable to the orbital
period and which obey Gaussian statistics, are superposed. If that
were the case, it would be very difficult to affect the long term
trend determined by the average laminar disk because of the large
ratio of disk lifetime to orbital period. An additional recent
investigation by Nelson (2005) suggests that turbulent fluctuations
occur on all timescales up to the global diffusion or viscous
time. Furthermore, the background perturbed by the planet is very
different from laminar, with large density fluctuations encountering
the planet and providing stochastic torques even when unperturbed by
it. Consider a density fluctuation of order unity with length scale of
order $H$ a distance of order $H$ from the planet.  The characteristic
specific torque acting on it is $G\Sigma R.$ Given the stochastic
nature of turbulence, one might expect the specific torque acting on
the planet to oscillate between $\pm G\Sigma R.$ Note that this
exceeds the net specific torque implied by equation (\ref{ttw}) by a
factor $N_t \sim [(M_{\odot}/M_p)(H/R)^3] (R/H).$ From the discussion in
section~\ref{sec:typeI}, the first factor should exceed unity for an
embedded planet. Thus, such an object is inevitably subject to large
torque fluctuations. The strength of the perturbation of the planet on
the disk is measured by $( M_p/M_{\odot})(R/H)^3.$ This perturbation
only needs to produce a bias in the turbulent fluctuations resulting
in a non zero mean, corresponding to the typical fluctuation reduced
by a gactor $N_t,$ for it to become comparable to the laminar type~I
value.

Also, global magnetic fields may be present and have significant
effects on migration (Terquem 2003), or the disk may have long lived
global distortions corresponding to non zero eccentricity (Papaloizou
2002) which may also significantly affect migration.  The migration
rates may also be affected by opacity transitions in the disk (Menou
and Goodman 2004).  Under some circumstances, significant fluctuations
that can affect migration may occur in the non magnetic case for
planets in the Saturn mass range (Koller \etal 2003) through the
production of instabilities in inviscid gap edges.

\subsubsection{Type~I migration in a turbulent disk: \\ \\}

\label{sec:turbulence}

\begin{figure*}
\centerline{
\epsfig{file=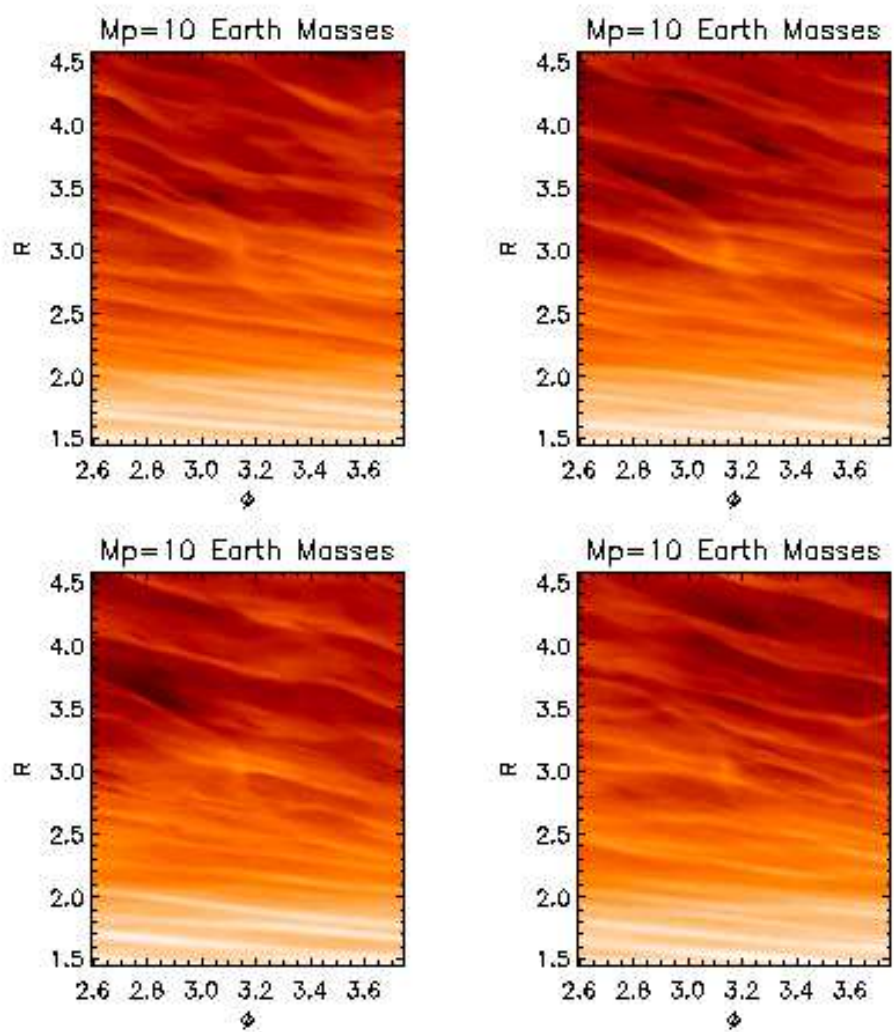,width=\textwidth} }
\caption[] {This figure shows midplane density contours in the region
close to the protoplanet. The protoplanet is located at ($R$,$\phi$)=
(3,$\pi$) in a disk with fully developed MHD turbulence with $H/R =
0.07$ and no net magnetic flux. The panels correspond to times
429.0687 ({\em top left}), 496.8698 ({\em top right}), 542.0710 ({\em
bottom left}) and 634.4136 ({\em bottom right}). The orbital period at
$R=1$ is $2\pi$ in these units. Note that the presence of the
protoplanet of mass 10~M$_{\oplus}$ is just detectable, although the
perturbation it makes to the disk is of lower amplitude than the
perturbations generated by the turbulence. (These plots are taken from
Nelson \& Papaloizou~2004).}
\label{MHD_fig11}
\end{figure*}

To illustrate stochastic migration when fully developed MHD turbulence
is present in a disk with with $H/R = 0.07$ and no net magnetic flux,
we show midplane density contours in figure~\ref{MHD_fig11}, taken from
Nelson \& Papaloizou (2004).  The presence of the protoplanet of mass
10~M$_{\oplus}$ is just detectable, although the perturbation it makes
to the disk is of lower amplitude than the perturbations generated by
the turbulence.

\begin{figure*}
\centerline{
\epsfig{file=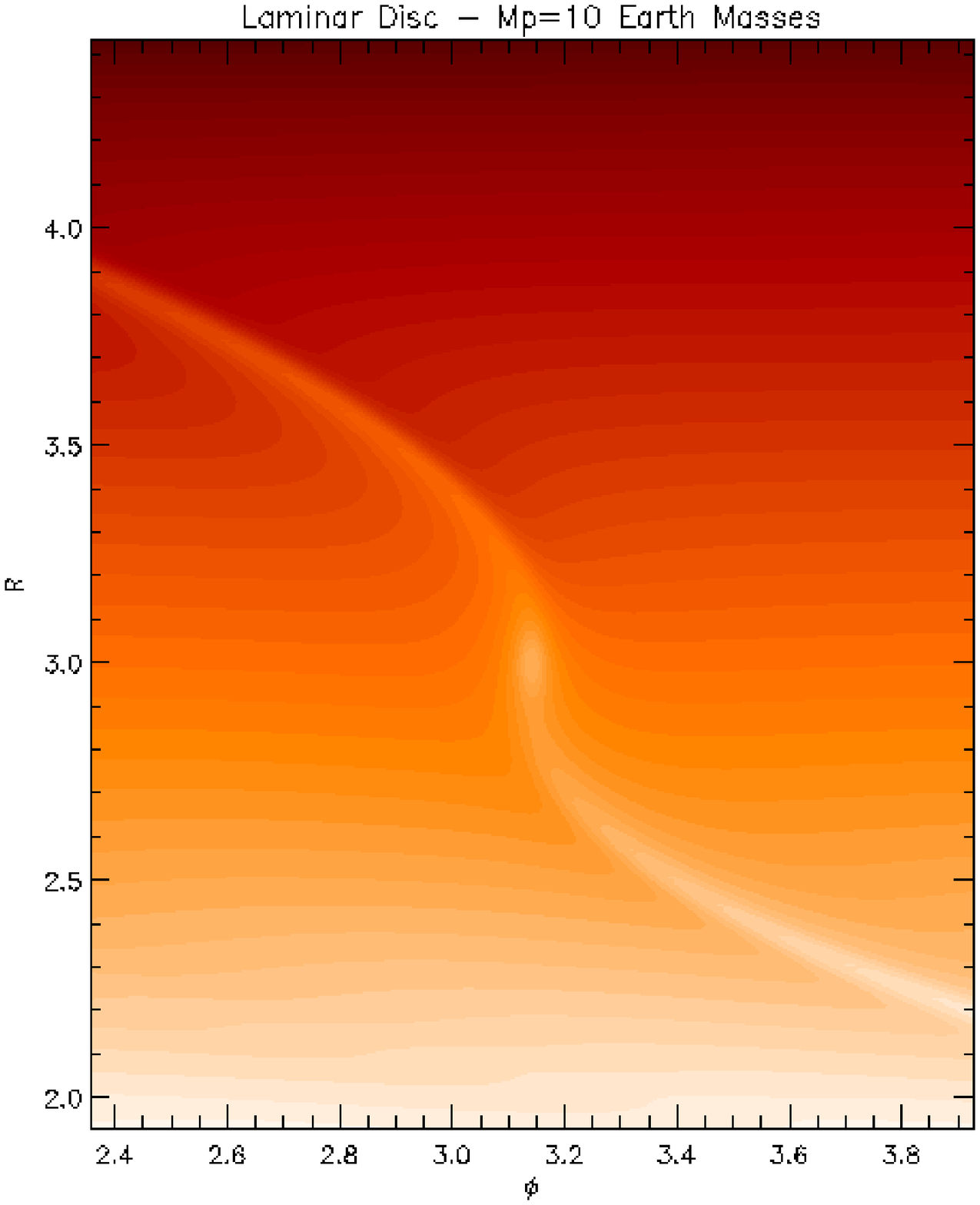,width=10.cm} }
\vspace*{-1.cm}
\caption[] {This figure shows midplane density contours for a laminar
disk run with disk and protoplanet parameters otherwise identical to
the simulation shown in figure~\ref{MHD_fig11}.  (This plot is taken
from Nelson \& Papaloizou~2004)}
\label{LAM_fig12}
\end{figure*}

\noindent Midplane density contours for the same protoplanet in a
laminar disk with identical parameters is given in figure~\ref{LAM_fig12}.
The disk response to the protoplanet is clearly much better defined in
this case.

\begin{figure}
\begin{minipage}[b]{\textwidth}
\parbox[t]{8cm}{
\includegraphics[width=8cm]{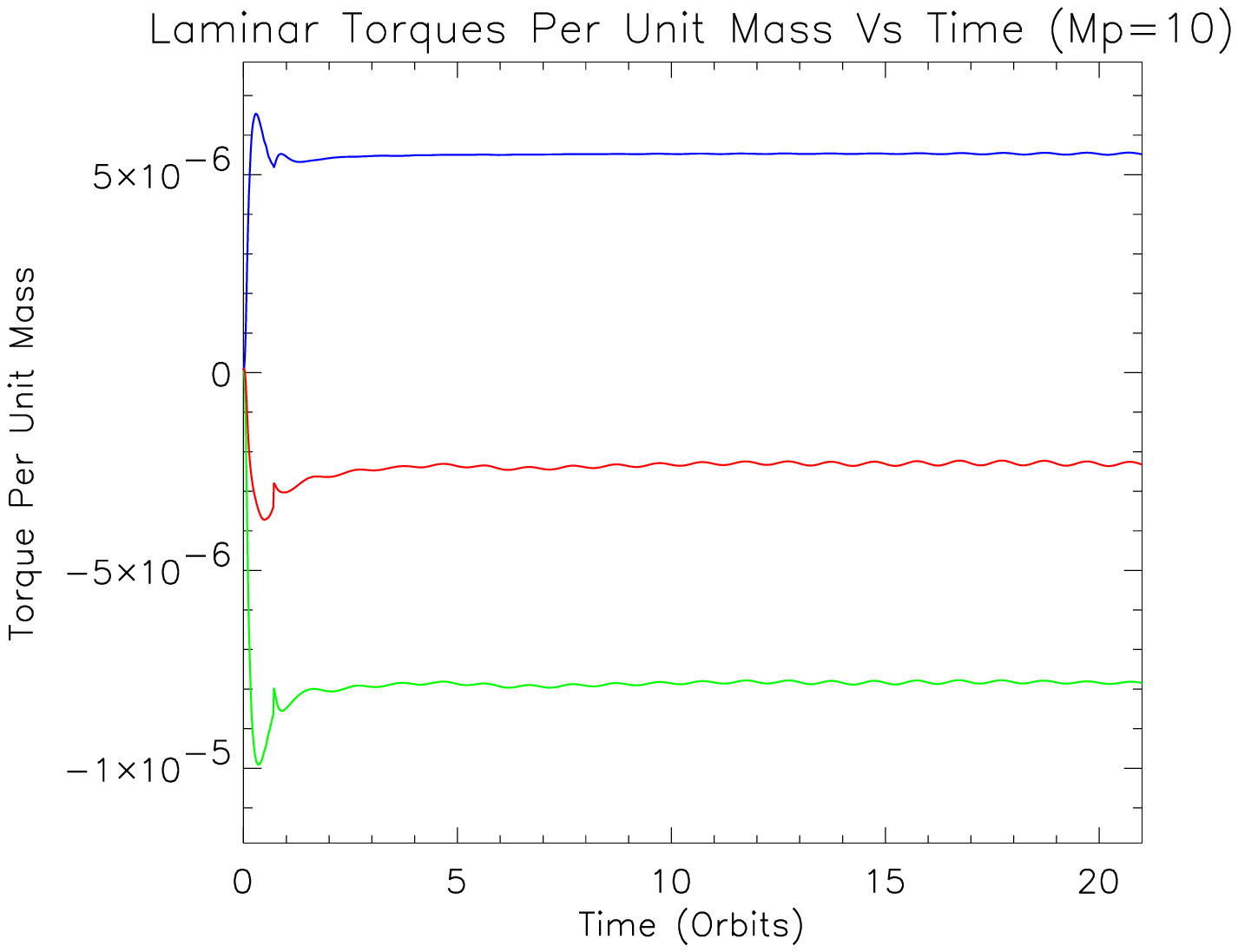}
}
\hfill
\parbox[t]{8cm}{
\includegraphics[width=8cm]{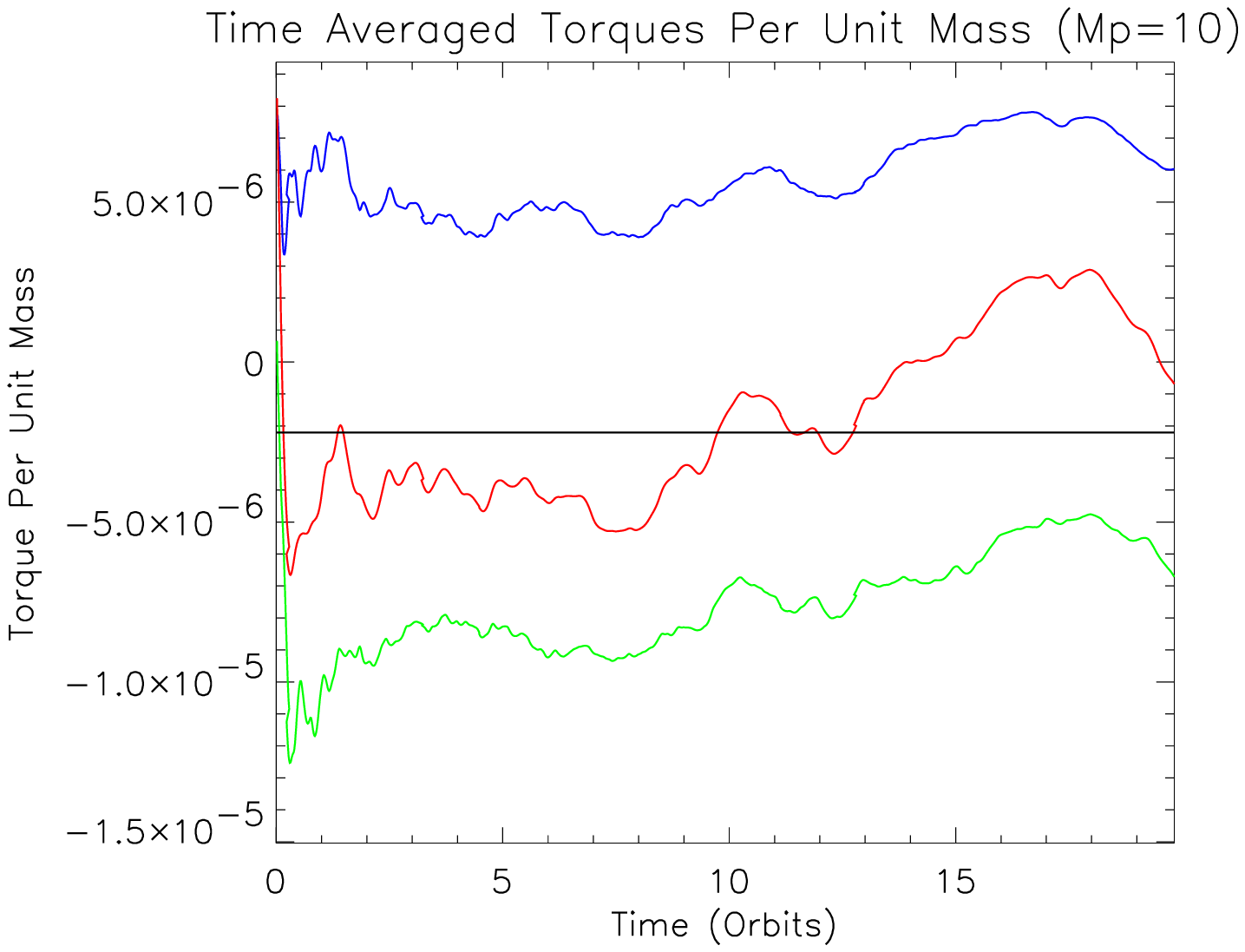}
}
\\ \mbox{}
\caption{\label{MHD_torque} {\em Left panel: \/} Torque per unit mass
exerted by the disk on the protoplanet for the laminar disk simulation
shown in figure~\ref{LAM_fig12}. The upper line shows the torque due
to the inner disk, the lower line shows the torque due to the outer
disk, and the middle line shows the total torque. It is clear that a
well defined torque is produced, with an associated migration time
scale.  \\ 
{\em Right panel: \/} Running time average of the torque per unit mass
exerted by the disk for the simulation shown in
figure~\ref{MHD_fig11}.  The upper line is the running time average of
the torque acting on the planet due to the inner disk.  The lower line
is that due to the outer disk. The middle (not straight) line is the
running time average of the total torque. The straight line is the
total torque exerted on the protoplanet in the comparable laminar disk
run. We note that the total time averaged torque does not converge to
a well defined value.  (These plots are taken from Nelson \&
Papaloizou~2004).}
\end{minipage}
\end{figure}

\noindent The behavior of the torque per unit mass exerted by the
disk on the protoplanet in the laminar case is illustrated in the left
panel of figure~\ref{MHD_torque}, taken from Nelson \&
Papaloizou~2004.  The upper line shows the torque due to the inner
disk, the lower line shows the torque due to the outer disk, and the
middle line shows the total torque. A well defined torque is produced,
with an associated inward migration timescale in agreement with the
standard linear theory.

\noindent By way of contrast, the right panel of
figure~\ref{MHD_torque} shows the running time average of the torque
per unit mass exerted by the disk in the turbulent case. The upper
line is the running time average of the torque acting on the planet
due to the inner disk.  The lower line is that due to the outer
disk. The middle (not straight) line is the running time average of
the total torque. The straight line is the total torque exerted on the
protoplanet in the laminar disk run.  The total time averaged torque
does not converge to a well defined value over the simulation run time
of 20 orbits.

\noindent These simulations indicate that, if the disk is not strictly
laminar, type~I migration may be significantly modified to the extent
that the survival of protoplanetary cores may occur (this is also the
case if the disk is magnetic or/and with gas not strictly on circular
orbits).  Furthermore, the enhanced mobility throughout the disk of a
core undergoing stochastic migration may act to counteract the effects
of core isolation (Rice \& Armitage~2003). However, considerable
theoretical developments need to be made in order to confidently
assess the outcome of these phenomena.

\subsection{Coorbital Torques and type~III migration}\label{coorb}

By coorbital torque, we mean a torque exerted on the protoplanet by
disk material flowing through the orbit. This has been discussed
recently by Masset \& Papa\-loizou~(2003) and Artymowicz~(2004)
in a global context and
by Papaloizou~(2005) in a local shearing box context.

\noindent One starts by considering a migrating protoplanet.  When gap
formation is only partial, material will flow through the orbit.  As
it does so, a torque will be exerted on the protoplanet.  This may
either act to assist the migration (positive feedback) or retard it
(negative feedback).

\noindent D'Angelo \etal (2005) have recently studied coorbital
torques acting on a Saturn mass planet in a massive disk under a
similar set up to Masset \& Papaloizou (2003). They used a multigrid
system for which the resolution could be increased close to the
planet, {\it but not elsewhwere}.  They report only slow migration
differing only marginally from type~II migration.  This was claimed to
result from important torques acting within the Roche lobe.  We
firstly comment that their computational set up allowed mass flow into
the Roche lobe, effectively increasing the protoplanet mass by more
than a factor of three. However, the increase in mass was not
incorporated in the action of the protoplanet on the disk to determine
its response, making the calculation inconsistent. Furthermore the
results, checked with the highest resolution, shown in their figure
14, which differ from those used to draw final conclusions, indicate a
fast inward migration time of $\sim 600$ orbits. This would not
indicate important contributions from inside the Roche lobe as one
would expect when there is not an increasing accumulation of mass
there (see the appendix).  However, the study of the migration of
partially gap forming planets of modest mass, with a disk mass
comparable to their own in the coorbital region, is difficult
numerically and clearly further work on the problem is required.

\noindent Masset \& Papaloizou~(2003) argue for positive feedback from
coorbital torques because, for an outwardly migrating protoplanet,
material traverses the coorbital zone from outside to inside.  As it
does so, it moves along the outer boundary of the coorbital zone
occupied by material librating around the coorbital equilibrium
(Lagrange) points passing close to the protoplanet.  As this passage
occurs, angular momentum is transferred to the protoplanet, a process
which acts to assist the migration of the protoplanet and which
accordingly gives a positive feedback.  We now look at this process in
more detail for an outwardly migrating protoplanet.  Similar
considerations apply with appropriate sign reversals to an inwardly
migrating protoplanet.

\noindent The expected force exerted on the protoplanet by material
flowing through the coorbital region is estimated as follows. The rate
at which the mass flows through is $ 2\pi R \Sigma dR/dt.$ The outward
momentum per unit mass imparted to the protoplanet when the disk
matter moves across the coorbital region of width $w$ is $w \omega
/2,$ where $\omega$ is the rotation rate of the protoplanet. 
We adopt $w=2H$, being appropriate for marginal gap formation.  Then,
the rate of transfer of  momentum to the planet is $F_{cr} =
2\pi R \omega H \Sigma (dR/dt).$ This can be expressed as:

\begin{equation}
F_{cr} = 2 \pi R \Sigma \omega^2 H^2 \frac{1}{c_s} \frac{dR}{dt} ,
\end{equation}

\noindent where $c_s$ is the sound speed.  Using $c_s \simeq \omega
H$, it may also be written in the form:

\begin{equation}
F_{cr} =  {1\over 2} M_d \omega  \frac{dR}{dt} , \label{crt00}
\end{equation}

\noindent where $ M_d = 4\pi R \Sigma H$ is the disk mass that would
fill the coorbital zone of width $2H$ were it to do so at the
background surface density.

\noindent However, torques on the protoplanet do not only arise from
material passing through the coorbital zone. Material that is forced
to comove with the protoplanet, either because it has been accreted by
it, or because it librates about coorbital equilibrium points, has to
be acted on by the same force per unit mass as the protoplanet so as
to maintain its migration.  This results in an additional force acting
on the protoplanet given by:

\begin{equation}
F_{crb} =  -{1\over 2} M_b \omega \frac{dR}{dt} ,
\end{equation}

\noindent where $ M_b$ is the coorbital bound mass.  The total force
so far is thus:

\begin{equation}
F_{cr} =  {1\over 2}( M_d - M_b) \omega \frac{dR}{dt} . \label{crto}
\end{equation}

\noindent However, there are other forces. The forces acting due to
density waves from the two sides will be affected by the migration and
flow through and thus differ from the non migrating case. In this
context, note that there is an associated asymmetry in the surface
density profile. Thus, we should expect a wave torque component which
is proportional to the migration speed. The indication from our
simulations is that this acts as a drag on the protoplanet, as does
$M_b.$ Accordingly, we shall consider the effects of asymmetric wave
torques as modifying the value of $M_b$ and continue to use
equation~(\ref{crto}).

\noindent Suppose now that the protoplanet is acted on by some
external torque $T_{ext}.$ The equation of motion governing the
migration of the protoplanet of mass $M_p$ with speed $dR/dt$, obtained
by considering the conservation of angular momentum, is:

\begin{equation}
{1\over 2} M_p R \omega \frac{dR}{dt} = {1\over 2}( M_d - M_b) R \omega
\frac{dR}{dt} + T_{ext}.
\label{crto1}\end{equation}

\noindent Accordingly, we can consider the planet to move with an
effective mass:

\begin{equation}
M_{eff} = M_p- ( M_d - M_b). \end{equation}

\noindent The quantity $(M_d - M_b)$ has been called the coorbital
mass deficit by Masset \& Papaloizou~(2003).  When this is positive,
there is an effective reduction in the inertia of the protoplanet.  If
there were no asymmetry in wave torques, the coorbital mass deficit
would be the amount of mass evacuated  from  the gap region
were it to be initially filled with the background density.  Gap
filling accordingly reduces the coorbital mass deficit.  It is also
clear that, because at least a partial gap is required, the
protoplanet must be massive enough to produce a nonlinear response in
the disk.

\noindent Masset \& Papaloizou~(2003) and Papaloizou~(2005) indeed
found the coorbital mass deficit to be positive, resulting in positive
feedback from coorbital torques. They also found that, in some
circumstances, the coorbital mass deficit could become as large as the
planet mass, so reducing the effective inertia to zero. Under these
circumstances, a very fast or runaway migration may occur. Because the
coorbital mass deficit is proportional to the disk surface density,
the phenomenon requires a massive disk, typically an order of
magnitude larger than the minimum mass solar nebula.

\noindent An interesting feature of this type of migration is that it
is most effective for protoplanets that bare marginally gap forming,
and thus in the mass range 0.1--0.3~M$_{\rm J}.$ Such forming
protoplanets may undergo a period of fast migration that brings them
into the inner regions of the disk, where gap formation is easier,
while they are still of sub--Jovian mass.  It may thus form an
explanation of why the 'hot Jupiters' are in general sub--Jovian in
mass (see Masset \& Papaloizou~2003 and Papaloizou \& Nelson~2005).
But note in this context that this explanation does not necessarily
require type III~migration, as migration rates tend to be maximal in
the mass range 0.1--0.3~M$_{\rm J}$ for low mass disks too (see
D'Angelo \etal 2003) while lower masses may have migration inhibited
by turbulence (Nelson \& Papaloizou 2004).

\begin{figure*}
\centerline{\epsfig{file=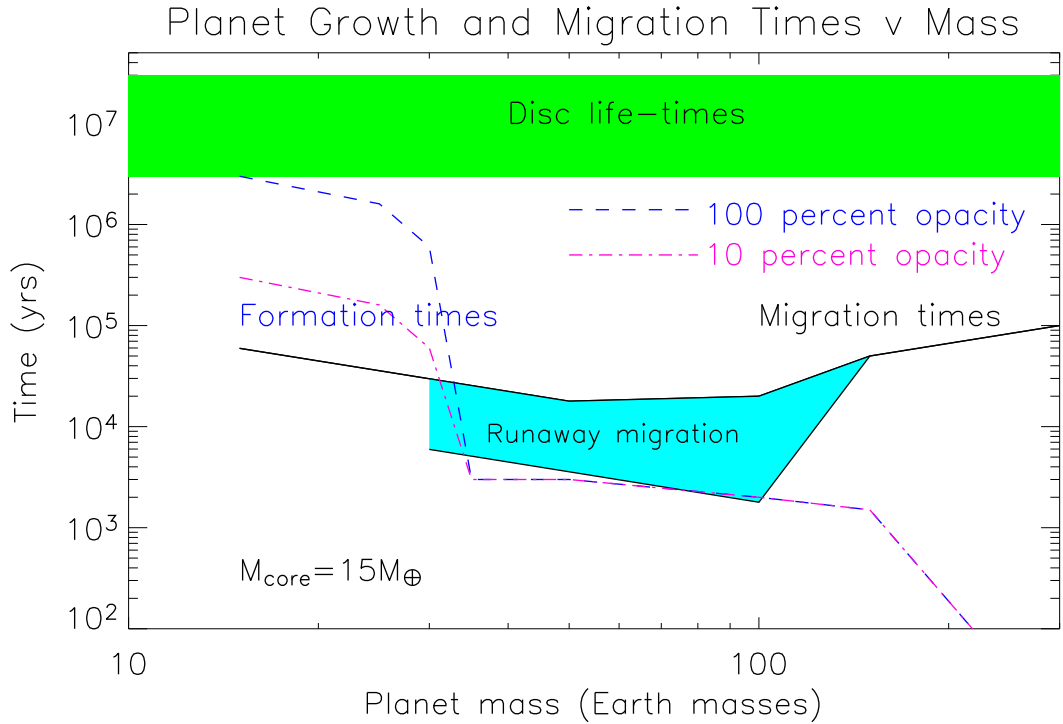,width=\textwidth, angle=0}}
\caption{\label{SUM_fig8} This diagram shows the migration time
estimated for a minimum mass solar nebula (solid line) and the mass
growth time for protoplanet models with 15~M$_{\oplus}$ cores.  The
lowest runaway migration times, forming the lower boundary to the
lower shaded region, are estimated for a model five times more massive
than the minimum mass solar nebula.  The dashed line shows the
protoplanet growth time as a function of mass for models with standard
opacity. The dashed--dotted line shows the growth time for models with
3 percent opacity.  (This plot is taken from Papaloizou \&
Nelson~2005) }
\end{figure*}

\noindent A graphical summary of the different types of migration is
shown in figure~\ref{SUM_fig8}.  This shows the migration timescale as
a function of protoplanet mass assuming a core of a constant mass of
15~M$_{\oplus}.$ It allows for the possibility of type~I, type~II and
type~III migration. It indicates that in order for a planet to form,
type~I migration, which produces radial infall timescales much shorter
than the protoplanet growth time, has to be strongly suppressed
relative to the prediction derived from a non magnetic, laminar and
circular disk.  However, beyond type~I migration, survivability, at
least to enable the formation of the planet, is not threatened on
account of the shortened growth times. This is the case even if
type~III migration operates.

\section{Multiplanet systems and their interactions}
\label{multiplanet}

The presence of high orbital eccentricities amongst extrasolar planets
is suggestive of a strong orbital relaxation or scattering process.
For this to happen, formation must occur on a timescale short enough
that strong dynamical interactions may take place subsequently.  The
gaseous environment of a disk may act to inhibit such interactions
until it is removed.  Gas free dynamical interactions of coplanar
protoplanets formed on neighboring circular orbits have been
considered by Weidenschilling \& Marzari~(1996) and Rasio \&
Ford~(1996).  These may produce close scatterings and high
eccentricities, but the observed distribution of extrasolar planets is
not reproduced.

\subsection{Orbital relaxation in young planetary systems}

Papaloizou \& Terquem~(2001, see also Adams \& Laughlin 2003)
investigate a scenario in which $5 \le N \le 100$ planetary objects in
the range of several Jupiter masses are assumed to form rapidly
through fragmentation or gravitational instability occurring in a disk
or protostellar envelope on a scale of $R_{max}= 100$~AU.  If these
objects are put down in circular orbits about a solar mass star, at
random in a volume contained within a spherical shell with inner and
outer radii of $0.1R_{max}$ and $R_{max}$ respectively, a strong
relaxation on a timescale $\sim 100$ orbits occurs, leading to
independence of details of initial conditions.

For the range of values of $N$ considered, the evolution is similar to
that of a star cluster.  Most objects escape, leaving at most 3 bound
planets, the innermost with semi--major axis in the range
$0.01R_{max}$--$0.1R_{max}.$ However, close encounters or collisions
with the central star occur for about 10\% of the cases.  Tidal
interaction, giving orbital circularization at fixed pericenter
distance and leading to the formation of a very closely orbiting giant
planet, is then a possibility.

\begin{figure*}
\centerline{
\epsfig{file=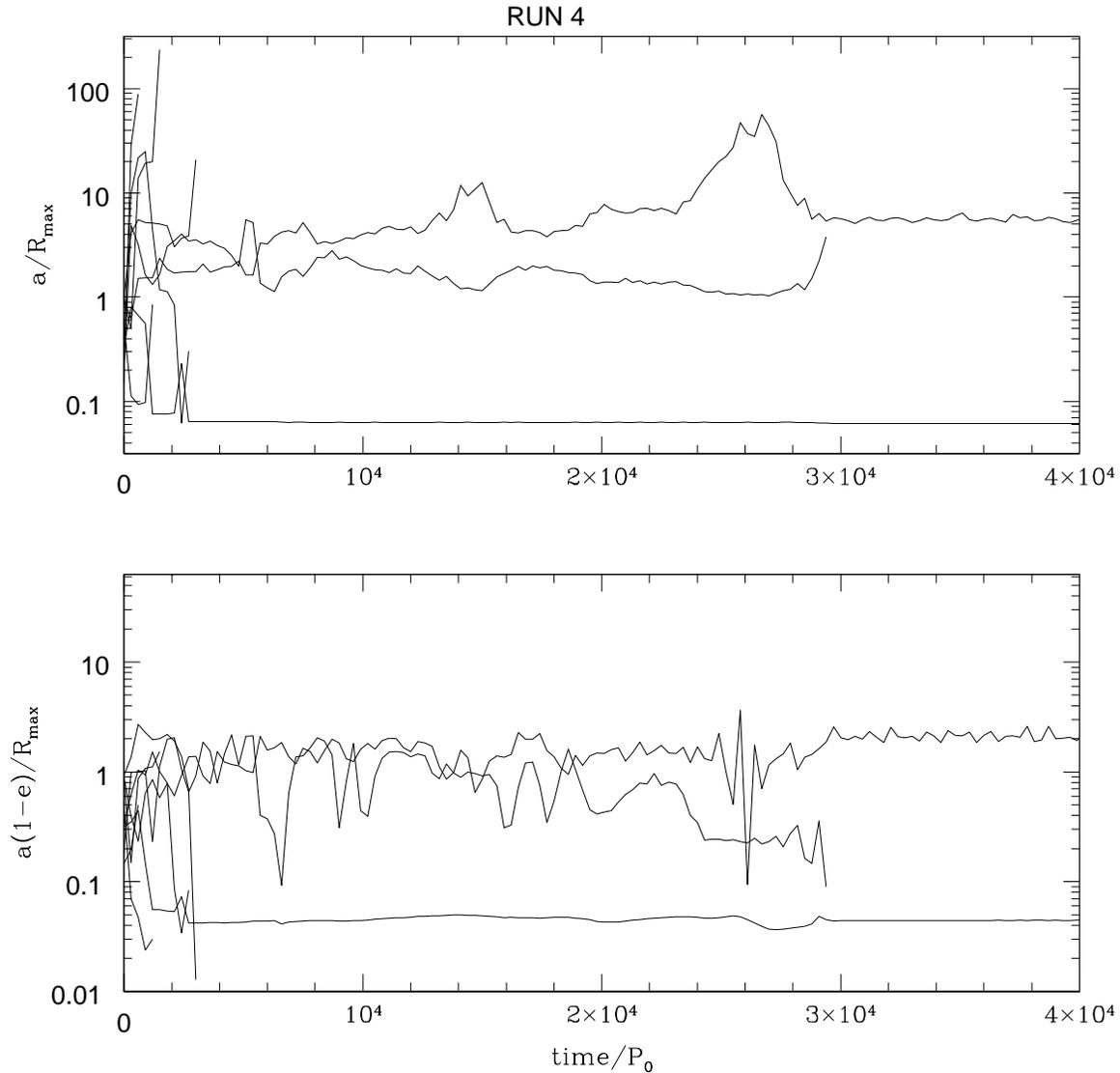,width=\textwidth} }
\caption[]{This figure shows the evolution of the semi-major axes
({\em upper plot}) and pericenter distances ({\em lower plot}) of the
$N=8$ planets in the system versus time (measured in units of $P_0,$
being the orbital period at $R_{max}$).  The lines correspond to the
different planets.  A line terminates just prior to the escape of a
planet. (These plots are taken from Papaloizou \& Terquem~2001)}
\label{fig5t}
\end{figure*}

An example of a run with $N=8$ masses selected uniformly at random in
the interval $(0 , 5\times 10^{-3} {\rm M}_{\odot})$ and with central
stellar radius $R_\star= 1.337 \times 10^{-4} R_{max}$ is illustrated
in figure \ref{fig5t}.  At the end of this run, only 2~planets remain
bound to the central star.

The relaxation processes discussed above are more likely to apply to
the more massive extrasolar planets exceeding 1~M$_{\rm J}$.  Their
observed number increases with distance (Zucker \& Mazeh~2002 and
section~\ref{mass_sep} above), as found in the simulations.  So far,
there are 23 isolated candidate objects with mass $M_p \sin i >
4.5$~M$_{\rm J}$, semi--major axes $a$ in the range 0.3--3.9~AU and
eccentricities $e$ in the range 0.03--0.76. Amongst the 'hot Jupiters'
detected so far, $\tau$~Boo is a possible candidate, being unusually
massive with $M_p \sin i \sim 4$~M$_{\rm J}.$ Note that, in multiple
systems, only 3~planets have been detected with a mass larger than
4.5~M$_{\rm J}$.  They are actually significantly more massive than
this limit, being between 6 and 13~M$_{\rm J}$.  One of these systems
contains a 7.7~M$_{\rm J}$ planet with $a=0.29$~AU and $e=0.529$, and
a 16.9~M$_{\rm J}$ brown dwarf with $a=2.85$~AU and $e=0.228$.  Such a
system could well have formed through the process described above.
The two other systems will be discussed below.

\subsection{Effect of an outer relaxing distribution of protoplanets 
on a low mass planet formed in an inner disc}

Terquem \& Papaloizou~(2002) have considered the effects of an outer
relaxing distribution of protoplanets on an inner disk in which one
inner planet with a mass $m_p=0.3$~M$_{\rm J}$ is formed on a
timescale of $10^6$~yr.  During the formation process, the
eccentricity is assumed to be damped by tidal interaction with the disk
while the planet is built up progressively with orbital radius $a$ in
the range 0.3--10~AU.

\begin{figure*}
\centerline{
\epsfig{file=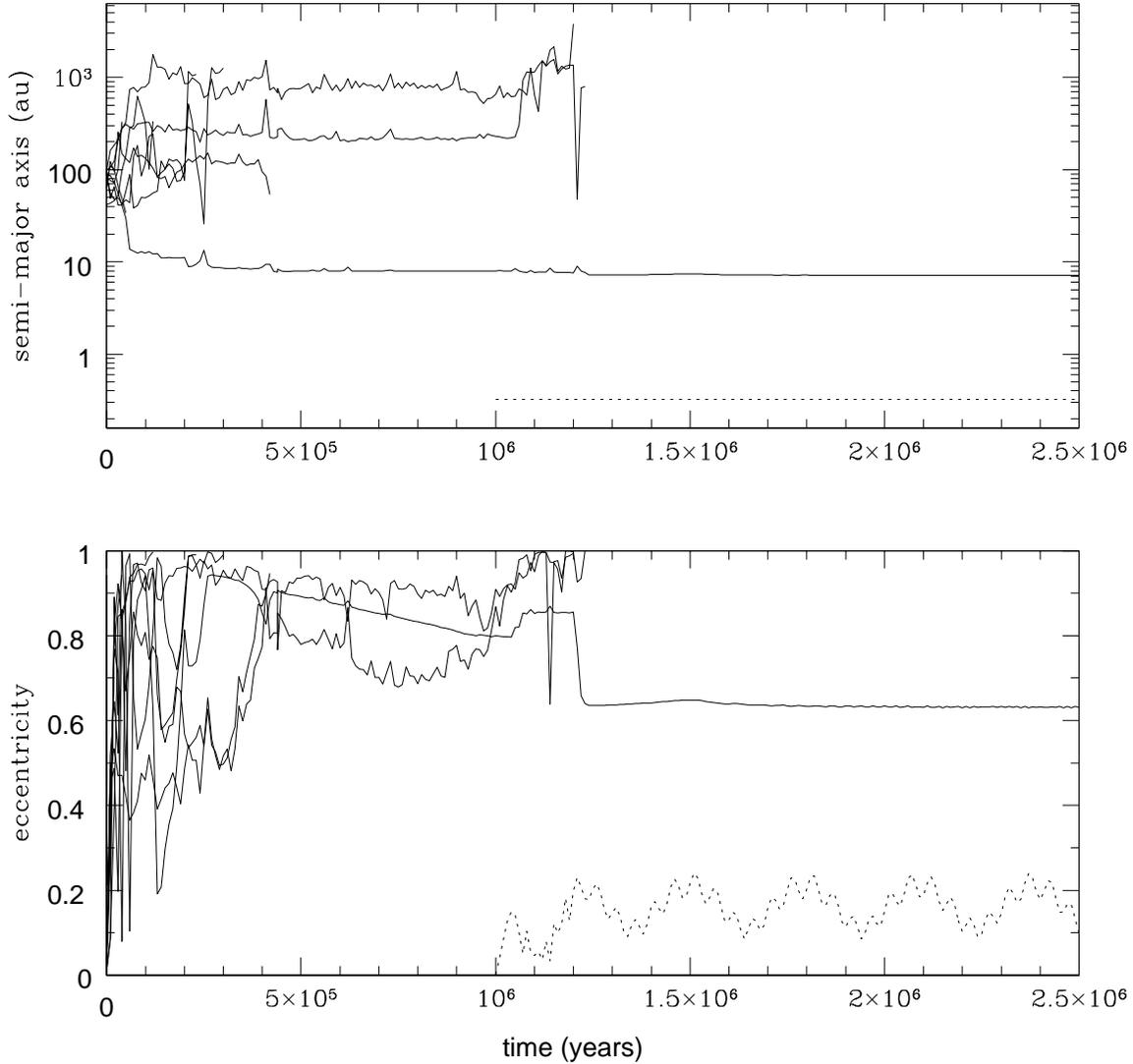,width=\textwidth} }
\caption[]{ Semi-major axes (in AU, {\em upper plot}) and
eccentricities ({\em lower plot}) of the $N=9$ outer planets ({\em
solid lines}) and low--mass planet ({\em dotted lines}) versus time
(in yr).  Here, for the innermost low--mass planet planet, $a=0.3$~AU
initially.  The eccentricity of the low--mass planet varies between
0.1 and 0.24 while $a$ is almost constant. (These plots are taken from
Terquem \& Papaloizou~2002)}
\label{fig5tt}
\end{figure*}

An example of a run with $N=9$ outer planets of mass 8~M$_{\rm J}$ and
$R_{max} = 100$~AU is illustrated in figure~\ref{fig5tt}.  After $\sim
10^6$~yr, one outer planet with $a \sim 8$~AU and $e \sim 0.6$
remains.  The inner planet enters into a cycle in which $e$ varies
between 0.1 and 0.24.  The mutual inclination oscillates between 0 and
30$^\circ$. More extreme cycles have been produced in other examples.

\begin{figure*}
\centerline{
\epsfig{file=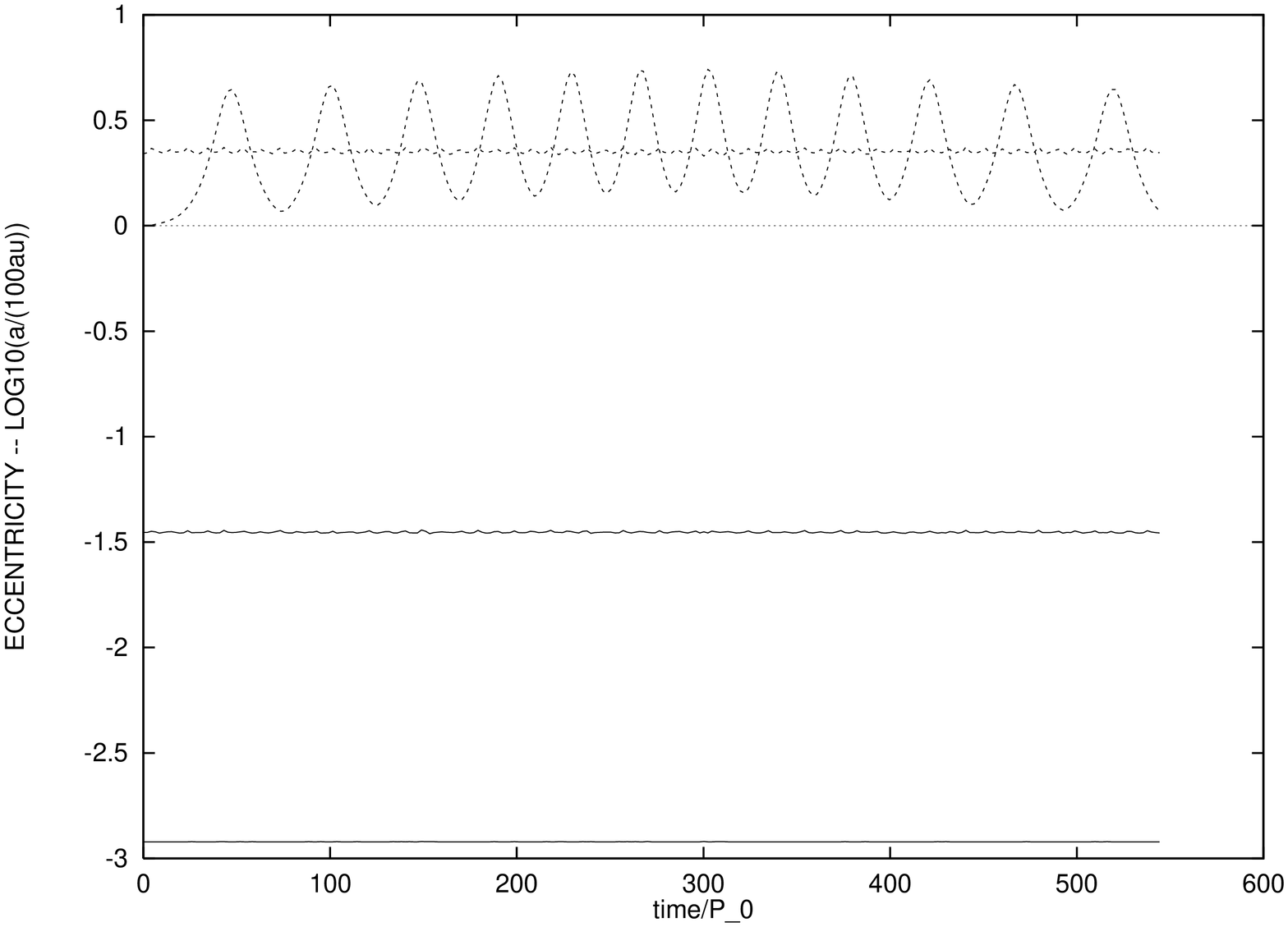, width=8.cm, angle=270} }
\caption{Motion in HD~35829 for mutual inclinations of 0 and
60$^\circ$. The unchanging $\log_{10}(a/R_{max})$ for the two planets
are given by the lower curves. The eccentricities by the upper curves.
The oscillatory curve indicates a Kozai like cycle for the inner
planet in the high inclination case. In the low inclination case the
eccentricity of the inner planet remains close to zero.}
\label{REL_fig6}
\end{figure*}

Thus, an outcome of an outer relaxing system could be an inner lower
mass protoplanet with high orbital eccentricity.  Among the observed
planets, there are several candidates for which no companion has been
detected so far.  If the above scenario applies, there should be an
outer massive planet with high eccentricity.  Among the candidate
systems with large radial velocity drifts potentially due to a
companion selected by Fischer \etal (2001) was HD~38529, which has a
planet with $m_p \sin i= 0.76$~M$_{\rm J}$, $a=0.13$~AU and $e=0.27.$
This is similar to the system illustrated in figure~\ref{fig5tt}.
More recently, the discovery of a companion with $M_p \sin i=
12.7$~M$_{\rm J}$, $a=3.7$~AU and $e=0.36$ has been announced (Fischer
\etal 2003).  If $\sin i=0.8$ is adopted for this system, a
significant eccentricity is not excited in an initially circular inner
planet orbit if the system is assumed coplanar.  But, as illustrated
in figure~\ref{REL_fig6}, for a mutual orbital inclination of
60$^\circ$, a cycle is found in which the inner eccentricity reaches
its observed value.

As mentioned above, three~of the observed multiple systems have a planet
with a mass larger than 4.5~M$_{\rm J}$.  In two of these systems
(HD~74156 and HD~38529), the massive planet has a semi--major axis on
the order of 3.5~AU and a significant eccentricity.  The companion has
a lower mass, a semi--major axis in the range 0.1--0.3~AU and also a
significant eccentricity.  These systems may be explained by the
scenario presented here.

\section{Concluding remarks}

The recent observations of extrasolar planets strongly suggest that
gravitational interactions between forming planets and the nebula in
which they are embedded form a key element in the understanding of the
configurations of the systems.  Given that the hot Jupiters have
almost certainly not formed where they are observed today, they must
have been subject to some orbital migration resulting from planet/disk
tidal interaction.  This type of interaction also leads to
commensurable pairs of planets and apsidal resonances, features that
are also observed in some extrasolar systems.

A substantial amount of work aimed at gaining a better understanding
of planet/disk tidal interactions has been done in the last ten
years. Three dimensional numerical simulations incorporating
internally generated MHD turbulence rather than anomalous viscosity
prescriptions have begun.  They show that orbital migration is
strongly affected by such processes, especially when the mass is in
the Earth mass range.  Analytical and numerical work has also shown
that the presence of a large scale magnetic field or a finite orbital
eccentricity can have a significant effect on migration.

Strong gravitational interactions amongst a population of distant
giant planets formed early in the life of the protostellar disk
through gravitational fragmentation may also produce close orbiters
and planets on highly eccentric orbits. The latter may belong to the
original distribution or be formed by a later accumulation in an inner disk.

We are currently at a stage where the important processes
(gravitational instability, dust sedimentation, core accumulation, gas
accretion and disk/planet interaction) can be studied in detail
individually through numerical simulations.  However, it is still not
possible to take into account all of these processes acting
simultaneously in a system with many embryos, to try to model the
evolution of planetary systems from birth to final state. This would
require very complex global simulations that cannot yet be performed.
Also, for planetary systems, the evolutionary outcome, in the form of
planetary masses and orbital configuration, depends very much on the
initial conditions. The amount of mass present in the disk when
planets begin to form is important in relation to gravitational
instabilities, solid core masses that can be accumulated and migration
rates. Improved understanding will come as more powerful computers
become available and observations put more constraints on realisable
initial conditions.

\section*{Acknowledgments}

This work was supported in part by the European Community's Research
Training Networks Programme under contract HPRN--CT--2002--00308,
``PLANETS''.  CT acknowledges partial support from the Institut
Universitaire de France.

\section*{APPENDIX}

\section*{Disk-protoplanet torques and angular momentum conservation}

We here consider a protoplanet in, either, a fixed circular orbit, or
migrating in a circular orbit with slowly varying radius, that is
embedded in and interacting with a gaseous disk.  We show that under
very general assumptions, the torque exerted between protoplanet and
disk can be expressed as a combination of advective wave and viscous
fluxes evaluated far from the protoplanet, together with a
contribution accounting for the rate of change of angular momentum of
the disk domain in the neighbourhod of the protoplanet when conditions
are non stationary. In this way, the contributions relevant to
different types of migration can be identified.

\noindent In order to proceed, we need to consider the problem set up
and distinguish between disk and protoplanet. It has been a common
practice to treat the protoplanet as providing a forcing of the disk
with a softened point mass potential. In that case, one may regard all
the gas as belonging to the disk and identify the protoplanet mass as
that associated with the potential, but only as long as the gas mass
within a softening lengthscale of the protoplanet centre is negligible
compared with that of the protoplanet.  Another possibility would be
to regard the protoplanet as residing interior to some domain within
its Roche lobe.  In that case, we consider the outer disk material to
be unable to enter this domain or transmit angular momentum interior
to it. Thus, the protoplanet itself does not change its mass or angular
momentum through interacting with disk material at its boundary. But
note that when this domain is well inside the Roche lobe, mass and
angular momentum can still accumulate inside the Roche lobe but
exterior to the protoplanet.

\noindent Adopting cylindrical coordinates $(r, \varphi, z)$ with
origin at the central star of mass $M_\star,$ we consider a
protoplanet orbiting in a disk with uniform angular velocity $\Omega
{\bf {\hat k}},$ where ${\bf {\hat k}}$ is the unit vector in the
vertical $z$ direction, relative to an inertial frame.

\noindent The basic equations governing the disk-protoplanet
interaction are the conservation of mass and momentum in the form:

\begin{equation} {{\partial \rho }\over {\partial t}} = -\nab 
\cdot(\rho {\bf v}), \end{equation}

\noindent and

\begin{equation} {{\partial {\bf v}} \over {\partial t}}+ {\bf v}\cdot 
\nab {\bf v} +2\Omega {\bf {\hat k}} \tim {\bf v} = -{\nab P \over
\rho} - {\bf f} -\nab \psi_0 -\nab \psi_1 , \label{AMOT}
\end{equation}

\noindent respectively.  Here, $\rho$ and ${\bf v} = (v_r,
v_{\varphi}, v_z)$ are the disk density and velocity, respectively.
The pressure is $P$, $\psi_0 = -GM_\star/\sqrt{r^2+z^2} -\Omega^2
r^2/2 $ is the gravitational potential due to the central object
together with the centrifugal potential and $\psi_1$ is the
gravitational potential due to the orbiting protoplanet.  

\noindent The force per unit volume $\rho {\bf f} = \rho( f_r,
f_{\varphi}, f_z)$ is taken to be the divergence of a stress tensor,
$t_{ij}.$ Thus, in Cartesian coordinates ${\bf r} \equiv (x_1, x_2,
x_3)$ for which ${\bf f} = (f_1,f_2,f_3),$ $\rho f_i = \partial
t_{ij}/ \partial x_j,$ where we have used the Einstein summation
convention.  Different effects may contribute to $t_{ij},$ including
Navier Stokes viscosity in a laminar disk, self--gravity in a massive
disk, or Maxwell stresses in a magnetised disk.

\subsection*{Conservation of angular momentum:}

We may derive the relationship between the total torque acting between
disk and protoplanet and wave and advected angular momentum fluxes in
the disk by considering global angular momentum conservation. To do
this, we start from the local conservation expressed by the azimuthal
component of the equation of motion~(\ref{AMOT}) in the form:

\begin{equation} 
{{\partial (r v_{\varphi})} \over {\partial t}}+ {\bf v} \cdot \nab(r
v_{\varphi}) +2\Omega r v_{r} = -{1\over \rho}{\partial P \over
\partial \varphi} - r f_{\varphi} -{\partial \psi_1\over \partial
\varphi}. \label{AMOT00} \end{equation} 

\noindent On multiplying by $\rho r$ and integrating over a fixed
domain ${\cal H},$ including the protoplanet but with a non
interacting core cut out if needed, we obtain a statement of the total
angular momentum conservation which, for an assumed steady state,
takes the form:

\begin{equation}
T = \Delta (F_{Re}+F_{str}). \label{jcon0} \end{equation} 

\noindent
Here, $T$ is the total torque acting between protoplanet and disk:

\begin{equation} 
T = -\int_{\cal H}\rho {\partial \psi_1\over \partial
\varphi}rdzd\varphi dr .\end{equation} 

\noindent $\Delta (F_{Re}+F_{str}) $ denotes the difference of the
total flux $F_{Re} + F_{str}$ (see below) evaluated on cylindrical
bounding surfaces exterior to and interior to the protoplanet orbit,
respectively.

\noindent $F_{Re}$ is the anglar momentum flux associated with
Reynolds stresses:

\begin{equation} F_{Re} = \int_{\Sigma}\rho v_r (v_{\varphi}+r\Omega)
r^2 dz d\varphi, \label{FADV0} \end{equation}

\noindent with the integral being taken over a cylindrical bounding
surface, $\Sigma$, assumed distant from the protoplanet. $ F_{str}$ is
the contribution from the stress term:

\begin{equation} F_{str} =-\int_{\Sigma}t_{r \varphi }r^2dz d\varphi .
\end{equation}

\noindent The mass flow through a cylindrical surface $\Sigma',$
normal to a particular radius $r$ in the disk, is given by:

\begin{equation}{\dot M} = \int_{\Sigma'}\rho v_r rdz d\varphi  . 
\end{equation}

\noindent Using this in equation~(\ref{FADV0}), we obtain:

\begin{equation} F_{Re} = \int_{\Sigma}\rho v_r 
(v_{\varphi}-{\overline v_{\varphi}})r^2dz d\varphi 
+{\dot M}({\overline v_{\varphi}}+r\Omega)r.
\label{FADV01} \end{equation}

\noindent Here the overline denotes azimuthal and vertical averaging
with weight $\rho.$ The first term can be identified as a wave or
viscous flux (see for example Papaloizou and Lin 1984), as can
$F_{str}.$ The second term may also be written as ${\dot M}j,$ with
$j$ being the averaged specific angular momentum as viewed from the inertial frame.  
We shall assume the boundaries are far enough away from the protoplanet
that $j$ may be approximated as the Keplerian value. Note that, in a
steady state, ${\dot M}$ is a constant.
Further. turbulent or fluctuating states may be dealt with through
appropriate time averaging and we assume this to be carried out where needed.  Equation~(\ref{jcon0}) then
gives the torque exerted on the disk by the protoplanet as:

\begin{equation}
T = {\dot M} \Delta ( j) + \Delta ( F_{Re} + F_{str})   .
\end{equation} 

\noindent In this case, the torque can be expressed as the difference
between boundary wave and viscous flux terms together with the
difference between direct boundary output and input associated with
the constant mass flux.  Note that, when the protoplanet mass is
reduced to zero, $T = 0$ and we have ${\dot M} \Delta ( j) + \Delta (
F_{str0}) = 0.$ Here $F_{str0}$ is the boundary viscous stress in the
limit of vanishing protoplanet mass.  This is just a statement that
the advected angular momentum flow in a standard steady state 
disk is balanced by torques exerted through viscous, magnetic or gravitational stresses.  Thus,
when a protoplanet is present, we have:

\begin{equation}
T = \Delta ( F_{Re} + F_{str} - F_{str0}) .
\end{equation}   

\noindent We see that, in this case, the torque exerted by the
protoplanet on the disk can be expressed entirely in terms of boundary
flux terms.

\subsection*{ A migrating protoplanet:}

To consider a migrating protoplanet, we suppose now that the angular
velocity $\Omega$ is a function of time, as is the orbital radius of
the protoplanet $R(t).$ We adopt a coordinate system such that the
protoplanet appears stationary. Thus, this rotates with the time
dependent angular velocity $\Omega,$ while the other coordinate
directions scale with the orbital radius $R(t).$ When conditions are
quasi-steady, this corresponds to a self--similar solution.

\noindent We define new dimensionles coordinates $x_i' = x_i/R(t).$ 
 We also adopt a scaled dimensionless time coordinate $\tau,$ which is
such that $d\tau = \Omega dt.$

\noindent Further, we  adopt new dimensionless  velocity components given by: 

\begin{equation} {\bf v'} = \frac{{\bf v}}{R\Omega} - 
{{\bf  r}\over R^2\Omega}{dR\over dt}. \end{equation}

\noindent In terms of the new variables, the continuity equation can
be written:

\begin{equation} {1\over \rho_3}\left({{\partial \rho_3 }\over 
{\partial \tau}}+ {\bf v'}\cdot \nab' \rho_3 \right)= 
{1\over \rho_3}{D\rho_3 \over D\tau}  = - \nab' \cdot({\bf v'}) ,
\label{simcont}\end{equation}

\noindent where $\rho_3 = \rho R^3/R_0^3 ,$ where $R_0$ is a constant
with the dimensions of length that
can be taken to be the value of $R$ at some time $t = t_0$
that can be chosen for convenience  and the prime in $\nab'$
denotes differentiation with respect to $x_i'.$
 
 \noindent Similarly, the eqation of motion becomes:

\begin{eqnarray}
{1\over R^2\Omega}{D\over D\tau}\left(R^2\Omega {\bf v}'\right)
  +  & & 2({\bf {\hat k}} \tim {\bf v}')  +  
{({\bf {\hat k}}\tim {\bf {\hat
 r}})r'\over \Omega^2 R^2} {d(R^2\Omega) \over dR}{dR\over dt} + {{\bf
 r}' \over \Omega^2 R}{d^2 R\over dt^2} = \nonumber \\
& & -{1 \over R^2 \Omega^2}\left(\nab'(\psi_0+\psi_1)+{1\over \rho} 
\nab' P +{\bf f}{ R}\right).
\label{simmot} \end{eqnarray}

\noindent Note that we here assume that $\Omega$ is a function of $R.$
In fact, we shall use the Keplerian form $\Omega^2 = GM_\star/R^3.$ In
addition, we shall adopt $dR/dt = \epsilon R\Omega,$ where $\epsilon$
is a constant of small magnitude corresponding to a slow radial
migration.  In that case, when the pressure and ${\bf f}$ can be
scaled appropriately, it is possible to seek a steady solution of
equations~(\ref{simcont})--(\ref{simmot}) for the primed state variables
that is independent of $\tau.$ When the equation of state is $P =\rho
c_s^2,$ with $c_s$ being the sound speed, assumed to be $\propto
r^{-1/2},$ the stress tensor corresponding to ${\bf f}$ scales as $P,$
and the gravitational softening length scales as $R$ if it is being
used, $R$ may be scaled out of the basic equations thus enabling a
steady state to be sought.

\noindent We comment that the last two terms on the left hand side of
equation~(\ref{simmot}) arise from the transformation to an
expanding/contracting rotating frame. In this frame, they appear as the
additional forces/torques required to induce the apparent modified disk
flow.

\subsection*{Conservation of angular momentum:}

As in the fixed orbit case, we formulate the global conservation of
angular momentum derived from equation~(\ref{simmot}). On multiplying
by $\rho_3 r'$ and integrating over a fixed domain ${\cal D}$, which is
similar to ${\cal H}$ but is defined in the primed coordinates, the
azimthal component of equation~(\ref{simmot}) gives the total angular
momentum conservation in the form:

\begin{equation}
\Omega{d J\over d\tau} +\Delta (F_{Re}+F_{str})= T. \label{jcon}
\end{equation} 

\noindent Here the total torque acting between protoplanet and disk is:

\begin{equation} T = -R_0^3\int_{\cal D}\rho_3 {\partial \psi_1\over 
\partial \varphi'}r'dz'd\varphi'dr' .\end{equation}

\noindent The angular momentum content within ${\cal D},$ as seen in
the inertial frame, is:

\begin{equation} J ={R^2 \Omega  R_0^3} 
\int_{\cal D}\rho_3(v_{\varphi}'+r')r'^2dr'dz'd\varphi' ,
\label{Jang} \end{equation}

\noindent while the anglar momentum flux associated with Reynolds
stresses is:

\begin{equation} F_{Re} ={R^2 \Omega^2  R_0^3} \int_{\cal S}
\rho_3v_r'(v_{\varphi}'+r')r'^2dz'd\varphi', \label{FADV}
\end{equation}

\noindent with the integral being taken over a bounding surface $S,$
assumed distant from the protoplanet amd fixed in the primed coordinates.
 The contribution from the
stress term is:

\begin{equation} F_{str} =-{R^3} 
\int_{\cal S}t_{r',\varphi'}r'^2dz'd\varphi' . \end{equation}

\noindent As before, the mass flow through a cylindrical surface
normal to a particular radius, $r',$ in the disk
 which remains fixed in the primed coordinates, is given by:

\begin{equation}{\dot M} = \Omega R_0^3\int_{\cal S}
\rho_3 v_r' r' dz' d\varphi'. \end{equation}

\noindent Using this in equation~(\ref{FADV}), we obtain:

\begin{equation} F_{Re} ={R^2 \Omega^2  R_0^3} \int_{\cal S}
\rho_3v_r'(v_{\varphi}'-{\overline v_{\varphi}}')r'^2dz'd\varphi' 
+{\dot M}{R^2 \Omega }({\overline v_{\varphi}}'+r')r'.
\label{FADV1} \end{equation}

\noindent Here the overline denotes azimuthal and vertical averaging
with weight $\rho_3.$ The first term can be identified as a wave or
viscous flux, as can $F_{str}.$ The second term may also be written as
${\dot M}j,$ with $j$ being the averaged specific angular momentum.
Again we assume the boundaries to be sufficiently far from
the protoplanet that $j$ may be approximated as Keplerian.
In general, ${\dot M}$ is a function of both $r'$ and $\tau$ or time. However, in
a putative steady state, as seen in the corotating
comoving coordinate system (primed variables independent of $\tau$),  it is a function of time only.

\noindent Using the above, equation~(\ref{jcon}) gives the torque
exerted on the disk by the protoplanet as:

\begin{equation}
T = \Omega{d J\over d\tau} +\Delta ({\dot M} j) + \Delta F_{wv}   ,
\end{equation} 

\noindent where we have combined the viscous and wave fluxes in
$F_{wv}.$ Thus the torque can be expressed as the difference between
boundary wave and viscous flux terms together with the difference
between direct output and input associated with the mass throughput in
combination with the rate of growth of the angular momentum contained
in the domain ${\cal D}.$ It is the latter two terms that are
responsible for the coorbital torques.  When the angular momentum
change associated with mass throughput is not accounted for by an
increase or decrease in the angular momentum content of the domain traversed, it
is accounted for by the protoplanet disk torque, which means that
there has to be angular momentum exchange with the protoplanet.

\noindent Let us look at this more closely. Although conclusions do not depend
on this, for simplicity  we consider an assumed  steady state
self--similar case, then ${\dot M}$ does not vary in space.  But note that, from
equation~(\ref{Jang}), because of the dependence on $R$ and $\Omega,$
$\Omega dJ/d\tau = J/(2R)(dR/dt) \ne 0.$ In this case we have:

\begin{equation}
T = { J\over 2R}{dR\over dt} + {\dot M} (\Delta j) + \Delta F_{wv}.
\end{equation}

\noindent Suppose now viscous effects are small, so that ${\dot M}$ as
viewed from the comoving coordinate frame we use is due to the flow of
matter induced by the migration. When the mass of the protoplanet is
negligible, but the migration maintained, we expect that, as then
there are no excited waves and $T=0$:

\begin{equation}
{ J_0\over 2R}{dR\over dt} + {\dot M} (\Delta j) =0. \label{BALO}
\end{equation} 

\noindent Here $J_0$ is the angular momentum content of ${\cal D}$ in
this limit. Using this in equation~(\ref{BALO}) gives:

\begin{equation}
T = { (J - J_0)\over 2R}{dR\over dt} + \Delta F_{wv}.
\end{equation}

\noindent Now, conservation of angular momentum of the planetary orbit
gives:

\begin{equation}
T = - { J_p\over 2R}{dR\over dt},
\end{equation} 

\noindent where $J_p$ is the angular momentum of the protoplanet.
Thus we have:

\begin{equation}
{( J_p -(J_0-J))\over 2R}{dR\over dt} =  -\Delta F_{wv}.
\end{equation} 

\noindent Thus, as far as the orbital evolution is concerned, the
protoplanet orbital angular momentum is reduced by the {\it coorbital
angular momentum deficit} $J_0 - J.$ Other things being equal, this is
expected to be positive if the protoplanet generates a partial gap, and
in that case,  once the effect of any asymmetric wave torques is taken into account, is clearly related to the {\it coorbital mass deficit}
considered in section~(\ref{coorb}.  There, it leads to a
positive feedback enhancing the migration. On the other hand, if $J_0
-J$ is negative on account of excess material accreted in the vicinity
of the planet increasing the angular momentum content interior to the
Roche lobe, then the reverse effect of negative feedback occurs.

\noindent Note further that, if the structure interior to the Roche
lobe is quasi steady, and if the total mass of material there,  not
counted as part of the protoplanet,  is small compared to
that of the protoplanet, only a small contribution to the
disk--protoplanet torque is expected from these regions.

\References

\item[] Adams F C and Laughlin G 2003 {\it Icarus} {\bf 163} 290

\item[] Alibert Y, Mordasini C and Benz W 2004 {\it
Astron. Astrophys.} {\bf 417} L25

\item Alonso R, Brown T M, Torres G \etal 2004 {\it Astrophys. J.}
{\bf 613} L153

\item[] Artymowicz P 1992 {\it PASP} {\bf 679} 769

\item[] Artymowicz P 1993 {\it Astrophys. J.} {\bf 419} 155

\item[] Artymowicz P 2004 {\it Astr. Soc.  Pac.} {\bf 324}  39

\item[] Backer D C, Foster R S and Sallmen S 1993 {\it Nature} {\bf
365} 817

\item[] Balbus S A and Hawley J F 1991 {\it Astrophys. J.} {\bf 376}
214

\item[] \dash 1998 {\it Rev. Mod. Phys.} {\bf 70} 1

\item[] Barge P and Sommeria J 1995 {\it Astron. Astrophys.} {\bf 295}
L1	

\item[] Bate M R, Lubow S H, Ogilvie G I and Miller K A 2003 {\it
Mon. Not. R. Astron. Soc.}  {\bf 341} 213

\item[] B\'ejar V J S, Mart\'{\i}n E L, Zapatero Osorio M R, Rebolo R,
Barrado y Navascu\'es D, Bailer--Jones C A L, Mundt R, Baraffe I,
Chabrier C and Allard F 2001 {\it Astrophys. J.} {\bf 556} 830

\item[] Bell K R and Lin D N C 1994 {\it Astrophys. J.} {\bf 427} 987

\item[] Bennett D P and Rhie S H 1996 {\it Astrophys. J.} {\bf 472}
660

\item[] Binney J and Tremaine S 1987 {\it Galactic Dynamics}
(Princeton University Press)

\item[] Bodenheimer P, Hubickyj O and Lissauer J J 2000 {\it Icarus}
{\bf 143} 2 

\item[] Bodenheimer P and Pollack J B 1986 {\it Icaru} {\bf 67} 391

\item[] Bond I A, Udalski A, Jaroszy\'nski M \etal 2004 {\it
Astrophys. J.} {\bf 606} L155

\item[] Boss A P 1998 {\it Astrophys. J.} {\bf 503} 923

\item[] \dash 2000 {\it Astrophys. J.} {\bf 536} L101

\item[] Bryden G, Chen X, Lin D N C, Nelson R P and Papaloizou J C B
1999 {\em Astrophys. J.} {\bf 514} 344

\item[] Burgasser A J, Kirkpatrick J D, McGovern M R, McLean I S,
Prato L, Reid I N 2004 {\it Astrophys. J.} {\bf 604} 827

\item[] Cameron A G W 1978 {\it Moon Planets} {\bf 18} 5

\item[] Chabrier G, Saumon D, Hubbard W B and Lunine J I 1992 {\it
Astrophys. J.}  {\bf 391} 817

\item[] Chambers J E, Wetherill G W and Boss A P 1996 {\it Icarus}
{\bf 119} 261

\item[] Chambers J E and Wetherill G W 1998 {\it Icarus} {\bf 136} 304

\item[] Charbonneau D, Brown T M, Latham D W and Mayor M 2000 {\it
Astrophys. J.}  {\bf 529} L45

\item[] Charbonneau D, Brown T M, Noyes R W and Gilliland R L 2002
{\it Astrophys. J.}  {\bf 568} 377

\item[]  Charbonneau D \etal 2005 {\it
Astrophys. J.} {\bf 626} 523

\item[] Cox J P and Giuli R T 1968 {\it Principles of Stellar
Structure: Physical Principles} (New York: Gordon and Breach)

\item[] Cuzzi J N, Dobrovolskis A R and Champney J M 1993 {\it Icarus}
{\bf 106} 102

\item[] Cuzzi J N, Hogan R C, Paque J M., Dobrovolskis A. R., 2001,
 ApJ,  546, 496.

\item[] D'Angelo G,  Bate  M and  Lubow S 2005 {\it Mon. Not. R. Astr. Soc.}
{\bf 358} 316

\item[] D'Angelo G, Kley W and Henning T 2003 {\it Astrophys. J.} 
{\bf 586} 540

\item[] Deming D, Seager S, Richardson L J and Harrington J {\em
Nature} 2005 {\bf 434} 740

\item[] Durisen R H 2001 {\it Proc. IAU Symp. 200 The Formation of
Binary Stars (Potsdam)} ASP Conf. Series vol~200 ed H~Zinnecker and
R~D~Mathieu (San Francisco) p~381

\item[] Durisen R H, Cai K, Mej\'{\i}a A C and Pickett M K 2005 {\it
Icarus} {\bf 173} 417

\item[] Eggenberger A, Udry S and Mayor M 2004 {\it
Astron. Astrophys.}  {\bf 417} 353

\item[] Fischer D A, Marcy G W, Butler R P, Vogt S S, Frink S and Apps
K 2001 {\it Astrophys. J.}  {\bf 551} 1107

\item[] Fischer D A, Marcy G W, Butler R P, Vogt S S, Henry G W,
Pourbaix D, Walp B, Misch A A and Wright J T 2003 {\it Astrophys. J.}
{\bf 586} 1394

\item[] Gaudi B S, Albrow M D, An J \etal 2002 {\it Astrophys. J.}
{\bf 566} 463

\item[] Gilliland R L, Brown T M, Guhathakurta P,
Sarajedini A, Milone E F \etal 2000 {\it Astrophys. J.} {\bf 545} L47

\item[] Goldreich P 1965 {\it Mon. Not. R. Astron. Soc.}  {\bf 130}
159

\item[] Goldreich P and Lynden--Bell D 1965 {\it
Mon. Not. R. Astron. Soc.}  {\bf 130} 125

\item[] Goldreich P Lithwick Y and Sari R 2004 {\it Ann. Rev. Astron. Astrophys.}{\bf 42} 549

\item[] Goldreich P and Tremaine S 1979 {\it Astrophys. J.} {\bf 233}
857

\item[] Goldreich P and Tremaine S 1980 {\it Astrophys. J.} {\bf 241} 425

\item[] Goldreich P and Ward W R 1973 {\it Astrophys. J.} {\bf 183}
1051

\item[] Gonzalez G 1997 {\it Mon. Not. R. Astron. Soc.}  {\bf 285} 403

\item[] Goodman J and Pindor B 2000 {\it Icarus} {\bf 148} 537

\item[]	Gould A and Loeb A 1992 {\it Astrophys. J.} {\bf 396} 104

\item[] Halbwachs J L, Mayor M and Udry S 2005 {\it Astron. Astrophys.}
{\bf 431} 1129

\item[]
Hayashi C 1981 {\it Prog. Theor. Phys. Suppl.} {\bf 70} 35

\item[] Heemskerk M H M, Papaloizou J C and Savonije G J 1992 
{\it Astron. Astrophys.} {\bf 260} 161

\item[] Ida S and Makino J 1993 {\it Icarus} {\bf 106} 210

\item[] Ikoma M, Nakazawa K and Emori H 2000 {\it Astrophys. J.} {\bf
537} 1013

\item[] Ivanov P B, Papaloizou J C B and Polnarev A G 1999 {\it
Mon. Not. R. Astron. Soc.}  {\bf 307} 79

\item[] Joshi K J and Rasio F A 1997 {\it Astrophys. J.} {\bf 479} 948

\item[] Kenyon S J  and Bromley B C 2004  {\it Astron. J.} {\bf 127} 513 

\item Klahr H and Henning T 1997 {\it Icarus} {\bf 128} 213

\item[] Kley W 1999 {\it Mon. Not. R. Astron. Soc.}  {\bf 303} 696

\item[] Kley W, Peitz J and Bryden G 2004 {\it Astron. Astrophys.} {\bf 414} 735

\item[]Kokubo E and  Ida S 1998 {\it Icarus} {\bf 131} 171

\item[] Koller J, Li H and Lin D N C 2003 {\it Astrophys. J.} {\bf
596} L91

\item[] Korycansky D G and Pollack J B 1993 {\it Icarus} {\bf 102} 105

\item[] Kuiper G P 1949 {\it Astrophys. J.} {\bf 109} 308

\item[] Laughlin G and Adams F C 1997 {\em Astrophys. J.} {\bf 491}
L51

\item[] Laughlin G and Bodenheimer P 1994 {\it Astrophys. J.} {\bf ApJ
436} 335

\item[] Laughlin G, Steinacker A and Adams F C 2004 {\em
Astrophys. J.} {\bf 608} 489

\item[] Lee M H and  Peale S J 2002 {\it  Astrophys. J.}  {\bf 567} 59

\item[] Lin D~N~C 1997 {\it Proc. IAU Colloq. 163: Accretion phenomena
and related outflows} ASP Conf. Series vol~121 ed D~T Wickramasinghe
\etal (San Francisco: ASP) p~321

\item[] Lin D N C, Bodenheimer P and Richardson D C 1996 {\it Nature}
{\bf 380} 606

\item[] Lin D~N~C and Papaloizou J C B 1979 {\it
Mon. Not. R. Astron. Soc.}  {\bf 186} 799

\item[] \dash 1986 {\it Astrophys. J.} {\bf 309} 846

\item[] \dash 1993 {\it Protostars and Planets III} ed E~H Levy and
J~I Lunine (Tucson: University of Arizona Press) p~749

\item[] Lin D~N~C, Papaloizou J C B, Terquem C, Bryden G and and Ida S
2000 {\em Protostars and Planets IV} ed V. Mannings \etal (Tucson:
University of Arizona Press) p~1111

\item[] Lissauer J J 1993 {\it Annu. Rev. Astron. Astr.} {\bf 31} 129

\item Low C and Lynden--Bell D 1976 {\it Mon. Not. R. Astron. Soc.}
{\bf 176} 367

\item[] Lubow S H, Seibert M and Artymowicz P 1999 {\it Astrophys. J.}
{\bf 526} 1001

\item[] Lucas P W and Roche P F 2000 {\it Mon. Not. R. Astron. Soc.}
{\bf 314} 858

\item[] Lynden--Bell D and Pringle J E 1974 {\it
Mon. Not. R. Astron. Soc.}  {\bf 168} 60

\item[] Malhotra R 1993 {\it Nature} {\bf 365} 819

\item[] Mao S and Paczynski B 1991 {\it Astrophys. J.} {\bf 374} L37

\item[] Marcy G W and Butler R P 1995 {\it 187th AAS Meeting} BAAS
vol~27 p~1379

\item[] \dash 1998 {\it Annu. Rev. Astron. Astr.} {\bf 36} 57

\item[]
Marcy G W, Butler R P, Fischer D, Vogt S S,  Lissauer J J and Rivera E J  2001
{Astrophys. J.} {\bf 556} 296

\item[]
Marcy G W, Butler R P, Fischer D, Vogt S S, Wright J T, Tinney C G and Jones H R A  2005
astro-ph/0505003

\item[] Masset F and Papaloizou J C B 2003 {\it Astrophys. J.} {\bf
588} 494

\item[] Masset F and Snellgrove M D 2001 {\it Mon. Not. R. Astron. Soc.}
{\bf 320} L55

\item Masunaga H and Inutsuka S I 1999 {\it Astrophys. J.} {\bf 510}
822

\item[] Mayer L, Quinn T, Wadsley J and Stadel J 2004 {\it
Astrophys. J.} {\bf 609} 1045

\item[] Mayor M and Queloz D 1995 {\it Nature} {\bf 378} 355

\item[]
Mayor  M,  Udry S, Naef D,  Pepe F, Queloz D, Santos N C, and Burnet M
2004 {\it Astron. Astrophys.} {\bf 415} 319

\item[]
McArthur B E \etal
2001 {\it Astrophys J.} { \bf 614}  L296

\item[]
Menou K and Goodman J 2004 {\it Astrophys. J.} {\bf 606} 520

\item[] Mizuno H 1980 {\it Prog. Theor. Phys.} {\bf 64} 544

\item[] Motte F and Andr\'e P 2001 {\it Astron. Astrophys.}  {\bf 365}
440

\item[] Murray N, Hansen B, Holman M and Tremaine S 1998 {\it Science}
{\bf 279} 69

\item[] Nakagawa Y, Nakazawa K and Hayashi C 1981 {\it Icarus} {\bf
45} 517

\item[] Nelson A F, Benz W and Ruzmaikina T V 2000 {\it Astrophys. J.}
{\bf 529} 357

\item[] Nelson R P 2005 {\it Astron. Astrophys. } at press
astro-ph/0508486

\item[] Nelson R P and Papaloizou J C B 2004 {\it
Mon. Not. R. Astron. Soc.}  {\bf 350} 849

\item[] Nelson R P, Papaloizou J C B, Masset F and Kley W 2000 {\it
Mon. Not. R. Astron. Soc.}  {\bf 318} 18

\item[] Papaloizou J~C~B 2002 {\it Astron. Astrophys.}  {\bf 388} 615

\item[] \dash 2003 {\it Celest. Mech. Dyn. Astr.}  {\bf 87} 53

\item[] \dash 2005 {\it Celest. Mech. Dyn. Astr.}  {\bf 91} 33

\item[] Papaloizou J~C~B and Larwood J D 2000 {\it
Mon. Not. R. Astron. Soc.}  {\bf 315} 823

\item[] Papaloizou J~C~B and Lin D N C 1984 {\it Astrophys. J.} {\bf
285} 818

\item[] Papaloizou J C B and Nelson R P 2005 {\it Astron. Astrophys.}
{\bf 433} 247

\item[] Papaloizou J C B, Nelson R P and Snellgrove M D 2004 {\it
Mon. Not. R. Astron. Soc.}  {\bf 350} 829

\item[] Papaloizou J C B and Savonije G J 1991 {\it
Mon. Not. R. Astron. Soc.}  {\bf 248} 353

\item[] Papaloizou J C B and Terquem C 1999 {\it Astrophys. J.} {\bf
521} 823

\item[] \dash 2001 {\it Mon. Not. R. Astron. Soc.}  {\bf 325} 221

\item[] Perri F and Cameron A G W 1974 {\it Icarus} {\bf 22} 416

\item[] Perryman M and Hainault O 2005 Extra-solar planets,
ESA-ESO working groups report, chaired by M Perryman and O Hainault,
published by ESA and ESO  http:$//$www.eso.org$/$gen-fac$/$pubs$/$esaesowg$/$espwg$_{-}$report.pdf

\item[] Pickett B K, Cassen P, Durisen R H and Link R 1998 {\it
Astrophys. J.} {\bf 504} 468

\item[] \dash 2000a {\it Astrophys. J.} {\bf 529} 1034 and Erratum
{\it Astrophy\textbf{\textbf{\textbf{}}}s. J.} {\bf 530} 1106

\item[] Pickett B K, Durisen R H, Cassen P and Mejia A C 2000b {\it
Astrophys. J.} {\bf 540} L95

\item[] Pinsonneault M H, DePoy D L and Coffee M 2001 {\it
Astrophys. J.} {\bf 556} L59

\item[] Pollack J B, Hubickyj O, Bodenheimer P, Lissauer~J~J,
Podolak~M and Greenzweig~Y 1996 {\it Icarus} {\bf 124} 62

\item[]  Rafikov  R 2002 {\it Astrophys J.} 
{\bf	572} 566 

\item[] Rasio F A and Ford E B 1996 {\it Science} {\bf 274} 54    

\item[] Rice W K M and Armitage P J 2003 {\it Astrophys. J.} {\bf 598}
L55

\item[] Rice W K M, Lodato G, Pringle J E, Armitage P J and Bonnell I
A 2004 {\it Mon. Not. R. Astron. Soc.}  {\bf 355} 543

\item[] Sadakane K, Ohkubo M, Takeda Y, Sato B, Kambe E and Aoki W
2002 {\it Publ. Astron. Soc. Jpn} {\bf 54} 911

\item[] Safronov V S 1969 {\it Evoliutsiia doplanetnogo oblaka}
(Moscow: Nauka) English translation 1972 {\it Evolution of the
protoplanetary cloud and formation of the Earth and planets} (IPST
Jeerusalem)

\item[] Sandquist E L, Taam R, Lin D N C, Burkert A 1998
{\it Astrophys. J.} {\bf 506} L65

\item[] Sandquist E L, Dokter J J, Lin D N C, Mardling R A 2002
{\it Astrophys. J.} {\bf 572} 1012

\item[] Santos N C, Garc\'{\i}a L\'opez R J, Israelian G, Mayor M,
Rebolo R, Garc\'{\i}a--Gil A, P\'erez de Taoro M R and Randich S 2002
{\it Astron. Astrophys.} {\bf 386} 1028

\item[] Santos N C, Israelian G and Mayor M 2001 {\it Astron. Astrophys.}
{\bf 373} 1019

\item[] Santos N C, Israelian G, Mayor M, Rebolo R and Udry S 2003 {\it 
Astron. Astrophys.} {\bf 398} 363

\item[] Sekiya M 1998  {\it Icarus} {\bf 133} 298

\item[] Shakura N I and Sunyaev R A 1973 {\it Astron. Astrophys.} {\bf 24}
337

\item[] Snellgrove M, Papaloizou J C B and  Nelson R P 2001
{\it Astron.   Astrophys.}  {\bf 374} 1092

\item[] Sozzetti A 2004 {\it Mon. Not. R. Astron. Soc.}  {\bf 354}
1194

\item[] Stapelfeldt K R, Krist J E, Menard F, Bouvier J, Padgett D L
and Burrows C J 1998 {\it Astrophys. J.}  {\bf 502} 65

\item[] Stevenson D J 1982 {\it Plan. Space. Sc.} {\bf 30} 755

\item[] Syer D and Clarke C J 1995 {\it Mon. Not. R. Astron. Soc.}
{\bf 277} 758

\item[] Tanaka H, Takeuchi T and Ward W R 2002 {\it Astrophys. J.}
{\bf 565} 1257

\item[] Tanaka H and Ward W R 2004 {\it Astrophys. J.}  {\bf 602} 388

\item[] Tanga P, Babiano A, Dubrulle B and Provenzale A 1996 {\it
Icarus} {\bf 121} 158

\item[] Terquem C E J M L J 2003 {\it Mon. Not. R. Astron. Soc.}  {\bf
341} 1157

\item[] Terquem C and Papaloizou J C B 2002 {\it
Mon. Not. R. Astron. Soc.}  {\bf 332} 39

\item[] Terquem C, Papaloizou J C B and Nelson R P 2000 {\it From Dust
to Terrestrial Planets} ed W. Benz \etal (ISSI Space Sciences Series 9
reprinted from Space Science Reviews 92) p~323

\item[] Thommes E W, Duncan M J and Levison H F 2002 {\it Astron. J.}
{\bf 123} 2862

\item[] Thorsett S E, Arzoumanian Z, Camilo F and Lyne A G 1999
{\it Astrophys. J.} {\bf 523} 763

\item[] Trilling D E, Benz W, Guillot T, Lunine J I, Hubbard W B and
Burrows A 1998 {\it Astrophys. J.}  {\bf 500} 428

\item[] Trilling D~E, Lunine J I and Benz W 2002 {\em Astron. Astrophys.}
{\bf 394} 241

\item[] Udry S, Mayor M and Santos N C 2003 {\em Astron. Astrophys.}
{\bf 407} 369

\item[] Vidal--Madjar A, D\'esert J M, Lecavelier des Etangs A,
H\'ebrard G, Ballester G E, Ehrenreich D, Ferlet R, McConnell J C,
Mayor M and Parkinson C D 2004 {\it Astrophys. J.}  {\bf 604} L69

\item[] Vidal--Madjar A, Lecavelier des Etangs A, D\'esert J M,
Ballester G E, Ferlet R, H\'ebrard G and Mayor M 2003 {\em Nature}
{\bf 422} 143

\item[] Ward W~R 1986 {\it Icarus} {\bf 67} 164

\item[] \dash 1997 {\it Icarus} {\bf 126} 261

\item[] Ward W R and Hahn J M 1995 {\it Astrophys. J.}  {\bf 440} L25

\item[] Weidenschilling S J 1977 {\it Mon. Not. R. Astron. Soc.}  {\bf
180} 57

\item[] \dash 1980 {\it Icarus} {\bf 44} 172

\item[] \dash 1984 {\it Icarus} {\bf 60} 553

\item[] \dash 2003 {\it Icarus} {\bf 165} 438

\item[] Weidenschilling S J and Marzari F 1996 {\rm Nature} {\bf 384}
619

\item[] Wetherill GW and Stewart GR 1989 {\bf Icarus} {\bf 77} 330

\item[] Wolszczan A. 1996 {\it Proc. IAU Colloq. 160 Pulsars:
problems and progress (Sydney)} ASP Conf. Series vol~105 ed S~Johnston
\etal (San Francisco: ASP) p~91

\item[] Wolszczan A. and Frail D A 1992 {\it Nature} {\bf 355} 145

\item[] Wu Y and Murray N 2003 {\em Astrophys. J.}  {\bf 589} 605

\item[] Wuchterl G 1995 {\it Earth Moon and Planets} {\bf 67} 51

\item[] Youdin A N and Shu F H 2002 {\em Astrophys. J.}  {\bf 580} 494

\item[] Zapatero Osorio M R, B\'ejar V J S, Mart\'{\i}n E L, Rebolo R,
Barrado y Navascu\'es D, Bailer--Jones~C~A~L and Mundt R 2000 {\it
Science} {\bf 290} 103

\item[] Zapatero Osorio M R, B\'ejar V J S, Mart\'{\i}n E L, Rebolo R,
Barrado y Navascu\'es D, Mundt R, Eisl\"offel J and Caballero J A 2002
{\em Astrophys. J.}  {\bf 578} 536

\item[] Zucker Shay and Mazeh T 2002 {\em Astrophys. J.}  {\bf 568}
L113

\endrefs

\end{document}